\RequirePackage{fixltx2e}
\documentclass[english,journal=jacsat, manuscript=article, layout=onecolumn]{achemso}
\usepackage[T1]{fontenc}
\usepackage[utf8]{inputenc}
\usepackage{color}
\usepackage{amsmath}
\usepackage{amssymb}
\usepackage{graphicx}

\makeatletter

\title{Coupling, lifetimes and ``strong coupling'' maps for single molecules
at plasmonic interfaces}

\author{Monosij Mondal}

\affiliation{Department of Chemistry, University of Pennsylvania, Philadelphia
PA 19104, USA}

\email{monosij@sas.upenn.edu}

\author{Maicol A. Ochoa\footnote{Present addresses: Nanoscale Device Characterization Division, National
Institute of Standards and Technology, Gaithersburg, Maryland 20899,
USA and Department of Chemistry and Biochemistry, University of Maryland,
College Park, Maryland 20742, USA.}}

\affiliation{Department of Chemistry, University of Pennsylvania, Philadelphia
PA 19104, USA}

\email{maicol@umd.edu}

\author{Maxim Sukharev}

\affiliation{College of Integrative Sciences and Arts, Arizona State University,
Mesa, AZ 85212, USA}

\altaffiliation{Department of Physics, Arizona State University, Tempe, AZ 85287,
USA}

\email{maxim.sukharev@asu.edu}

\author{Abraham Nitzan}

\affiliation{Department of Chemistry, University of Pennsylvania, Philadelphia
PA 19104, USA}

\altaffiliation{School of Chemistry, Tel Aviv University, Tel Aviv 69978, Israel}

\email{anitzan@sas.upenn.edu}

\newcommand{\lyxdot}{.}

\SectionNumbersOn

\usepackage{achemso}
\usepackage{dcolumn}
\usepackage{amsmath}
\usepackage{mathrsfs}
\usepackage{hyperref}
\usepackage{svg}
\usepackage{bm}        
\usepackage{amssymb}   
\usepackage{float}
\usepackage{color}
\usepackage{enumitem}
\usepackage[normalem]{ulem}
\usepackage{comment}
\usepackage{afterpage}
\usepackage{braket}

\usepackage[tableposition=below]{caption}
\usepackage{xr}
\usepackage{indentfirst}

\DeclareMathOperator\csch{csch}

\newcommand*{\citen}[1]{%
  \begingroup
    \romannumeral-`\x 
    \setcitestyle{numbers}%
    \cite{#1}%
  \endgroup   
}

\interfootnotelinepenalty=\@M

\@ifundefined{showcaptionsetup}{}{%
 \PassOptionsToPackage{caption=false}{subfig}}
\usepackage{subfig}
\makeatother

\usepackage{babel}
\begin{document}
\begin{abstract}
The interaction between excited states of a molecule and excited states
of metal nanostructure (e.g. plasmons) leads to hybrid states with
modified optical properties. When plasmon resonance is swept through
molecular transition frequency an avoided crossing may be observed,
which is often regarded as a signature of strong coupling between
plasmons and molecules. Such strong coupling is expected to be realized
when $2|U|/\hbar\Gamma>1$, where $U$ and $\Gamma$ are the molecule-plasmon
coupling and the spectral width of the optical transition respectively.
Because both $U$ and $\Gamma$ strongly increase with decreasing
distance between a molecule and a plasmonic structure it is not obvious
that this condition can be satisfied for any molecule-metal surface
distance. In this work we investigate the behavior of $U$ and $\Gamma$
for several geometries. Surprisingly, we find that if the only contributions
to $\Gamma$ are lifetime broadenings associated with the radiative
and nonradiative relaxation of a single molecular vibronic transition,
including effects on molecular radiative and nonradiative lifetimes
induced by the metal, the criterion $2|U|/\hbar\Gamma>1$ is easily
satisfied by many configurations irrespective of the metal-molecule
distance. This implies that the Rabi splitting can be observed in
such structures if other sources of broadening are suppressed. Additionally,
when the molecule-metal surface distance is varied keeping all other
molecular and metal parameters constant, this behavior is mitigated
due to the spectral shift associated with the same molecule-plasmon
interaction, making the observation of Rabi splitting more challenging. 
\end{abstract}

\section{Introduction \label{sec:introduction}}

Plasmonic excitations occurring in small nanoparticles provide an
interesting avenue for light and energy manipulation below the diffraction
limit. Collective excitations of the electrons in the nanoparticle
can couple to nearby molecules, and applications in nano-optics\cite{benz2014nanooptics,lin2015nanooptics,karnetzky_towards_2018,kamp_cascaded_2020},
nanosensing\cite{sekhon2011optimal,jeong2016dispersion,anker2010biosensing,choi2011plasmonic,verschueren_nano-optical_2019},
nanocatalysis\cite{de2017plasmonic,tesema2019plasmon,shaik2018plasmon,wu2019photocatalytic,yu2018plasmonic,ochoa2013single,baffou2014nanoplasmonics,huang2019synergy,gelle_applications_2020,zhang_plasmon-driven_2019,nazemi_plasmon-enhanced_2019},
plasmonic photosynthesis\cite{zeng2018plasmon,yu_selective_2019,kontoleta_self-optimized_2019},
quantum information\cite{DelPino2014,bogdanov_overcoming_2019} and
energy-harvesting technologies\cite{kamat2007meeting,mackowski2010hybrid,boghossian2013application,Herrera2016,jiang2016light,li_plasmon-induced_2015}
have emerged.

In standard molecular spectroscopy and photochemistry, the molecule-photon
encounter is essentially described as a scattering process that results
in both species changing their quantum states (eigenstates of their
respective Hamiltonians), with the added provision that the photon
species is not conserved and can be converted to other forms of molecular
energy. The molecule-photon coupling weighted by the density of photon
states is weak, as exemplified by the relatively slow spontaneous
emission rates ($\sim10^{9}$ $\text{s}^{-1}$ for allowed electronic
transitions), although using a high-intensity radiation field can
lead to a strong and often non-linear molecular response. Different
situations are encountered when the molecule interacts with one or
a few localized photon mode(s), as encountered in optical cavities\cite{walther_cavity_2006,aspelmeyer_cavity_2014,ebbesen_hybrid_2016,Ribeiro2018a,frisk_kockum_ultrastrong_2019,basov_polariton_2021}.
Coupling to such modes is strong because of their localized nature,
and the system response is best described in terms of hybrid radiation-matter
states (polaritons) that diagonalize the strongly coupled part of
the Hamiltonian, which include the molecule, these localized modes
and their mutual interaction. The most prominent manifestation of
this hybridization is the avoided crossing (Rabi splitting, $\Omega_{\text{R}}$),
observed in the optical response of such systems when the molecular
transition frequency approaches and crosses a cavity mode frequency.
Importantly, because the cavity photon interacts through the transition
dipole of the whole molecular system, the coupling strength reflected
in the observed splitting depends also on the number of molecules
$N$ and is given by $\Omega_{\text{R}}\sim2g\sqrt{N}$ if the molecular
aggregate is much smaller than the cavity mode wavelength. Here $g$
is the coupling parameter of a dipole ${\bf D}$ (representing the
transition dipole of a single molecule) to the radiative mode, given
for a rectangular cavity of volume $V$ and dielectric constant $\varepsilon=\varepsilon_{r}\varepsilon_{0}$
by $g=D(\hbar\omega/2\varepsilon V)^{1/2}$. By this experimental
measure, “strong coupling” is often defined by the observability of
this spectral structure, namely by the condition $\Omega_{\text{R}}/\Gamma>1$,
where $\Gamma$ is the width of the polaritonic peaks. The possible
consequences of strong coupling for transport phenomena\cite{feist_extraordinary_2015,schachenmayer_cavity-enhanced_2015,garcia-vidal_long-distance_2017,saez-blazquez_organic_2018,groenhof_coherent_2018,Hagenmuller2017,hagenmuller_cavity-assisted_2018,rozenman_long-range_2018,yang_phonon_2020}
and chemical rates\cite{Ribeiro2018a,hutchison_modifying_2012,kowalewski_non-adiabatic_2016,kowalewski_cavity_2016,Herrera2016,feist_polaritonic_2018,herrera_theory_2018,munkhbat_suppression_2018,Martinez-Martinez2018,campos-gonzalez-angulo_resonant_2019,yuen-zhou_polariton_2019,Galego2019,shi_enhanced_2018,du_remote_2019,Thomas2016,Thomas2019,fregoni_strong_2020,lather_cavity_2019,mandal_investigating_2019,mandal_polariton-mediated_2020,Semenov2019,avramenko_quantum_2020,chupeau_optimizing_2020,davidsson_atom_2020,flick_ab_2020,herrera_molecular_2020}
have been subject to intense discussion over the last decade.

Along with Fabri-Pérot type cavities, where the localization length
is limited by the optical mode wavelength and is bounded by the cavity
size, strong coupling phenomena are also observed in interacting plasmon-exciton
systems\cite{barnes_surface_2003,torma_strong_2014,sukharev_optics_2017,vasa_strong_2018}.
In particular, metallic nanostructures offer alternative setups for
localization of electromagnetic modes, where considerably higher dissipation
rates due to losses in the metal constituents are compensated by the
considerably smaller, sub-wavelength localization volumes. Intermetallic
gaps between plasmonic structures are known to support particularly
strong localization. Such localities have been identified as “hotspots”
in studies of surface-enhanced spectroscopies\cite{halas_plasmons_2011,wei_hot_2013,kleinman_creating_2012,stockman_nanoplasmonics_2011},
and are now referred to as plasmonic cavities\cite{perez-gonzalez_optical_2010,Santhosh2016,Hugall2018,aguilar-galindo_electronic_2019,berghuis_lightmatter_2020,rousseaux_strong_2020}.
A cavity-like structure is not, however, an absolute requirement.
It was recently pointed out that strong coupling, as defined above,
may be realized even in the vicinity of single plasmonic particles\cite{trugler_strong_2008,delga_quantum_2014,balci_colloidal_2019,gros_near-field_2018}. 

In addition to the usual spectroscopic implications (i.e. the observed
Rabi splitting), the emergence of strong coupling in plasmonic cavities
and other localities characterized by strong focusing of the electromagnetic
field suggests, as pointed out above, that other dynamical processes
may be modified in such strong coupling situations. In particular,
while photochemical processes can be enhanced or induced at plasmonic
interfaces due to the focused nature of the local EM field\cite{nitzan_theoretical_1981,nitzan_can_1981}
or as a consequence of plasmon-induced generation of hot electrons\cite{narang_plasmonic_2016,zhang_surface-plasmon-driven_2018},
strong coupling induced modification of inter and intra-molecular
interactions may lead to new or modified chemistry even without incident
light\cite{Semenov2019,flick_atoms_2017,schafer_ab_2018,jestadt_light-matter_2019}.

With the latter possibility in mind, in this paper we study the emergence
of light-matter strong coupling near several prototype plasmonic structures,
aiming to map the phenomenon as a function of nanostructure geometry
and molecular position and orientation. This study is motivated by
three related considerations. First, while even in Fabri-Pérot-like
cavities coupling of molecules to the cavity mode(s) depends on the
molecular position in the cavity\cite{ahn_vibrational_2018}, this
dependence is expected to be much more pronounced near plasmonic structures.
Secondly, while in Fabri-Pérot cavities strong coupling can be observed
only with relatively large molecular ensembles, the much smaller volume
of their plasmonic counterparts makes it possible to observe this
phenomenon down to a single nanodot\cite{Santhosh2016,gros_near-field_2018,savasta_nanopolaritons_2010,stete_size-dependent_2018,park_tip-enhanced_2019}
and even a single molecule\cite{trugler_strong_2008,dvoynenko_revisiting_2013,kongsuwan_suppressed_2018,Chikkaraddy2016}.
In the latter case, averaging over molecular orientation is not an
option and orientation dependence should be addressed explicitly.
Yet another important factor is the geometry of the plasmonic structure
that can be used to tune the plasmonic resonance. Here we examine
the dependence of strong coupling as experienced by a single molecule,
represented by a point-dipole and positioned near metallic spheres,
ellipsoids and bispherical dimers, on these structural parameters. 

Finally, as defined, a practical manifestation of “strong coupling”
is a relative concept: The possibility to observe Rabi splitting in
the optical response of a metal-plasmonic structure composite reflects
not only the magnitude of the molecule-radiation field coupling, an
intrinsic property of this composite, but also the linewidth which
can have many origins including spectral connection within the molecule
as well as its interaction with the thermal environment. Limiting
ourselves to lifetime broadening, it is often the case that, at close
proximity to a metal structure, this attribute of the optical response
is also dominated by the molecule-metal interaction, namely by effect
of the metal structure on the radiative and non-radiative relaxation
of the excited molecule. It is therefore useful to examine the relative
magnitudes of the molecule-plasmon coupling strength and the metal
induced lifetime broadening as providing an intrinsic bound to the
presence of strong coupling as manifested by the observability of
strong coupling. A recent observation of Rabi splitting in the zero-phonon
line of a single molecule located in a plasmonic cavity at low temperature\cite{pscherer_single-molecule_2021}
suggests that “strong coupling” may be more pervasive in such systems
and can be observed if other sources of spectral line broadening are
eliminated.

While light-matter interaction has important quantum ramifications,
the information needed to establish the occurrence of strong coupling
can be obtained from classical considerations and we invoke this approach
in the present study. Furthermore, our focus on nanoparticles and
their close vicinity makes it possible to ignore retardation effects
and work in the electrostatic limit. Furthermore, we represent the
molecule as a point-dipole emitter. While such a model is useful for
mapping regions of strong coupling about the considered nanostructures,
it should be kept in mind that any realistic calculation on a specific
molecule should take its finite size and specific structure into account\cite{neuman_coupling_2018}.

The models used in this study and our method of calculation are described
in Section \ref{sec:Method}.\textcolor{red}{{} }Details of the calculations
performed for the different models are provided in Section \ref{sec:response_function_relaxation}
and in the\textcolor{red}{{} }SI. Some representative results are presented
and discussed in Section \ref{sec:results_and_discussions}. Section
\ref{sec:conclusion} concludes.

\section{Models and method \label{sec:Method}}

\begin{figure*}
\subfloat[]{\includegraphics[scale=0.42]{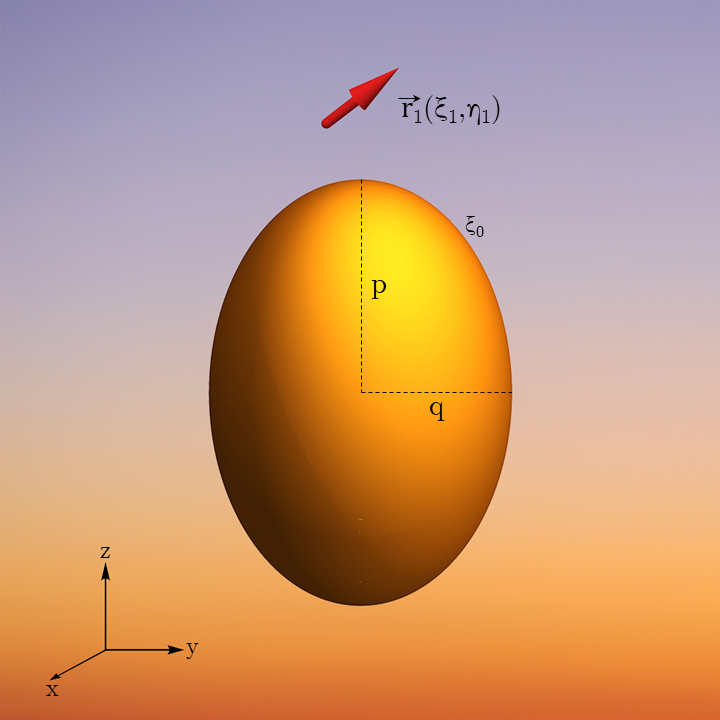}\quad{}

\label{fig:System_Spheroid_Schematic}}\subfloat[]{\includegraphics[scale=0.42]{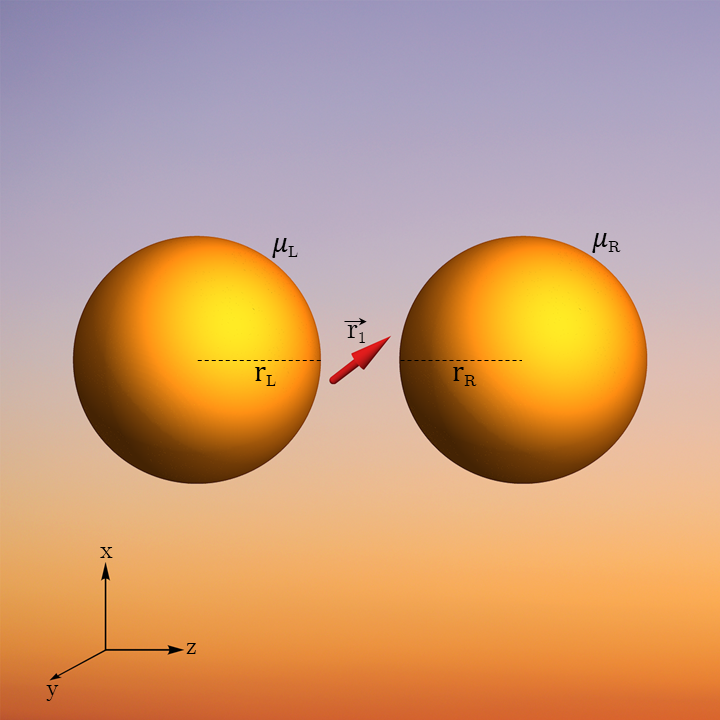}

\label{fig:System_Two_Sphere_Schematic}}

\caption{(a) An emitter, represented by an oscillating dipole, is positioned
(at $\boldsymbol{r_{1}}(\xi_{1},\eta_{1})$ in prolate spheroidal
coordinates; SI Section \ref{SI:spheroid_coord_sys}) near a prolate
spheroidal metal nanoparticle orienting in an arbitrary direction
with respect to the nanoparticle surface. The nanoparticle surface
is described by $\xi_{0}$. The spheroid aspect ratio is the ratio
between the semi-major and the semi-minor axis $p/q$. (b) An emitter
is positioned (at $\boldsymbol{r_{1}}(\mu_{1},\pi)$ in bispherical
coordinates; SI Section \ref{SI:coord_sys_bisph}) on the internuclear
(azimuthal) axis between two nanospheres of radii $r_{L}$ and $r_{R}$
with a gap $d$ in between them, oriented in an arbitrary direction
with respect to the azimuthal axis.}
\label{fig:spheroid_two_sphere_schematic}
\end{figure*}

Figure \ref{fig:spheroid_two_sphere_schematic} displays the models
used in this study. The plasmonic system comprises of either a metallic
ellipsoid (that includes a sphere as a limiting case), shown in Figure
\ref{fig:System_Spheroid_Schematic}, or a metallic bispherical dimer,
shown in Figure \ref{fig:System_Two_Sphere_Schematic} -- systems
for which analytical solutions to the relevant electrostatic boundary
value problem can be obtained. The metal is represented as a continuum
dielectric whose dielectric response function is given in the Drude
form
\begin{align}
\varepsilon_{\text{ns}}(\omega) & =\varepsilon_{b}\Bigg[1-\frac{\omega_{p}^{2}}{\omega(\omega+i\gamma_{p})}\Bigg]\label{eq:drude_function}
\end{align}
where $\varepsilon_{b}$ is the background permittivity from the bound
electrons in the metal, $\omega_{p}$ is the bulk plasma frequency
and $\gamma_{p}$ is a phenomenological damping constant. In the calculations
presented below the following parameters were used: for gold\cite{Derkachova2016}:
$\varepsilon_{b}=9.84$, $\omega_{p}=2.872$ eV$=4.363\times10^{15}$
rad/s, and $\gamma_{p}=0.072$ eV$=1.094\times10^{14}$ rad/s, and
for silver\cite{Yang2015}: $\varepsilon_{b}=5$, $\omega_{p}=3.98$
eV$=6.05\times10^{15}$ rad/s, and $\gamma_{p}=0.039$ eV$=5.93\times10^{13}$
rad/s. These metal particles are embedded in a uniform dielectric
environment with a dielectric constant $\varepsilon_{s}$, taken to
be 1. Other model dielectric functions can, of course, be used. In
particular, to include effects of interband transition in the metal
or the motion of the ionic core, one often considers Drude-Lorentz
type dielectric response functions\cite{Sukharev2017}. The molecule
is taken as a point-dipole characterized by its magnitude and its
position and orientation relative to the plasmonic structure. Finite
molecular size effects can be considered as described in Ref. (\citen{neuman_coupling_2018})
but are disregarded in the present calculation. 

The consequences of coupling between a dipole emitter and a plasmonic
nanostructure can be calculated in terms of the Dyadic Green's function
${\bf \underline{G}}({\bf r},{\bf r^{\prime}},\omega)$ that describes
the response at ${\bf r}$ to a point-current positioned at a given
location ${\bf r^{\prime}}$ and orientation relative to this plasmonic
structure\cite{novotny_principles_2006}. In particular, for a dipole
oscillating at frequency $\omega$ this Green's function yields the
field associated with the oscillating dipole, leading to the interaction
between the dipole and the plasmonic system in the form $\overline{U}=-{\bf D}\cdot{\bf E}$
where ${\bf D}={\bf D}_{12}$ is the transition dipole between the
relevant states 1 and 2 of the molecular emitter and its direction
reflects the molecular orientation in space, and ${\bf E}$ is the
electric field associated with the response of the plasmonic structure
to this dipole, calculated at the dipole position. In the quantum
case ${\bf D}$ and ${\bf E}$ are replaced by their quantum operators
counterparts, but classical considerations are known to be sufficient
for the present purpose\cite{agarwal_quantum_1975,wijnands_greens_1997,Gersten1981,ruppin_lifetime_2004,marocico_effect_2011,Sukharev2014}.
The relaxation induced by the plasmonic nanostructure can be obtained
from the same calculation: at steady state, the energy dissipation
rate can be obtained from the average work per unit time that this
field does on the oscillating dipole. Furthermore for a dipole oscillating
in an environment characterized by a real dielectric response function,
the spontaneous radiative relaxation rate can be described by the
golden rule, $\gamma_{\text{R}}=\frac{2\omega}{3\hbar\varepsilon_{0}}|{\bf D}|^{2}\rho({\bf r},\omega)$
in terms of the local density of states $\rho({\bf r},\omega)=\frac{2\omega}{\pi c^{2}}\text{Im}\left\{ {\bf Tr}\left[{\bf \underline{G}}({\bf r},{\bf r},\omega)\right]\right\} $,
providing an elegant framework for analyzing the effect of an inhomogeneous
dielectric environment on the emission rate (Purcell effect).

In the electrostatic limit and point-dipole emitter model considered
here, this formulation can be made significantly simpler and can take
two forms. First, following Gersten and Nitzan\cite{Gersten1981},
the point-dipole, positioned at ${\bf r}_{\text{1}}$, is assumed
to oscillate with constant amplitude ${\bf D}(\omega)$ (taken as
the transition dipole of the molecular emitter) and frequency $\omega$,
thereby driving the nearby plasmonic structure. The Laplace equation
is solved for given $\varepsilon_{\text{ns}}(\omega)$ (Eq. (\ref{eq:drude_function}))
and $\varepsilon_{s}$ and standard boundary conditions at the surface
of the dielectric nanostructure to yield the electrostatic potential
$\Phi({\bf r},\omega)$ and electric field ${\bf E}({\bf r},\omega)=-\boldsymbol{\nabla}\Phi({\bf r},\omega)$
anywhere in the system (excluding the point occupied by the dipole)
and in particular in the volume occupied by the nanostructure. Furthermore,
the reaction field -- the part ${\bf E}_{r}$ of the field ${\bf E}$
that arises from the polarization induced in the dielectric structure
is obtained as ${\bf E}_{r}({\bf r},\omega)={\bf E}({\bf r},\omega)-{\bf E}_{d}({\bf r},\omega)$
where ${\bf E}_{\text{d}}$ is the field of the bare dipole. Note
that these response functions depend on $\omega$ through the frequency
dependence of $\varepsilon_{\text{ns}}(\omega)$ and that retardation
effects are insignificant for the assumed small system. Also note
that the linear response nature of standard electrostatics implies
that these responses are linear in the driving dipole ${\bf D}$. 

The calculation then proceeds in the following way:
\begin{enumerate}
\item For the given dipole-nanostructure configuration, and a dipole oscillating
with a frequency $\omega$ according to 
\begin{align}
{\bf D}(t) & ={\bf D}(\omega)e^{-i\omega t}+{\bf D}(-\omega)e^{i\omega t}\label{eq:dipole_driving}
\end{align}
the local electric field ${\bf E}(\omega)={\bf E}(-\omega)^{*}$ associated
with the polarization induced by the dipole in the nanostructure (the
reaction field) is calculated. The component $E(\omega)$ of this
field, calculated at the dipole position and in the dipole direction,
is used to define the needed component $G$ of the Green's function:
\begin{align}
E(\omega) & =G(\omega)D(\omega)\label{eq:E_and_G}
\end{align}
This is essentially the electrostatic limit of the relevant component
of the Dyadic Green’s function mentioned above (calculated at the
dipole position) except that the term associated with the bare dipole
field is ignored.
\item The coupling between the dipole emitter and the plasmonic nanostructure,
at a frequency $\omega_{p}$ of a plasmon peak, is estimated using
the following approximate procedure. First, the interaction of the
dipole with the nanostructure is written in the form
\begin{align}
\overline{U} & =-MyD\label{eq:coupling_Ubar_M}
\end{align}
where $y$ is a dimensionless coordinate associated with the nanostructure,
whose deviation from zero describes the polarization of this structure,
and $My$ is the field associated with this polarization at the position
and direction of the dipole. If the deviation of $y$ from zero is
induced by the dipole, then $My=GD$. In the quantum analog of Eq.
(\ref{eq:coupling_Ubar_M}), $M\hat{y}$ is the electric field operator
at the molecular position and $\hat{D}$ is the molecular dipole operator.
The coupling is given by the matrix element between the state $|1,1\rangle$,
in which the molecule is in the ground state $1$ and the field has
one photon, and the state $|2,0\rangle$, where the molecule is in
the excited state $2$ and the field has $0$ photon (with the actual
field amplitude accounted by the parameter $M$):
\begin{align}
U & =-\langle1,1|\hat{U}|2,0\rangle\nonumber \\
 & =-M\langle1|\hat{y}|0\rangle\langle1|\hat{D}|2\rangle=-MD_{12}\label{eq:U_Eq}
\end{align}
Our task is to find $M$. In a vacuum system of volume $V$ 
\begin{align}
M & =\sqrt{\frac{\hbar\omega}{2\varepsilon_{0}V}}\label{eq:expression_M_vac}
\end{align}
It is worth pointing that $My$ is the classical analogue to the quantum
expression of the electric field $iM\left(\hat{a}-\hat{a}^{\dagger}\right)$.
To evaluate $M$ for a given molecule-nanostructure configuration
we assume that the underlying dynamics associated with the plasmonic
peak at $\omega_{p}$ is that of a damped harmonic oscillator 
\begin{align}
\ddot{y}+\gamma_{pl}\dot{y}+\omega_{p}^{2}y & =\frac{2\omega_{0}}{\hbar}MD\label{eq:M_damped_oscillator}
\end{align}
The term on the right\footnote{The factor $2\omega_{0}/\hbar$ on the right side enters to make $y$
dimensionless (distance is measured in units of $\sqrt{m\omega/\hbar}$).} is the force exerted on $y$ by its interaction with the dipole as
derived from Eq. (\ref{eq:coupling_Ubar_M}). $\omega_{0}$ is the
transition frequency of the free emitter. The long time solution of
Eq. (\ref{eq:M_damped_oscillator}) under driving in (\ref{eq:dipole_driving})
is $y(\omega)e^{-i\omega t}+y(-\omega)e^{i\omega t}$ where 
\begin{align*}
y(\omega) & =\frac{2\omega_{0}}{\hbar}\frac{MD(\omega)}{\omega_{p}^{2}-\omega^{2}-i\omega\gamma_{pl}}
\end{align*}
whence
\begin{align}
G(\omega) & =\frac{2\omega_{0}}{\hbar}\frac{M^{2}}{\omega_{p}^{2}-\omega^{2}-i\omega\gamma_{pl}};\nonumber \\
\Big|G(\omega)\Big|^{2} & =\left(\frac{2\omega_{0}}{\hbar}\right)^{2}\frac{M^{4}}{(\omega_{p}^{2}-\omega^{2})^{2}+(\omega\gamma_{pl})^{2}}\label{eq:G_freq_coupling}
\end{align}
The field-dipole coupling parameter $M$ is estimated by using the
Laplace equation to obtain the reaction field ${\bf E}(\omega)$ at
the position of a driving dipole ${\bf D}(\omega)$ near the metal
nanostructure, and fitting $\Big|G(\omega)\Big|^{2}=\Big|{\bf E}(\omega)\cdot{\bf D}(\omega)/D(\omega)^{2}\Big|^{2}$to
the form given by Eq. (\ref{eq:G_freq_coupling}). In Section \ref{sec:results_and_discussions}
we present the so obtained coupling parameter in terms of an effective
volume defined by
\begin{align}
M & =\sqrt{\frac{\hbar\omega_{p}}{2\varepsilon_{0}V_{\text{eff}}}}\label{eq:eff_volm}
\end{align}
\item The total dipole moment of the composite system, ${\bf D}_{\text{tot}}(\omega)$,
is calculated as ${\bf D}_{\text{tot}}(\omega)={\bf D}(\omega)+{\bf D}_{\text{ns}}(\omega)$
where ${\bf D}_{\text{ns}}(\omega)$ is the dipole induced on the
nanostructure which is obtained as the following integral over the
nanoparticle volume $V_{\text{ns}}$:
\begin{align}
{\bf D}_{\text{ns}}(\omega) & =\big(\varepsilon_{\text{ns}}(\omega)-\varepsilon_{s}\big)\varepsilon_{0}\int_{V_{\text{ns}}}{\bf E}({\bf r},\omega)d{\bf r}\label{eq:induced_dipole_ns}
\end{align}
\item The radiative decay rate, modified by the presence of the nanostructure
is obtained from \cite{gersten_theory_2005,Gersten1981}
\begin{align}
\Gamma_{\text{R}}(\omega) & =\frac{\sqrt{\varepsilon_{s}}\omega^{3}}{3\hbar c^{3}4\pi\varepsilon_{0}}\Big|{\bf D}_{\text{tot}}(\omega)\Big|^{2}\label{eq:rad_decay_expression}
\end{align}
Obviously, the ratio $|{\bf D}_{\text{tot}}|{}^{2}/|{\bf D}|^{2}$,
represents the Purcell lifetime modification factor for the present
situation. 
\item The non-radiative relaxation affected by the presence of the dielectric
nanostructure is calculated as the rate of Ohmic heat generation on
the nanostructure\cite{gersten_theory_2005}, given by the following
integral over the nanostructure volume:
\begin{equation}
\Gamma_{\text{NR}}^{\text{ns}}(\omega)=\frac{\varepsilon_{0}\text{Im}[\varepsilon_{\text{ns}}(\omega)]}{2\hbar}\int_{V_{\text{ns}}}\Big|{\bf E}({\bf r},\omega)\Big|^{2}d{\bf r}\label{eq:nonrad_decay_expression}
\end{equation}
In addition to the non-radiative relaxation rate associated with dissipation
in the dielectric nanostructure, an intrinsic molecular component,
denoted $\Gamma_{\text{NR}}^{\text{0}}$, results from the interaction
of the molecule with its thermal environment and, for large molecules,
also with intramolecular relaxation processes. This relaxation is
molecule specific and cannot be quantitatively accounted for in our
generic model. In the calculations reported below, unless otherwise
stated, we have taken this intrinsic molecular relaxation rate to
be equal to the radiative decay rate of the free molecule, implying
that the emission yield of the free molecule is $0.5$.
\item The emission quantum yield of the molecular emitter- plasmonic nanostructure
system is given by
\begin{align}
Y(\omega) & =\frac{\Gamma_{\text{R}}(\omega)}{\Gamma_{\text{R}}(\omega)+\Gamma_{\text{NR}}^{\text{ns}}(\omega)+\Gamma_{\text{NR}}^{\text{0}}(\omega)}\label{eq:q_yield}
\end{align}
\item The strong coupling condition on this level of treatment is given
by
\begin{align}
2U & >\hbar\Gamma=\hbar\left(\Gamma_{\text{R}}+\Gamma_{\text{NR}}^{\text{ns}}+\Gamma_{\text{NR}}^{\text{0}}\right)\label{eq:SC_condition}
\end{align}
with $U$ given by Eq. (\ref{eq:U_Eq}).
\end{enumerate}
The above calculation scheme provides a rather straightforward, albeit
approximate, procedure for mapping strong coupling regions in a system
of coupled nanostructure and molecular emitter, where the driving
frequency $\omega$ is naturally taken to be the relevant molecular
transition frequency $\omega_{0}$. It suffers, however from a drawback,
originating from the fact that since the dipole emitter is assumed
to drive the system, such calculation cannot account for any spectral
shift that is another manifestation of the molecule-radiation field
coupling. The criterion (\ref{eq:SC_condition}), in which the coupling
and relaxation rates are calculated for the driving frequency $\omega$,
may thus be viewed as an approximate criterion for the onset of strong
coupling, but may not fully account for the spectral structure and
dynamics beyond it.

To overcome the latter limitation the calculation should be done self-consistently
(see, e.g., Refs \citen{kirakosyan_surface_2016,pustovit_energy_2014,pustovit_plasmon-mediated_2010}).
Here the system, comprising the molecular emitter and the nanostructure,
is driven by an external oscillating electric field, ${\bf E}(t)={\bf E}(\omega)\cos(\omega t)={\bf E}(\omega)\text{Re}\,e^{-i\omega t}$
with a real amplitude ${\bf E}(\omega)$, that couples to the molecular
emitter. The dynamics of the latter can be represented as a driven
damped harmonic oscillator, which in Fourier space takes the form
\begin{align}
(\omega_{0}^{2}-\omega^{2}-i\omega\Gamma^{0}){\bf D} & =\boldsymbol{\alpha}\omega_{0}^{2}\Big[{\bf E}(\omega)+{\bf \underline{G}}({\bf r}_{1},\omega)\textbf{D}\Big]\label{eq:EOM_damped_HO}
\end{align}
 where $\boldsymbol{\alpha}$ is the static polarizability tensor
and ${\bf \underline{G}}({\bf r},\omega)$ was defined by Eq. (\ref{eq:E_and_G}).
Here $\omega_{0}$ is the transition frequency of the free emitter
while $\Gamma^{0}$ represents the combined radiative, $\Gamma_{\text{R}}^{0}$,
and non-radiative, $\Gamma_{\text{NR}}^{0}$, relaxation rates associated
with this free emitter. For simplicity, we assume that the molecular
polarizability tensor is diagonal, $\boldsymbol{\alpha}=\alpha_{x}{\bf e}_{x}+\alpha_{y}{\bf e}_{y}+\alpha_{z}{\bf e}_{z}$
with ${\bf e}_{j},j=x,y,z$ being unit vectors. Eq. (\ref{eq:EOM_damped_HO})
can then be written for each component of ${\bf D}$ in the form
\begin{align}
(\omega_{0}^{2}-\omega^{2}-i\omega\Gamma^{0})D_{j} & =\alpha_{j}\omega_{0}^{2}\big(E_{j}(\omega)\nonumber \\
 & +\sum_{j'}G_{jj'}({\bf r}_{1},\omega)D_{j}\big)\label{eq:EOM_damped_HO_j}
\end{align}
Here, the static polarizability of the 2-level molecular model is
given by
\begin{align}
\alpha_{j} & =2\frac{|D_{12}^{(j)}|^{2}}{\hbar\omega_{0}}\label{eq:polarizability_j}
\end{align}
where $D_{12}^{(j)}$ is the $j$-th component of the molecular transition
matrix element $\text{\textbf{D}}_{12}$. Making another simplifying
assumption that ${\bf \underline{G}}$ is diagonal (this will be true
for geometries considered in our calculations below), this leads to
\begin{align}
 & D_{j}=\nonumber \\
 & \Bigg[1-\Big(\frac{\omega}{\omega_{0}}\Big)^{2}-i\frac{\omega\Gamma^{0}}{\omega_{0}^{2}}-\alpha_{j}G_{jj}({\bf r}_{1},\omega)\Bigg]^{-1}\alpha_{j}E_{j}(\omega)\label{eq:induced_dipole_j_from_EOM}
\end{align}
The absorption lineshape $A(\omega)$ may be calculated as the work
done on the molecular dipole per unit time, averaged over the driving
period, $\overline{E_{j}(t)\dot{D}_{j}(t)}$. This yields
\begin{align}
 & A(\omega)=\nonumber \\
 & \frac{\omega\Bigg(\frac{\omega\Gamma^{0}}{\omega_{0}^{2}}+\alpha_{j}\text{Im}G_{jj}({\bf r}_{1},\omega)\Bigg)}{\Bigg[1-\Big(\frac{\omega}{\omega_{0}}\Big)^{2}-\alpha_{j}\text{Re}G_{jj}(r_{1},\omega)\Bigg]^{2}+\Bigg[\frac{\omega\Gamma^{0}}{\omega_{0}^{2}}+\alpha_{j}\text{Im}G_{jj}({\bf r}_{1},\omega)\Bigg]^{2}}\label{eq:A_omega}
\end{align}
The $\alpha G$ term in Eq. (\ref{eq:induced_dipole_j_from_EOM})
represents the effect of the nanostructure on the dynamics of the
molecular emitter. Its real and imaginary part lead respectively to
a spectral shift and to modifications in the radiative and nonradiative
relaxation rates. When $G(\omega)$ is characterized by a resonance
structure, e.g. a distinct plasmonic peak as seen in Eq. (\ref{eq:G_freq_coupling})\footnote{Such a peak appears when a pole of the response function is characterized
by an imaginary part that is smaller than its distance from other
poles. Such behavior is manifested by the dipolar plasmon of some
metals and semiconductors. In the electrostatic limit under consideration
these poles depend on the shape, but not the size of the dielectric
nanostructure. An example is shown in the SI, section \ref{subsec_SI:plasmon_freq_spheroids}
for the dipolar plasmons of a prolate spheroid.}, a split peak structure may appear in $A(\omega)$ and will indicate
the onset of strong coupling. The following points should be noted:

(a) The classically calculated absorption lineshape, Eq. (\ref{eq:A_omega})
can be used to calculate the optical response of the molecular dipole-nanostructure
composite as function of frequency, thereby directly looking at the
spectral manifestation of strong coupling, not just at the approximate
condition Eq. (\ref{eq:SC_condition}) for its onset. The needed response
function $G$ is obtained as function of $\omega$ from the solution
of the Laplace equation as described above Eq. (\ref{eq:E_and_G}).
The same calculation also yields the radiative and nonradiative decay
rates, and the quantum yield as functions of $\omega$, as described
above. 

(b) In setting up the model of Eq. (\ref{eq:EOM_damped_HO}), we aimed
to examine the effect of proximity of the molecular emitter to a plasmonic
nanostructure. We have therefore implemented the driving by an external
field as acting only on the molecular emitter. In reality, both the
molecule(s) and nanostructure comprising the system that responds
to the incident field, and the effect of directly driving the nanostructure
by this field should be included. In this case ${\bf E}(\omega)$
on the RHS of Eq. (\ref{eq:EOM_damped_HO}) will be replaced by the
vector sum of the incident field and the field created by the polarization
induced on the nanostructure by the incident field -- an important
ingredient in the electromagnetic theory of surface enhanced spectroscopies.
Consequently, after making the same simplifying assumptions as above,
$E_{j}(\omega)$ in Eq. (\ref{eq:induced_dipole_j_from_EOM}) will
be augmented by the corresponding additive field from that polarization.
This will not change the consequences regarding strong coupling as
determined by the denominator in Eq. (\ref{eq:induced_dipole_j_from_EOM}).

(c) The simple criterion Eq. (\ref{eq:SC_condition}) for strong coupling,
namely the appearance of split peak at the crossing of the molecular
transition and a plasmonic resonance frequency is derived from a model
in which the plasmon is represented, like the molecule in Eq. (\ref{eq:EOM_damped_HO}),
as a damped harmonic oscillator with bilinear coupling between these
oscillators\cite{Sukharev2017}. This is a reasonable approximation
where the distance between the molecular emitter and the plasmonic
structure is large enough to represent the response of the latter
by its dipolar plasmon alone. In the general case, while the condition
for strong coupling is again contained in the properties of the denominator
in Eq. (\ref{eq:induced_dipole_j_from_EOM}), its actual manifestation
depends on the properties of the response function $G$ and can be
determined only numerically. Note that $G$ accounts not only for
plasmonic effects but also for other electrostatic aspects of the
response such as the lightning rod effect.\cite{chen_study_2012}

Whichever level of analysis we aim at, a prerequisite of the calculation
is the evaluation of the response function $G$ as well as the radiative
and non-radiative relaxation rates of the molecular emitter when oscillating
in close proximity to the given nanostructure. The models depicted
in Figures \ref{fig:System_Spheroid_Schematic} and \ref{fig:System_Two_Sphere_Schematic}
admit analytical solutions for these functions. These are described
in Section \ref{sec:response_function_relaxation} and the SI. Using
these solutions we will analyze in Section \ref{sec:results_and_discussions},
for silver and gold particles (as represented by their dielectric
response function) in some select configurations, the emergence of
strong coupling and the molecular emission properties as functions
of geometrical parameters. Section \ref{sec:conclusion} concludes.

\section{Response function, relaxation rates and emission yield\label{sec:response_function_relaxation}}

In this section we present analytical results for the response function
$G$ as well as the radiative and non-radiative relaxation rates for
several geometries associated with the structures displayed in Figures
\ref{fig:System_Spheroid_Schematic} and \ref{fig:System_Two_Sphere_Schematic}
(The case of a molecule near a spherical particle is obviously a limit
of the spheroid results presented below). The solutions of the Laplace
equations are obtained as infinite sums that are computed numerically
by truncating the series while ensuring convergence beyond a desired
accuracy. For quick estimates, we focus on the procedure described
in Eqs. (\ref{eq:E_and_G})-(\ref{eq:SC_condition}). In Section \ref{subsec:self_consistent_method},
we provide some details pertaining to the calculation based on Eq.
(\ref{eq:A_omega}).

\subsection{A Molecular (point) dipole near a prolate spheroidal nanoparticle}

The solution of the Laplace equation for this geometry is best done
in the prolate spheroidal coordinate system (SI, Section \ref{SI:spheroid_coord_sys}).
In this coordinate system the nanoparticle surface is defined by $\xi=\xi_{0}$.
The potentials inside and outside the spheroidal particle (solutions
of the Laplace equation) are written in the form
\begin{align}
\Phi^{\text{in}}(\xi,\eta,\phi) & =\sum_{l=0}^{\infty}\sum_{m=0}^{l}A_{l,m}P_{l}^{m}(\xi)P_{l}^{m}(\eta)\cos m\phi;\nonumber \\
 & \qquad\qquad\qquad\qquad\qquad\qquad\;\xi<\xi_{0}\label{eq:phi_in_spheroid}
\end{align}
\begin{align}
\Phi^{\text{out}}(\xi,\eta,\phi) & =\sum_{l=0}^{\infty}\sum_{m=0}^{l}B_{l,m}Q_{l}^{m}(\xi)P_{l}^{m}(\eta)\cos m\phi\nonumber \\
 & +\frac{1}{4\pi\varepsilon_{0}\varepsilon_{s}}\frac{{\bf D}\cdot({\bf r}-{\bf r}_{1})}{|{\bf r}-{\bf r}_{1}|^{3}};\qquad\xi>\xi_{0}\label{eq:phi_out_spheroid}
\end{align}
$P_{l}^{m}(\xi)$ and $Q_{l}^{m}(\xi)$ are the associated Legendre
functions of the first and second kind respectively. The two foci
of the nanoparticle overlap with that of the coordinate system itself
at $\pm\frac{a}{2}$. Here and below, the dielectric response functions
of the nanoparticle and the environment are taken as $\varepsilon_{d}(\omega)$
and $\varepsilon_{s}$ (assumed frequency independent), respectively.
The dipole ${\bf D}$ is situated at ${\bf r}_{1}(\xi_{1},\eta_{1},0)$
(implying that it is on the $xz$ plane) with an arbitrary orientation
relative to the nanoparticle surface, ${\bf D}=D_{\xi}\hat{\boldsymbol{\xi}}+D_{\eta}\hat{\boldsymbol{\eta}}$
where $\hat{\boldsymbol{\xi}}$ and $\hat{\boldsymbol{\eta}}$ are
the unit vectors along the coordinates $\xi,\eta$ respectively. Following
Ref. (\citen{Gersten1981}), we impose continuity of the potential
$\Phi$ and of the component of the displacement field normal to the
spheroidal surface, $\varepsilon\frac{\partial\Phi(\xi,\eta,\phi)}{\partial\xi}$,
on this surface. To this end the dipole potential (last term of Eq.
(\ref{eq:phi_out_spheroid})) is written in prolate spheroidal coordinates
(Eq. (\ref{eq:phi_out_spheroid_SI}) in the SI)\textcolor{red}{{} }and
imposing these boundary conditions leads to the coefficients $A_{l,m}$
and $B_{l,m}$ as follows:
\begin{align}
A_{l,m} & =H_{l,m}(\xi_{1},\eta_{1})\nonumber \\
 & \times\frac{P_{l}^{m^{\prime}}(\xi_{0})Q_{l}^{m}(\xi_{0})-P_{l}^{m}(\xi_{0})Q_{l}^{m^{\prime}}(\xi_{0})}{\varepsilon_{d}P_{l}^{m^{\prime}}(\xi_{0})Q_{l}^{m}(\xi_{0})-\varepsilon_{s}P_{l}^{m}(\xi_{0})Q_{l}^{m^{\prime}}(\xi_{0})}\label{eq:soln_Alm_spehroid_general}
\end{align}
and
\begin{align}
B_{l,m} & =-H_{l,m}(\xi_{1},\eta_{1})\frac{\varepsilon_{d}-\varepsilon_{s}}{\varepsilon_{s}}\nonumber \\
 & \times\frac{P_{l}^{m}(\xi_{0})P_{l}^{m^{\prime}}(\xi_{0})}{\varepsilon_{d}P_{l}^{m^{\prime}}(\xi_{0})Q_{l}^{m}(\xi_{0})-\varepsilon_{s}P_{l}^{m}(\xi_{0})Q_{l}^{m^{\prime}}(\xi_{0})}\label{eq:soln_Blm_spehroid_general}
\end{align}
where
\begin{align}
 & H_{l,m}(\xi_{1},\eta_{1})\nonumber \\
 & =\frac{K_{l,m}}{2\pi\varepsilon_{0}a}\Bigg[\sqrt{\frac{\xi_{1}^{2}-1}{\xi_{1}^{2}-\eta_{1}^{2}}}Q_{l}^{m^{\prime}}(\xi_{1})P_{l}^{m}(\eta_{1})D_{\xi}\nonumber \\
 & +\sqrt{\frac{1-\eta_{1}^{2}}{\xi_{1}^{2}-\eta_{1}^{2}}}Q_{l}^{m}(\xi_{1})P_{l}^{m^{\prime}}(\eta_{1})D_{\eta}\Bigg]\label{eq:H_lm}
\end{align}
and
\begin{align}
K_{l,m} & =(-1)^{m}\frac{2}{a}(2l+1)(2-\delta_{m0})\left[\frac{(l-m)!}{(l+m)!}\right]^{2}\label{eq:K_lm}
\end{align}
The solution $\Phi^{\text{out}}(\xi,\eta,\phi)$ of Eq. (\ref{eq:phi_out_spheroid})
is related to the response Dyadic ${\bf \underline{G}}$ according
to $-\boldsymbol{\nabla}\Phi({\bf r})={\bf \underline{G}}\cdot{\bf D}$.
In particular, the component of ${\bf \underline{G}}$ in the dipole
direction is evaluated here. To exemplify, for the configuration where
the dipole is placed along the major axis of the nanoparticle and
is oriented perpendicular to the nanoparticle surface, i.e., $\eta_{1}=1$,
${\bf D}=D_{\xi}\hat{\boldsymbol{\xi}}$, the component $G_{\text{sphd}}^{\text{maj,}\perp}$
is obtained in the form (maj\textit{$\equiv$major}, sphd\textit{$\equiv$spheroid}):
\begin{align}
 & G_{\text{sphd}}^{\text{maj,}\perp}\nonumber \\
 & =\frac{2}{\pi\varepsilon_{0}\varepsilon_{s}a^{3}}\sum_{l=0}^{\infty}(2l+1)\nonumber \\
 & \times\frac{(\varepsilon_{d}-\varepsilon_{s})P_{l}(\xi_{0})P_{l}^{\prime}(\xi_{0})Q_{l}^{\prime}(\xi_{1})^{2}}{\varepsilon_{d}P_{l}^{\prime}(\xi_{0})Q_{l}(\xi_{0})-\varepsilon_{s}P_{l}(\xi_{0})Q_{l}^{\prime}(\xi_{0})}\label{eq:V_spheroid_major_perp}
\end{align}
For the same configuration, the non-radiative decay rate associated
with heat production in the spheroid is calculated from Eq. (\ref{eq:nonrad_decay_expression})
which leads to
\begin{align}
 & \Gamma_{\text{NR,sphd}}^{\text{maj,}\perp}\nonumber \\
 & =\frac{D_{\xi}^{2}\text{Im}[\varepsilon_{d}(\omega)]}{\pi\varepsilon_{0}a^{3}\hbar\left(\xi_{0}^{2}-1\right)}\nonumber \\
 & \times\sum_{l=0}^{\infty}\frac{(2l+1)P_{l}(\xi_{0})P_{l}^{\prime}(\xi_{0})Q_{l}^{\prime}(\xi_{1})^{2}}{\Big|\varepsilon_{d}P_{l}^{\prime}(\xi_{0})Q_{l}(\xi_{0})-\varepsilon_{s}P_{l}(\xi_{0})Q_{l}^{\prime}(\xi_{0})\Big|^{2}}\label{eq:NR_decay_rate_spheroid_major_perp}
\end{align}
Also, from Eq. (\ref{eq:induced_dipole_ns}) we calculate the total
dipole in the molecule-dielectric spheroid system. The result is
\begin{align}
 & D_{\text{net}}^{\text{maj,}\ensuremath{\perp}}\nonumber \\
 & =D_{\xi}\left[1-\frac{(\varepsilon_{d}-\varepsilon_{s})\xi_{0}Q_{1}^{\prime}(\xi_{1})}{\varepsilon_{s}\left[\varepsilon_{d}Q_{1}(\xi_{0})-\varepsilon_{s}\xi_{0}Q_{1}^{\prime}(\xi_{0})\right]}\right]\label{eq:dipole_net_spheroid_major_perp}
\end{align}
This can be used with Eq. (\ref{eq:rad_decay_expression}) to calculate
the radiative decay rate. Finally, the emission yield is calculated
from Eq. (\ref{eq:q_yield}). The detailed derivation of Eqs. (\ref{eq:soln_Alm_spehroid_general})-(\ref{eq:dipole_net_spheroid_major_perp})
is provided in the SI. Similar results for several other configurations
are also detailed in the SI.

\subsection{A Molecular (point) dipole between two nanospheres\label{subsec:therory_bisph}}

As in Ref. (\citen{MorsePhilipMcCordandFeshbach1946}), the Laplace
equation for this problem is solved in a bispherical coordinate system
$(\mu,\eta,\phi)$ (SI, Section \ref{SI:coord_sys_bisph}). The nanospheres
of radii $r_{\text{R}}$ and $r_{\text{L}}$ (their surfaces defined
by $\mu_{\text{R}}$ and $\mu_{\text{L}}$ respectively) are characterized
by the dielectric response functions $\varepsilon_{\text{R}}(\omega)$
and $\varepsilon_{\text{L}}(\omega)$ with their centers positioned
on the $z$-axis at $z=a\coth\mu_{i},i=\text{L,R},x=y=0$ (in Cartesian
coordinates) where the poles of the coordinate system are at $z=\pm a$
on the $z$-axis. The dipole is also positioned on this line at $(\mu_{1},\pi,0)$($x_{1}=y_{1}=0$
in Cartesian coordinates).

\textbf{Parallel configuration.} For a dipole oriented along the line
connecting the sphere centers, ${\bf D}=D\hat{\boldsymbol{\mu}}$
where $\hat{\boldsymbol{\mu}}$ is the unit vector corresponding to
the coordinate $\mu$. The azimuthal symmetry of this configuration
simplifies the solution. The solution of the Laplace equation for
$\mu>\mu_{R}>0$, namely inside the right sphere is written in the
form
\begin{align}
 & \Phi_{\text{R}}^{\text{in}}(\mu,\eta)\nonumber \\
 & =\sqrt{\cosh\mu-\cos\eta}\sum_{n=0}^{\infty}A_{R,n}^{\text{in}}e^{-(n+1/2)\mu}P_{n}(\cos\eta)\label{eq:phi_in_R_bisph_parallel}
\end{align}
While for $\ensuremath{\mu<\mu_{L}<0}$, namely inside the left sphere
we have
\begin{align}
 & \Phi_{\text{L}}^{\text{in}}(\mu,\eta)\nonumber \\
 & =\sqrt{\cosh\mu-\cos\eta}\sum_{n=0}^{\infty}B_{L,n}^{\text{in}}e^{(n+1/2)\mu}P_{n}(\cos\eta)\label{eq:phi_in_L_bisph_parallel}
\end{align}
The potential outside both the nanoparticles takes the form
\begin{align}
 & \Phi^{\text{out}}(\mu,\eta)=\sqrt{\cosh\mu-\cos\eta}\nonumber \\
 & \times\sum_{n=0}^{\infty}\Bigg[C_{n}^{\text{out}}e^{-(n+1/2)\mu}+D_{n}^{\text{out}}e^{(n+1/2)\mu}\Bigg]\nonumber \\
 & \times P_{n}(\cos\eta)+\frac{D}{4\pi\varepsilon_{0}\varepsilon_{s}}\hat{\boldsymbol{\mu}}\cdot\boldsymbol{\nabla}_{1}\frac{1}{|{\bf r}-{\bf r}_{1}|}\label{eq:phi_in_out_bisph_parallel}
\end{align}
where the dipole potential is written explicitly (last term in Eq.
(\ref{eq:phi_in_out_bisph_parallel})). The boundary conditions, given
by
\begin{align}
\Phi_{\text{R}}^{\text{in}}(\mu_{\text{R}},\eta) & =\Phi^{\text{out}}(\mu_{\text{R}},\eta)\label{eq:BC_1_bisph_general}\\
\Phi_{\text{L}}^{\text{in}}(\mu_{L},\eta) & =\Phi^{\text{out}}(\mu_{\text{L}},\eta)\label{eq:BC_2_bisph_general}\\
\varepsilon_{R}\frac{\partial\Phi_{\text{R}}^{\text{in}}(\mu,\eta)}{\partial\mu}\Big|_{\mu=\mu_{\text{R}}} & =\varepsilon_{s}\frac{\partial\Phi^{\text{out}}(\mu,\eta)}{\partial\mu}\Big|_{\mu=\mu_{\text{R}}}\label{eq:BC_3_bisph_general}\\
\varepsilon_{L}\frac{\partial\Phi_{\text{L}}^{\text{in}}(\mu,\eta)}{\partial\mu}\Big|_{\mu=\mu_{\text{L}}} & =\varepsilon_{s}\frac{\partial\Phi^{\text{out}}(\mu,\eta)}{\partial\mu}\Big|_{\mu=\mu_{\text{L}}}\label{eq:BC_4_bisph_general}
\end{align}
yield an infinite linear system of equations for the coefficients
(see SI, Section \ref{SI:sol_BC_bisph_para}) that can be solved numerically
after truncating at a desired order $n_{\text{max}}$ to give the
needed coefficients up this order . Expressions for the component
of $\underline{\textbf{G}}$ in the dipole direction and the radiative
and nonradiative rates in terms of these coefficients are provided
in the supplement\textcolor{red}{{} }(see SI, Section \ref{SI:subsec:Parallel-configuration}).

\textbf{Perpendicular configuration.} As before, the molecule is positioned
at $(\mu_{1},\pi,0)$ on the line connecting centers of the two nanoparticles
with parameters defined as in the parallel configuration case. The
environment is described by a frequency independent dielectric constant
$\varepsilon_{s}$ as usual. The molecular transition dipole is taken
to be oriented perpendicular to the line connecting the sphere centers,
given by ${\bf D}=D\hat{\boldsymbol{\eta}}$ where $\hat{\boldsymbol{\eta}}$
is the unit vector of the coordinate $\eta$. As this configuration
no longer has an azimuthal symmetry, the solutions of the Laplace
equation for $\mu>\mu_{R}>0$ (inside the right sphere) and $\ensuremath{\mu<\mu_{L}<0}$
(inside the left sphere) are written as:
\begin{align}
 & \Phi_{\text{R}}^{\text{in}}(\mu,\eta,\phi)\nonumber \\
 & =\sqrt{\cosh\mu-\cos\eta}\sum_{n=1}^{\infty}A_{R,n}^{\text{in}}e^{-(n+1/2)\mu}\nonumber \\
 & \times P_{n}^{1}(\cos\eta)\cos(\phi-\phi_{1})\label{eq:phi_in_R_bisph_perp}
\end{align}
and
\begin{align}
 & \Phi_{\text{L}}^{\text{in}}(\mu,\eta,\phi)\nonumber \\
 & =\sqrt{\cosh\mu-\cos\eta}\sum_{n=1}^{\infty}B_{L,n}^{\text{in}}e^{(n+1/2)\mu}\nonumber \\
 & \times P_{n}^{1}(\cos\eta)\cos(\phi-\phi_{1})\label{eq:phi_in_L_bisph_perp}
\end{align}
respectively. The potential outside both the nanoparticles is given
by:
\begin{align}
 & \Phi^{\text{out}}(\mu,\eta,\phi)=\sqrt{\cosh\mu-\cos\eta}\nonumber \\
 & \times\sum_{n=1}^{\infty}\Bigg[C_{n}^{\text{out}}e^{-(n+1/2)\mu}+D_{n}^{\text{out}}e^{(n+1/2)\mu}\Bigg]\nonumber \\
 & \times\cos(\phi-\phi_{1})P_{n}^{1}(\cos\eta)\nonumber \\
 & +\frac{D}{4\pi\varepsilon_{0}\varepsilon_{s}}\hat{\boldsymbol{\eta}}\cdot\boldsymbol{\nabla}_{1}\frac{1}{|{\bf r}-{\bf r}_{1}|}\label{eq:phi_out_bisph_perp}
\end{align}
The same boundary conditions given by Eqs. (\ref{eq:BC_1_bisph_general})-(\ref{eq:BC_4_bisph_general})
yield an infinite linear system of equations for the coefficients
(SI, Section \ref{SI:sol_BC_bisph_perp}) and are solved numerically
as done for the parallel configuration. The radiative and nonradiative
rates and the relevant component of $\underline{\textbf{G}}$ in terms
of these coefficients are provided in the supplement\textcolor{red}{{}
}(see SI, Section \ref{SI:sol_BC_bisph_perp}).

\subsection{The self-consistent procedure \label{subsec:self_consistent_method}}

For the configurations discussed above, once the needed component
of the Green’s dyadic has been evaluated, we can use Eqs. (\ref{eq:A_omega})
and (\ref{eq:polarizability_j}) and the given free molecule parameters
$\Gamma^{0}$ and $\omega_{0}$ to calculate the absorption lineshape
for the given dipole-metal nanostructure configuration. We emphasize
again that Eq. (\ref{eq:A_omega}) is an expression for the absorption
lineshape for a model in which the incident light interacts with the
molecular dipole only. In such a model, a distinct plasmon-dominated
peak accompanying the molecular response is a manifestation of ``strong
coupling''. Generalization to the case where the incident radiation
drives both the molecular dipole and the metal nanostructure is straightforward,
but more costly and the resulting lineshape will also show the signature
of interference between the dipole and the metal responses. 

\section{Results and discussions\label{sec:results_and_discussions}}

Here we present numerical results that map single molecule strong
coupling regions about spherical, spheroidal and bispherical structures.
Three comments are in order. First, since strong coupling as defined
by Eq. (\ref{eq:SC_condition}) or by the appearance of split peaks
at the overlap of the molecular and plasmon resonances as calculated
from Eq. (\ref{eq:A_omega}), it necessarily depends on properties
of the molecular resonance (non-radiative relaxation aside from that
due to coupling to the metal nanostructure as well pure dephasing)
that are not addressed in the present analysis. The results shown
below are obtained by assuming that the quantum yield for emission
by a free molecule is $0.5$ and that pure dephasing is absent. 

Second, our classical calculations disregard quantum mechanical effects,
mainly tunneling and non-locality of the dielectric response, which
are expected to become important at small molecule-metal distances\cite{neuman_coupling_2018}.
Classical estimates should be regarded reliable only at distances
larger than, say, $0.3$ nm where both effects are small.

Finally, as discussed in Section \ref{sec:Method}, estimates of strong
coupling based on Eq. (\ref{eq:SC_condition}) are made at a given
frequency, usually taken (as in Figures \ref{fig:QSC_sphere_perp_para_silver_gold_10nm}-\ref{fig:QSC_M_Gamma_bisph_para_gold_silver_10_10_dss}
below) as the transition frequency $\omega_{0}$ of the free molecule.
Because of the metal-induced spectral shift, satisfying the inequality
(\ref{eq:SC_condition}) at this frequency, while indicating strong
coupling, does not therefore guarantee the appearance of a split peak
in the actual plasmon-molecule system. The actual spectral shape is
obtained from Eq. (\ref{eq:A_omega}) as demonstrated in Figures \ref{fig:alter_treatment_SC_sphere_gold}-\ref{fig:alter_treatment_SC_two_sphere_para_gold}
below.

Figures \ref{fig:QSC_sphere_perp_para_silver_gold_10nm}-\ref{fig:QSC_M_Gamma_bisph_para_gold_silver_10_10_dss}
show the coupling strength $U$, the lifetime broadening $\Gamma$
and the coupling-broadening ratio (CBR) $2|U|/\hbar\Gamma$ calculated
at the resonance frequency of the free molecule for several geometries.
A value larger than $1$ of the CBR indicates strong coupling by the
criterion (\ref{eq:SC_condition}). For definiteness, in the calculation
shown below, the molecular transition dipole is taken to be $10$
D. However the free molecule transition frequency $\omega_{0}$ is
taken to be equal to the dipolar plasmon frequency of the metal corresponding
to geometry under consideration. For example, when considering a molecular
dipole near small gold and silver spheres, the molecular frequencies
are taken equal to corresponding dipolar plasmon frequencies, $2.62$
eV and $3.37$ eV for gold and silver spheres respectively. However,
the coupling strength, which is calculated by fitting the calculated
$G$ to the plasmon peak is, by this definition, frequency independent. 

The calculations of the radiative and metal-induced nonradiative contributions
to $\Gamma$ (Eqs. (\ref{eq:rad_decay_expression}), (\ref{eq:nonrad_decay_expression}))
have been extensively discussed before\cite{Gersten1981}, and typical
results showing the dependence of these rates on the molecular orientation
and its distance from the metal structure, and on the size and shape
of this structure are shown in the SI (Section \ref{SI_results_additional}).
We emphasize again, that the only contributions to the linewidth $\Gamma$
that are considered here are the lifetime (radiative and non-radiative)
of the free molecule, and the metal induced modifications of these
lifetimes, and that in the calculations reported below we have assumed
that a free molecule has an intrinsic non-radiative relaxation rate
that is equal to its radiative rate, implying that the emission yield
of a free molecule is $0.5$. Note that if the intrinsic non-radiative
relaxation rate of the free molecule is disregarded, the remaining
contributions to the linewidth are proportional to the square of the
molecular transition dipole $D_{12}$, whereas the coupling $U$ (Eq.
(\ref{eq:U_Eq})) is linear in this parameter. Consequently, and perhaps
counter-intuitively, the CBR satisfies $2|U|/\hbar\Gamma\propto1/D_{12}$
and decreases with increasing molecular transition dipole. The relaxation
rates (and magnitude of coupling) are normalized by the radiative
relaxation rate $\Gamma_{\text{R}}^{0}$ of a free molecule which
is calculated from $\sqrt{\varepsilon_{s}}\omega_{0}^{3}D_{12}^{2}/(3\hbar c^{3}4\pi\varepsilon_{0})$
at the respective molecular frequency. 

\begin{figure}[H]
\subfloat[]{\includegraphics[scale=0.5]{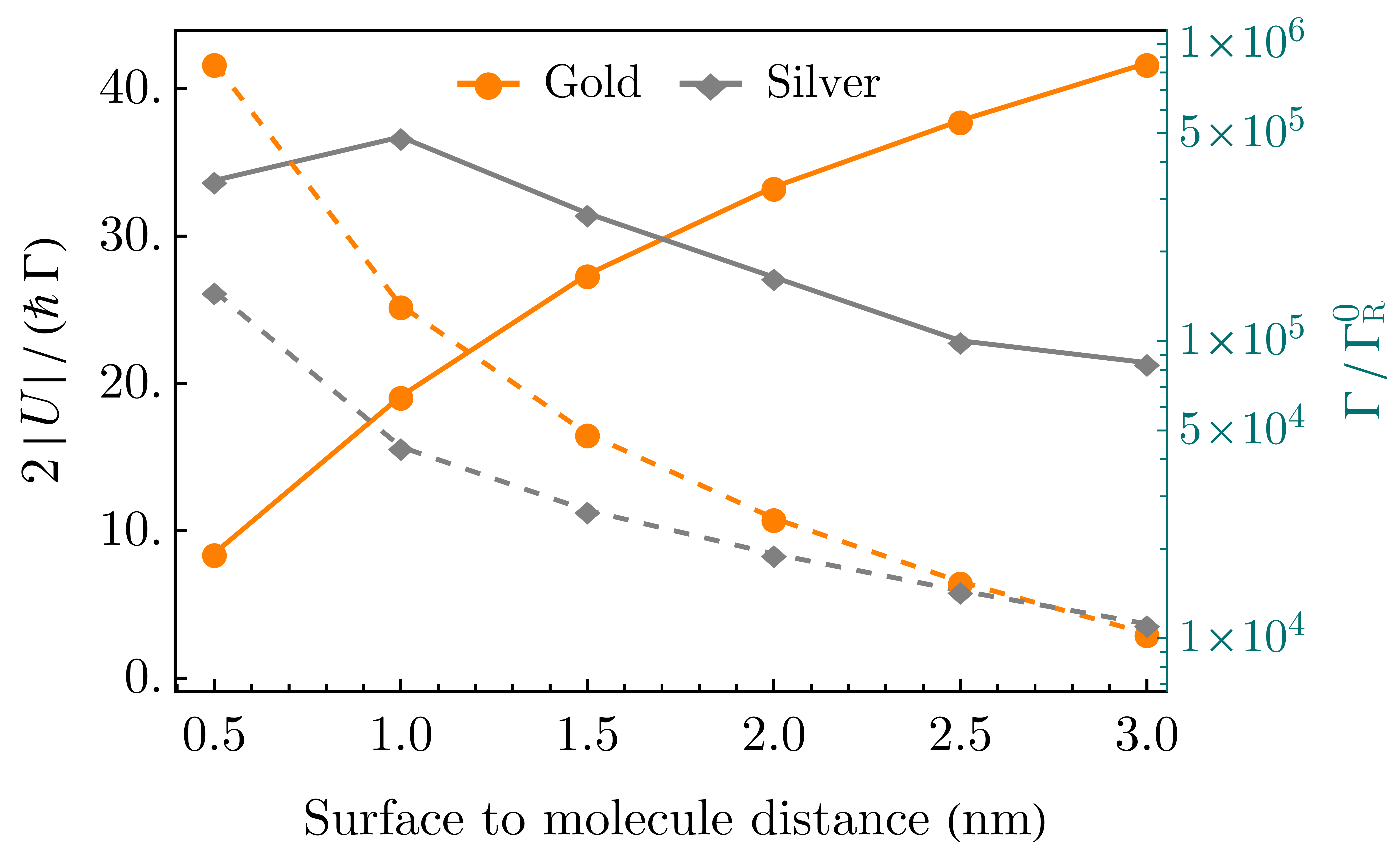}

\label{fig:QSC_sphere_perp_silver_gold_10nm}}\subfloat[]{\includegraphics[scale=0.5]{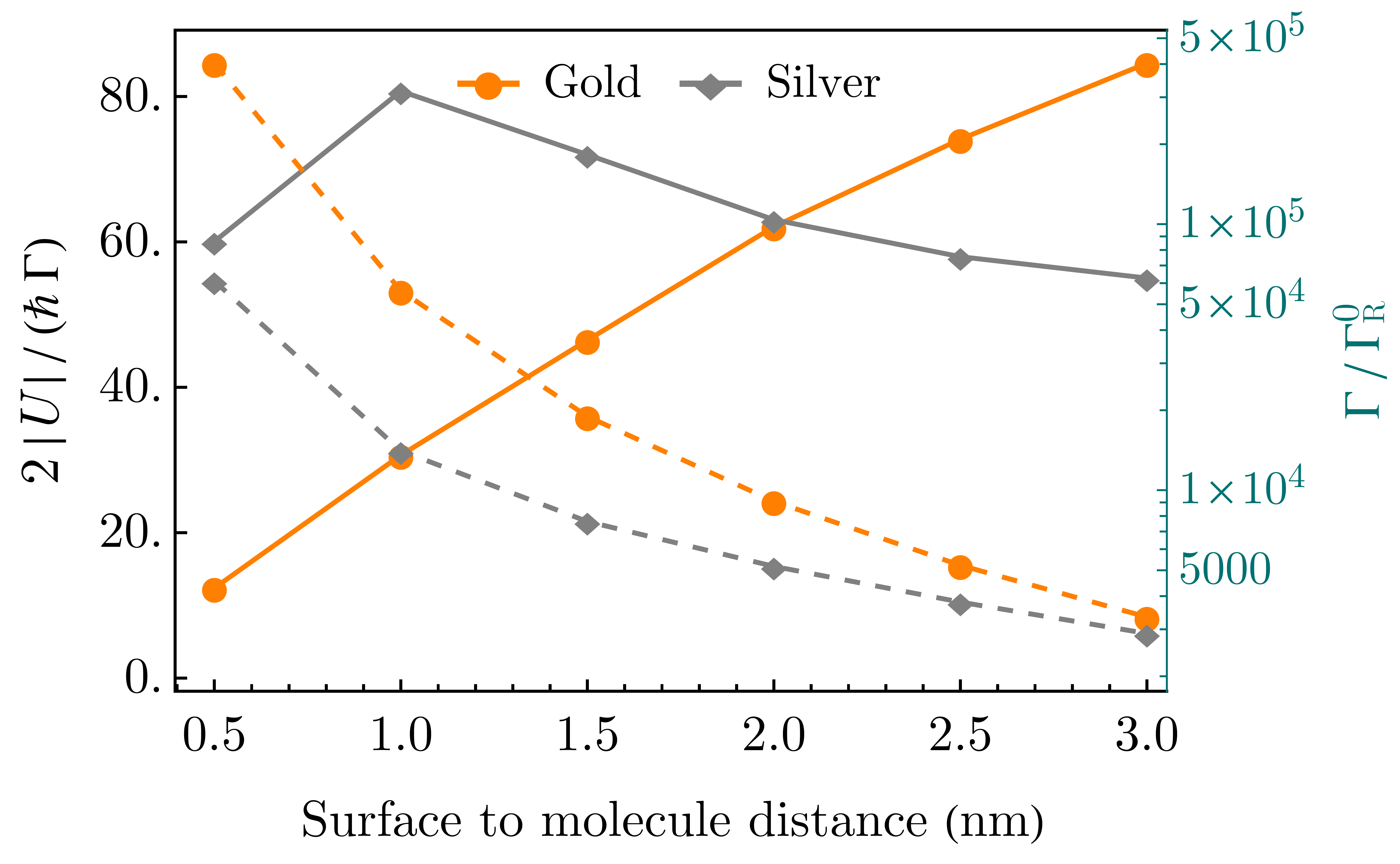}

\label{fig:QSC_sphere_para_silver_gold_10nm}

}

\caption{The CBR parameter, $2|U|/\hbar\Gamma$ (left axis, solid lines) and
total relaxation rate $\Gamma$ (right axis, dashed lines), plotted
as functions of the molecule-sphere surface distance, shown for gold
and silver nanospheres of radius $10$ nm for (a) normal and (b) parallel
molecular orientations relative to the sphere surface. $\omega_{0}$
is taken to be in resonance with the dipolar plasmon of the metal
particle ($2.62$ eV for gold and $3.37$ eV for silver). $\Gamma_{\text{R}}^{0}=7.38\times10^{7}$
$\text{s}^{-1}$ (gold) and $1.57\times10^{8}$ $\text{s}^{-1}$ (silver)
at the respective molecular frequency of the emitter.}
\label{fig:QSC_sphere_perp_para_silver_gold_10nm}
\end{figure}

Figure \ref{fig:QSC_sphere_perp_para_silver_gold_10nm} depicts the
CBR for a molecule near spherical gold and silver particles of radius
$10$ nm, in normal and parallel orientations relative to the sphere
surface, as a function of the molecule-surface distance. Also shown
is the combined relaxation rate $\Gamma$. It is seen that “strong
coupling” as defined by the $\text{CBR}>1$ criterion, is not necessarily
associated with close proximity. Both and $U$ and $\Gamma$ go quickly
to zero when the molecule-surface distance increases, however their
ratio varies relatively slowly, and remains larger than $1$ in all
the distance range studied. It is interesting to note that in both
the normal and parallel configurations this ratio goes through maximum
for silver and decreases beyond a distance $\sim1$ nm while for gold
it rises monotonically with increasing distance. The metal induced
nonradiative relaxation, being the major contribution to the combined
relaxation rates, is very large when close to the nanoparticle surface.
As a consequence, for silver there is an optimum distance of the molecule
from the surface where the CBR is maximum. For gold, on the other
hand, because the extent of the effect of nonradiative decay rate
is even more, for the distance considered, CBR is larger as the molecule
is distant from the surface. Qualitatively for both perpendicular
and parallel orientation, these observations are common.

\begin{figure}[H]
\subfloat[]{\includegraphics[scale=0.5]{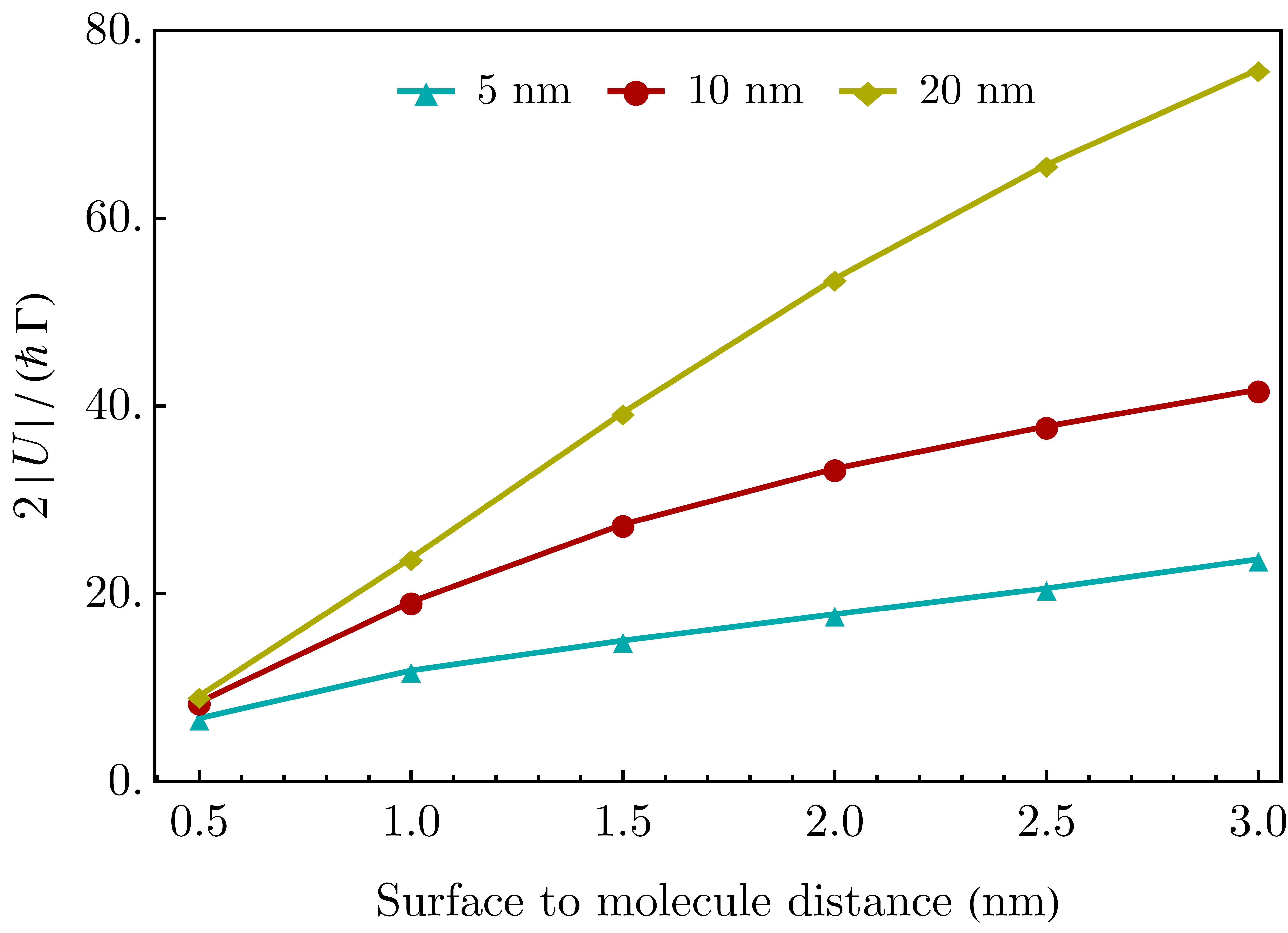}

\label{fig:QSC_sphere_perp_gold_5_10_20nm}}\subfloat[]{\includegraphics[scale=0.5]{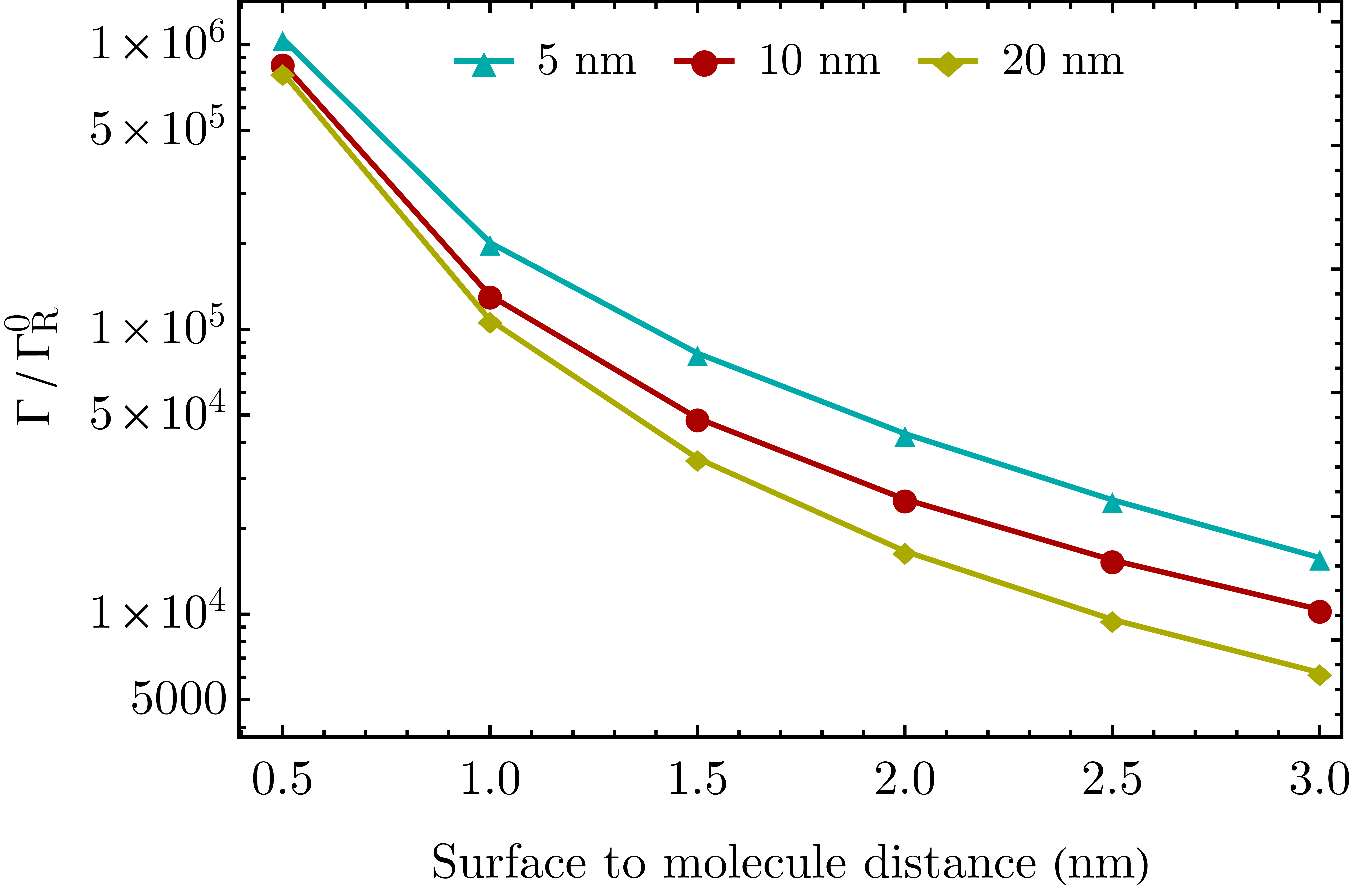}

\label{fig:Gamma_sphere_perp_gold_5_10_20nm}

}

\caption{(a) The CBR parameter, $2|U|/\hbar\Gamma$, and (b) total relaxation
rate $\Gamma$ plotted against molecule-surface distance for a molecule
near a gold nanoparticle in the perpendicular orientation. $\omega_{0}$
for the molecule is equal to the dipolar plasmon resonance frequency
of gold ($2.62$ eV). Results are shown for spheres of radii $5$
nm, $10$ nm and $20$ nm. Free emitter radiative decay rate, $\Gamma_{\text{R}}^{0}=7.38\times10^{7}$
$\text{s}^{-1}$.}
\label{fig:QSC_Gamma_sphere_perp_gold_5_10_20nm}
\end{figure}

These estimates of strong coupling show considerable dependence on
the size and shape of the dielectric nanostructure. Figure \ref{fig:QSC_Gamma_sphere_perp_gold_5_10_20nm}
shows the CBR parameter as well as the lifetime broadening $\Gamma$
plotted against the molecule-(spherical gold) nanoparticle surface
distance for different particle sizes in the perpendicular dipole
orientation, while Figures \ref{fig:QSC_U_Gamma_spheroid_major_perp_gold_aspect_ratio}
and \ref{fig:QSC_U_Gamma_spheroid_major_perp_silver_aspect_ratio}
show similar results for different particle shapes by considering
metal spheroids of constant volume ($=$ volume of a sphere of radius
$10$ nm) and different aspect ratios $p/q$ when the molecule is
positioned on the long axis with a normal orientation relative to
the nanoparticle surface.
\begin{figure}
\subfloat[]{\includegraphics[viewport=0bp 0bp 468bp 328bp,scale=0.33]{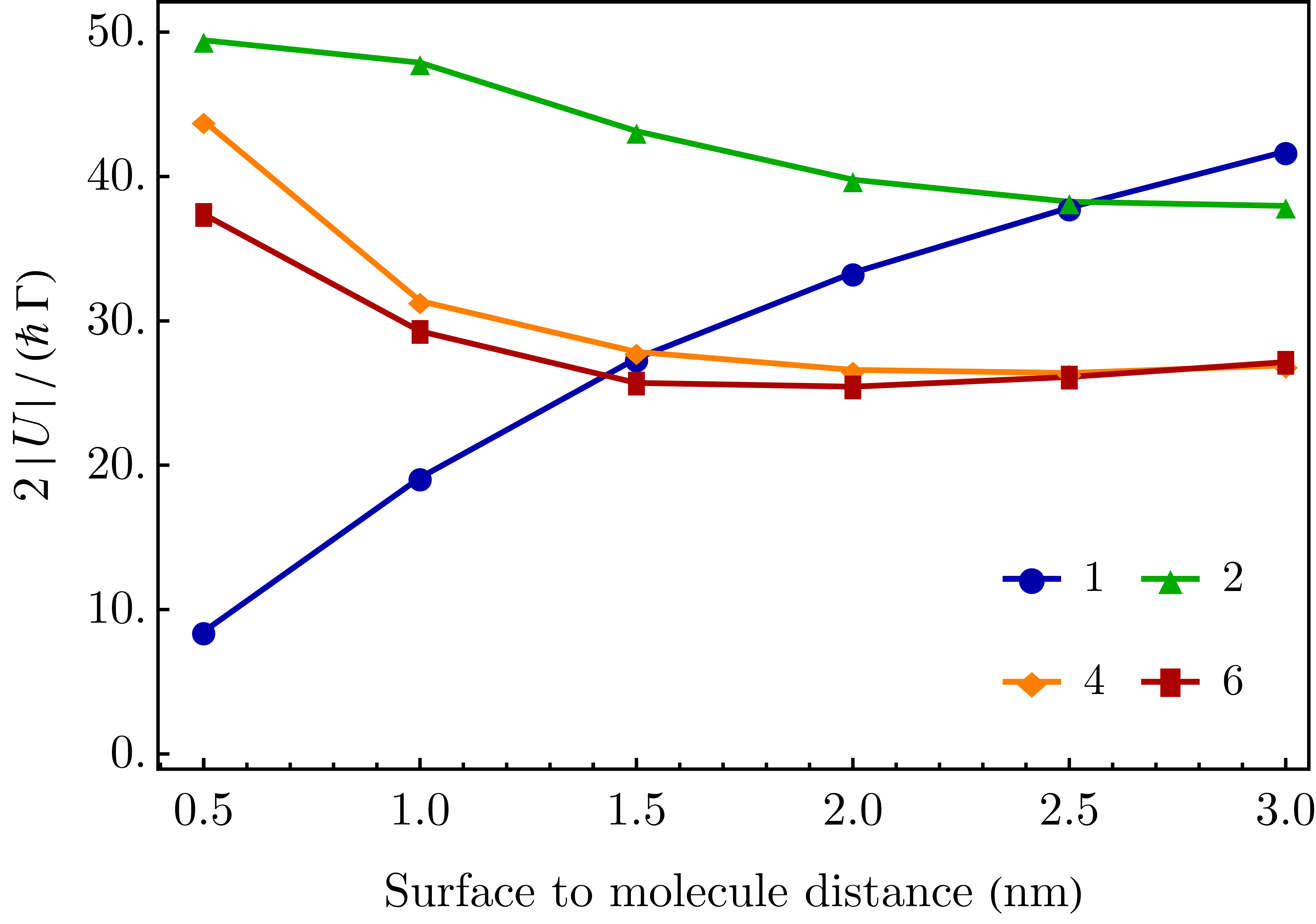}

\label{fig:QSC_spheroid_major_perp_gold_aspect_ratio}}\subfloat[]{\includegraphics[viewport=0bp 0bp 468bp 317.705bp,scale=0.33]{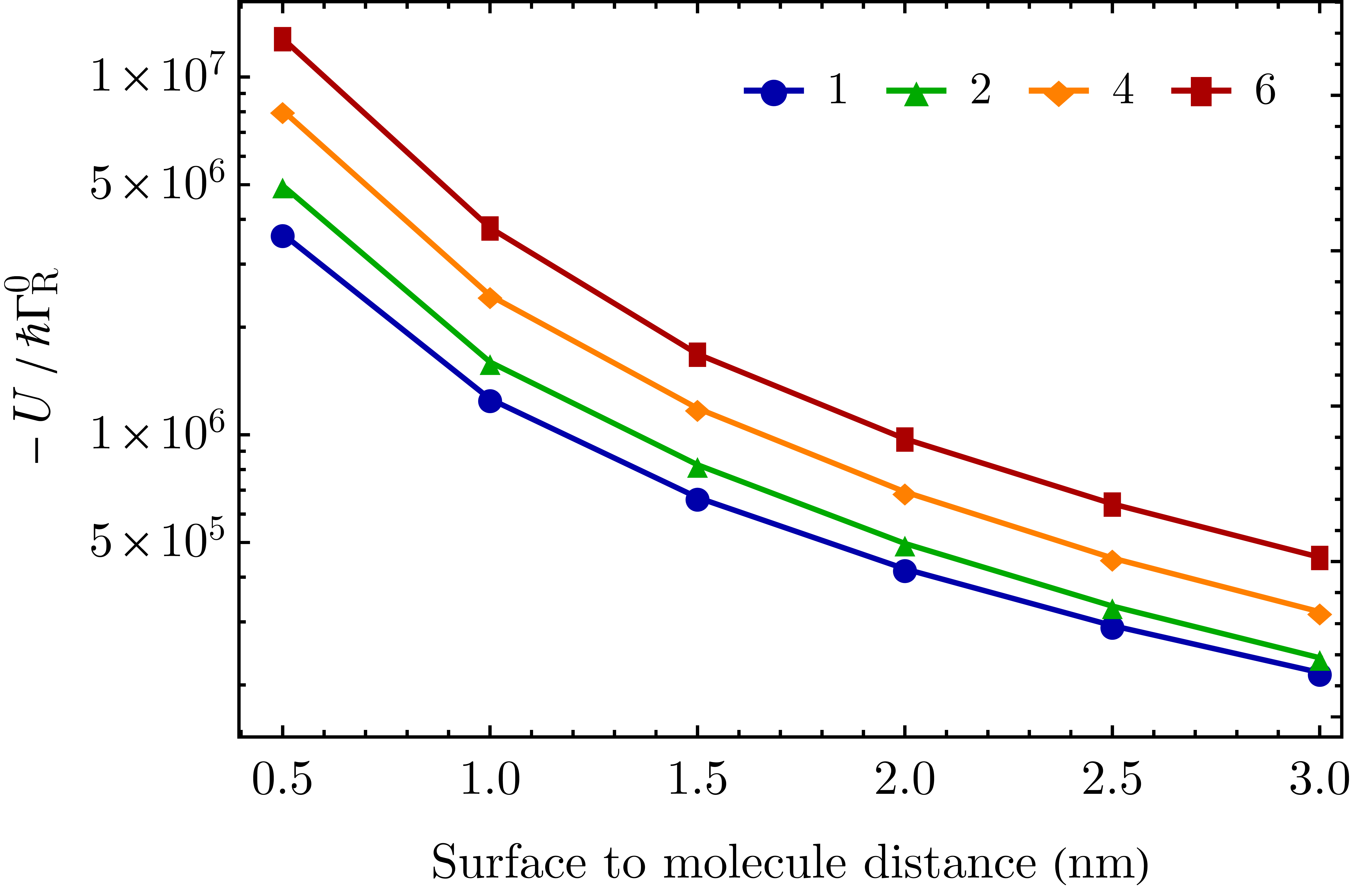}\label{fig:U_spheroid_major_perp_gold_aspect_ratio}}\subfloat[]{\includegraphics[viewport=0bp 0bp 468bp 309.4585bp,scale=0.33]{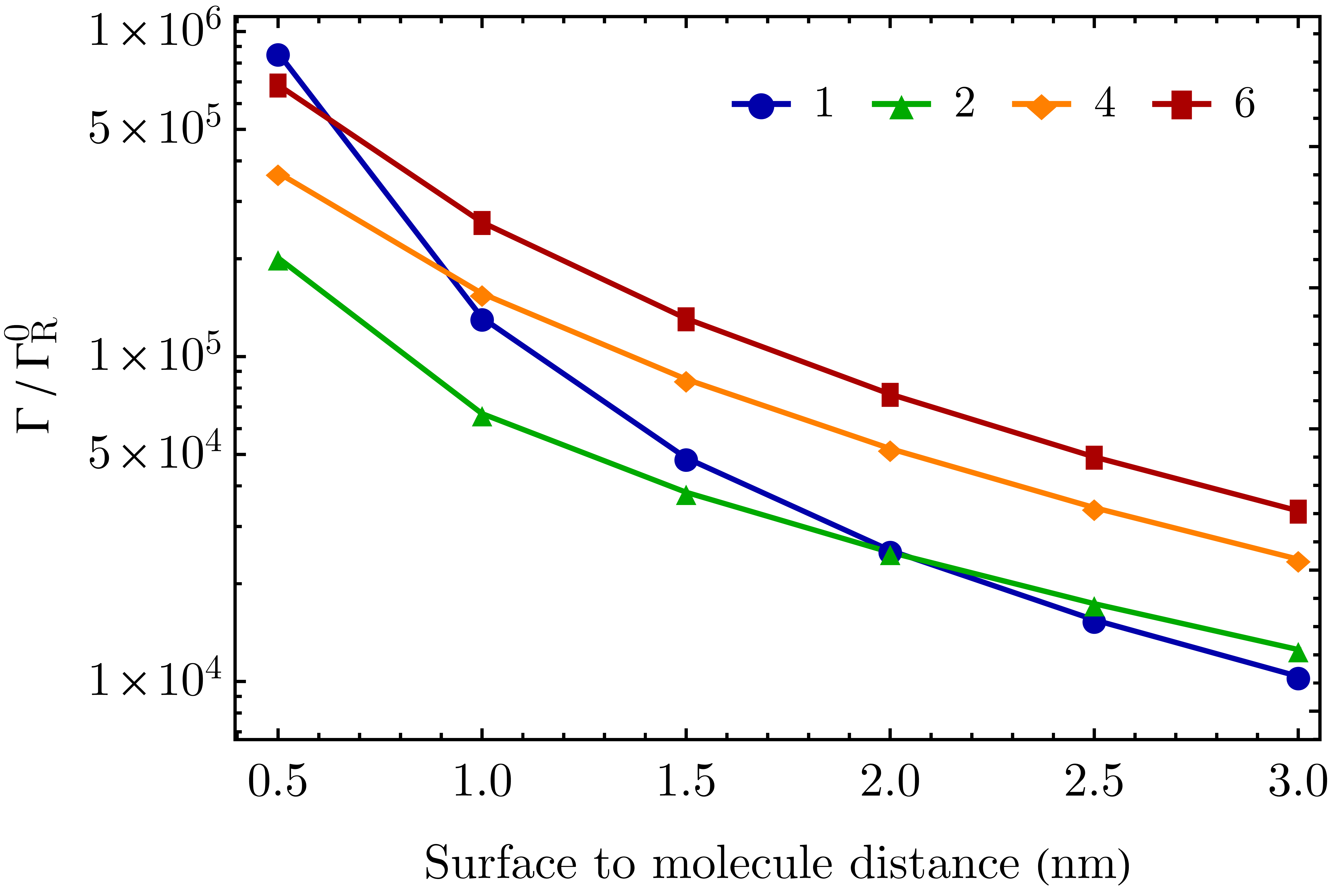}

\label{fig:Gamma_spheroid_major_perp_gold_aspect_ratio}}

\caption{(a) The CBR parameter (b) the magnitude of coupling, $-U/\hbar$,
and (c) the total relaxation rate $\Gamma$ as functions of the molecule
to the (spheroidal gold) nanoparticle surface distance when the molecule
is placed on the long axis oriented perpendicular to the spheroid
surface. The aspect ratio of the spheroids $p/q$ varies from $1$
to $6$ keeping the particle volume constant (all have the volume
of a sphere of radius $10$ nm). For each aspect ratio we take $\omega_{0}$
in resonance with the respective longitudinal dipolar plasmon (excited
along the particle long axis) and use the corresponding free molecule
radiative decay rate: $2.62$ eV and $7.38\times10^{7}$ $\text{s}^{-1}$
for $p/q=1$, $2.36$ eV and $5.39\times10^{7}$ $\text{s}^{-1}$
for $p/q=2$, $1.92$ eV and $2.89\times10^{7}$ $\text{s}^{-1}$
for $p/q=4$, and $1.59$ eV and $1.66\times10^{7}$ $\text{s}^{-1}$
for $p/q=6$.}
\label{fig:QSC_U_Gamma_spheroid_major_perp_gold_aspect_ratio}
\end{figure}
In the latter figures (\ref{fig:QSC_U_Gamma_spheroid_major_perp_gold_aspect_ratio}-\ref{fig:QSC_U_Gamma_spheroid_major_perp_silver_aspect_ratio})
we show the magnitude of coupling $U$ (panel b in these figures)
and total relaxation rates $\Gamma$ (panel c in the same figures).
More details on the radiative and non-radiative contributions to $\Gamma$
for a molecular dipole positioned near gold and silver nanospheroids,
are provided in the SI, Section \ref{subsec:spheroid_perp_results_gold_silver}.
\begin{figure}[H]
\subfloat[]{\includegraphics[scale=0.33]{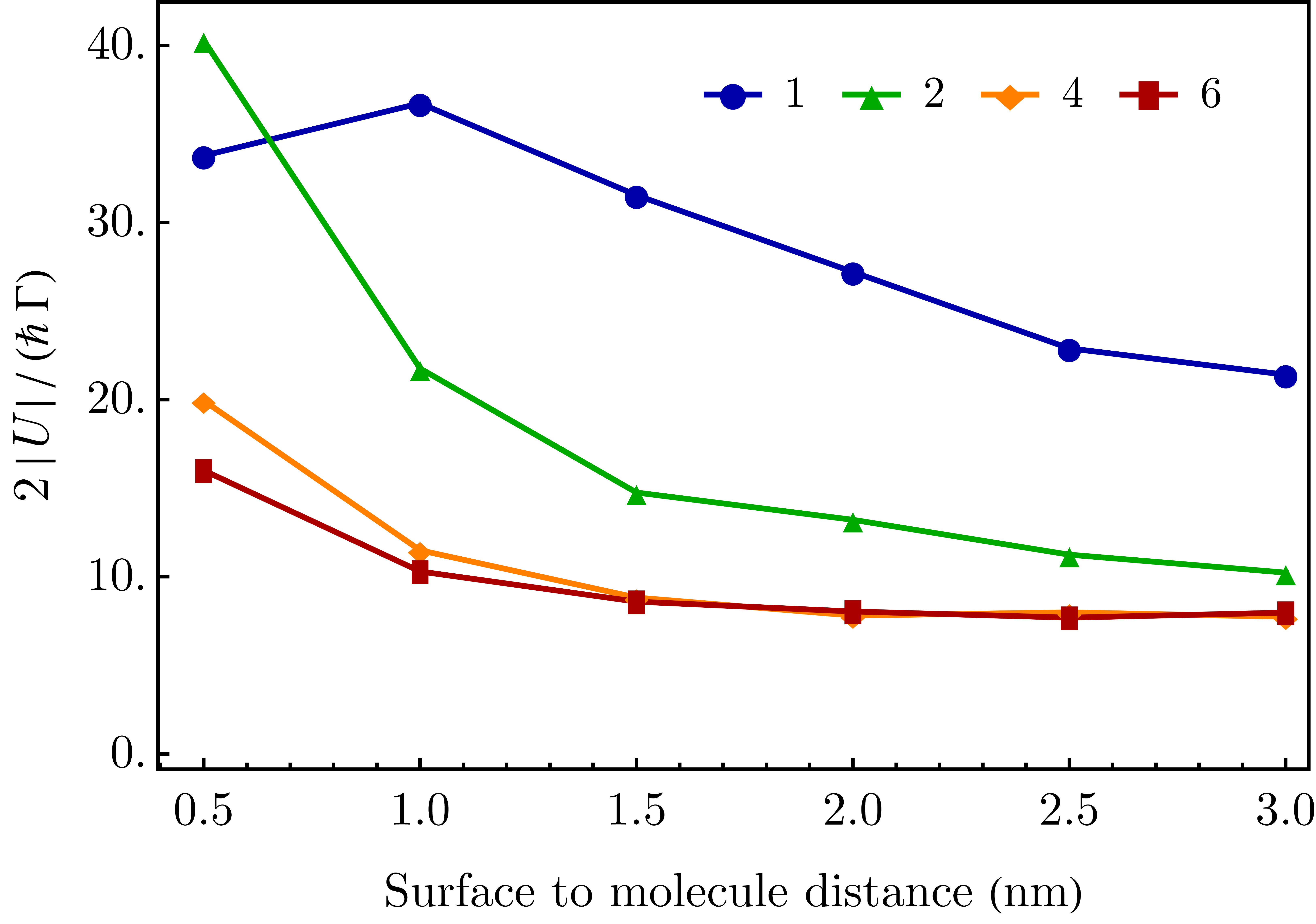}

\label{fig:QSC_spheroid_major_perp_silver_aspect_ratio}}\subfloat[]{\includegraphics[scale=0.33]{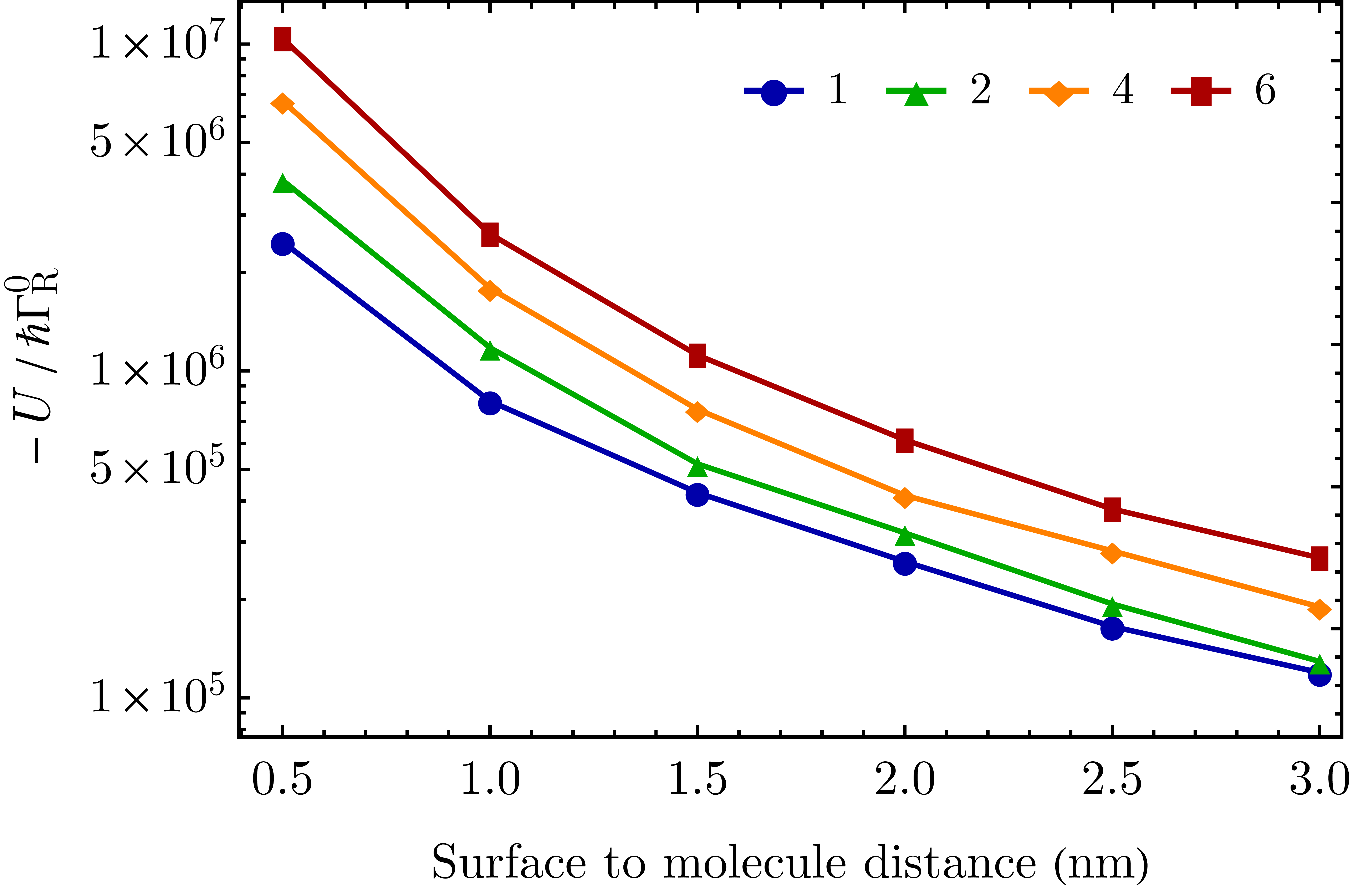}\label{fig:U_spheroid_major_perp_silver_aspect_ratio}}\subfloat[]{\includegraphics[scale=0.33]{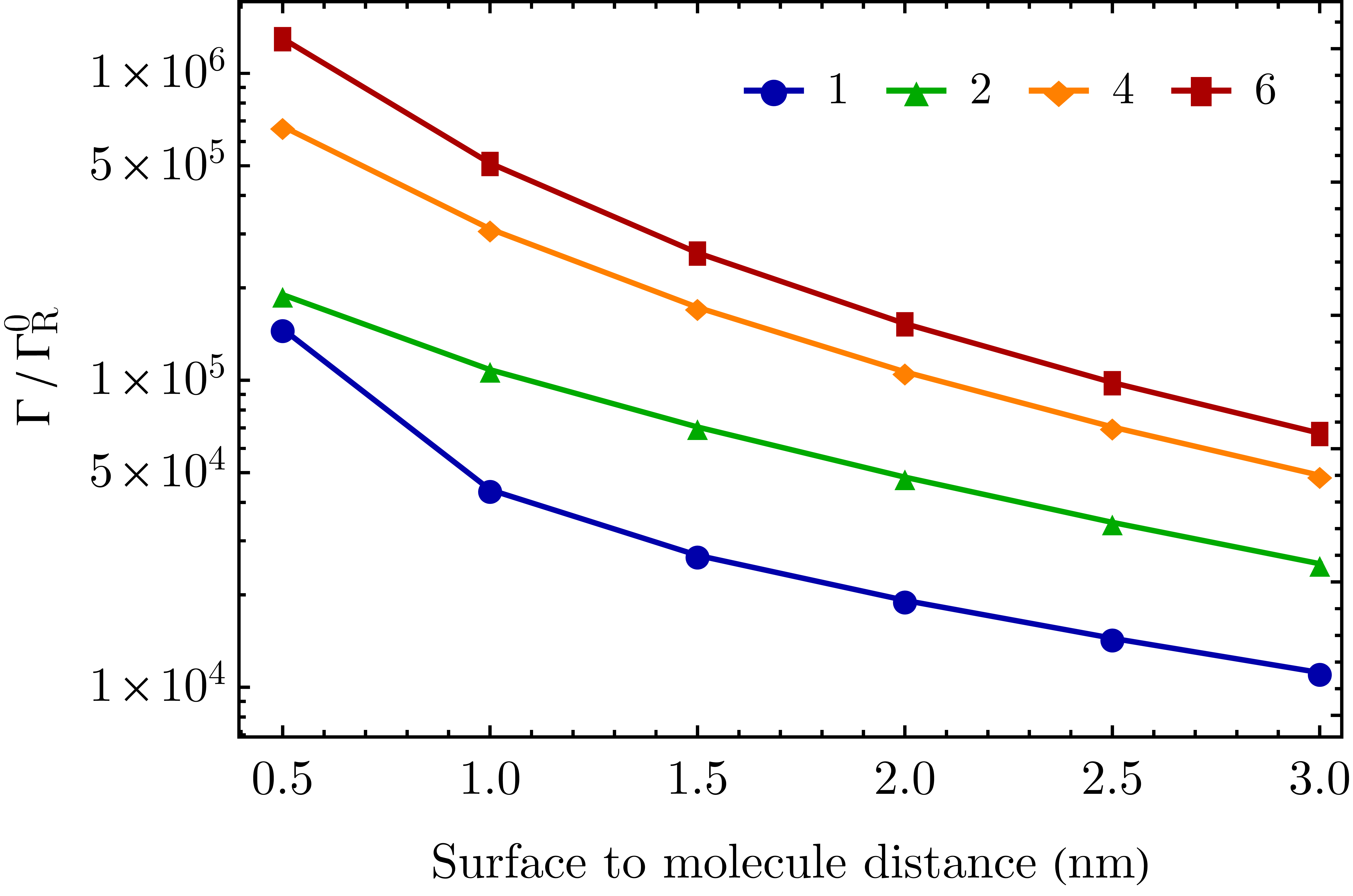}

\label{fig:Gamma_spheroid_major_perp_silver_aspect_ratio}}

\caption{Similar to Figure \ref{fig:QSC_U_Gamma_spheroid_major_perp_gold_aspect_ratio},
now with spheroidal silver nanoparticles. Molecular frequency and
the corresponding free molecule radiative decay rate are chosen similar
to Figure \ref{fig:QSC_U_Gamma_spheroid_major_perp_gold_aspect_ratio}:
$3.37$ eV and $1.57\times10^{8}$ $\text{s}^{-1}$ for $p/q=1$,
$2.85$ eV and $9.55\times10^{7}$ $\text{s}^{-1}$ for $p/q=2$,
$2.15$ eV and $4.06\times10^{7}$ $\text{s}^{-1}$ for $p/q=4$,
and $1.71$ eV and $2.06\times10^{7}$ $\text{s}^{-1}$ for $p/q=6$.}
\label{fig:QSC_U_Gamma_spheroid_major_perp_silver_aspect_ratio}
\end{figure}

Figures \ref{fig:QSC_M_Gamma_bisph_para_perp_gold_silver_10_10_2}
and \ref{fig:QSC_M_Gamma_bisph_para_gold_silver_10_10_dss} show similar
quantities for a molecular dipole positioned between two metal spheres.
The (point) dipole is placed on the axis connecting the sphere centers
(called the intersphere axis) and oriented parallel (and normal) to
this axis. The molecular transition frequency is taken to be in resonance
with the stronger plasmon peak associated with the corresponding configuration
(see Figure \ref{fig:SI_rad_decay_bisph_uni_field_gold_silver_para_perp}
for the optical response of gold and silver bispherical structures).

\begin{figure}[H]
\subfloat[]{\includegraphics[scale=0.33]{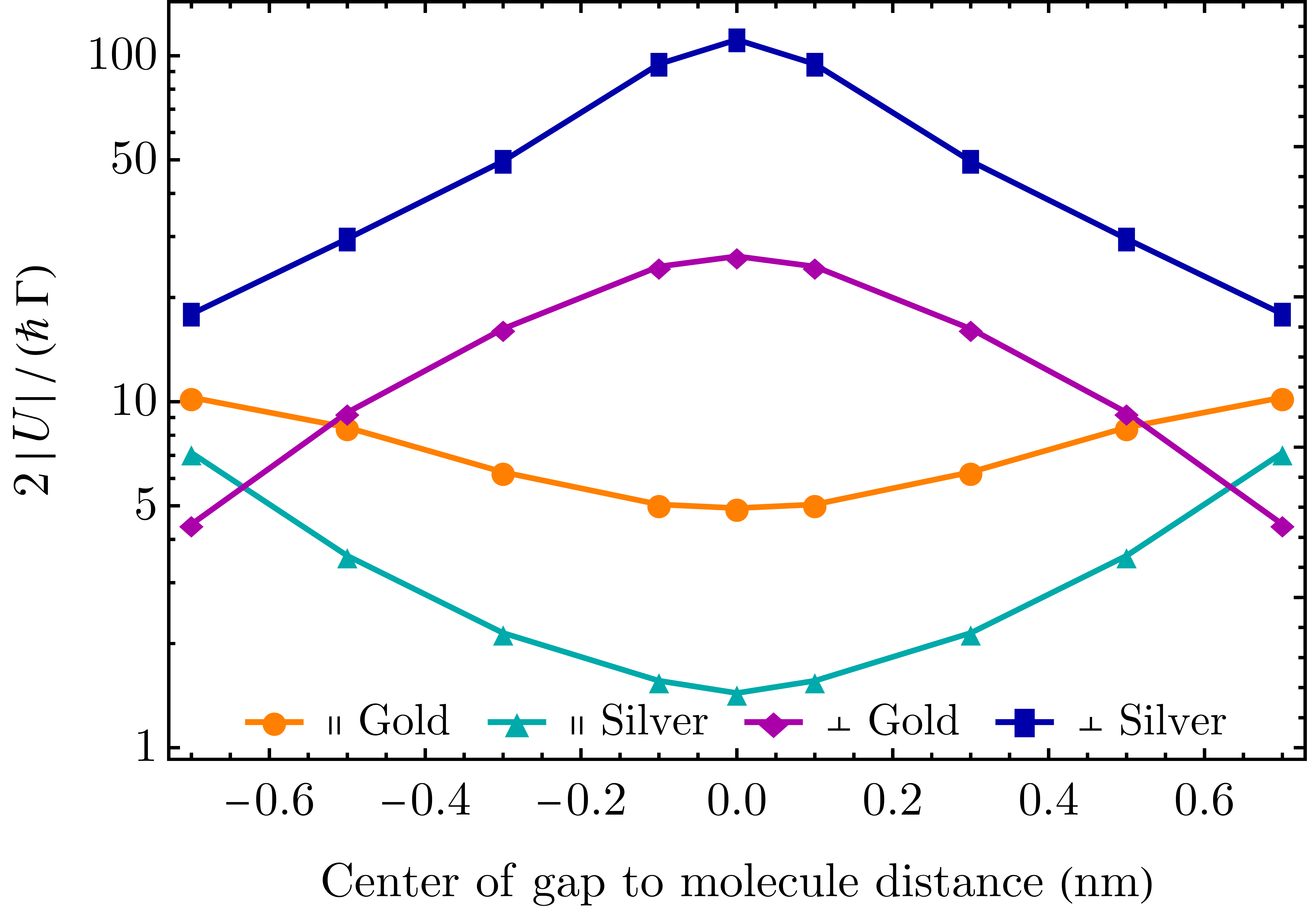}

\label{fig:QSC_bisph_para_perp_gold_silver_10_10_2}}\subfloat[]{\includegraphics[scale=0.33]{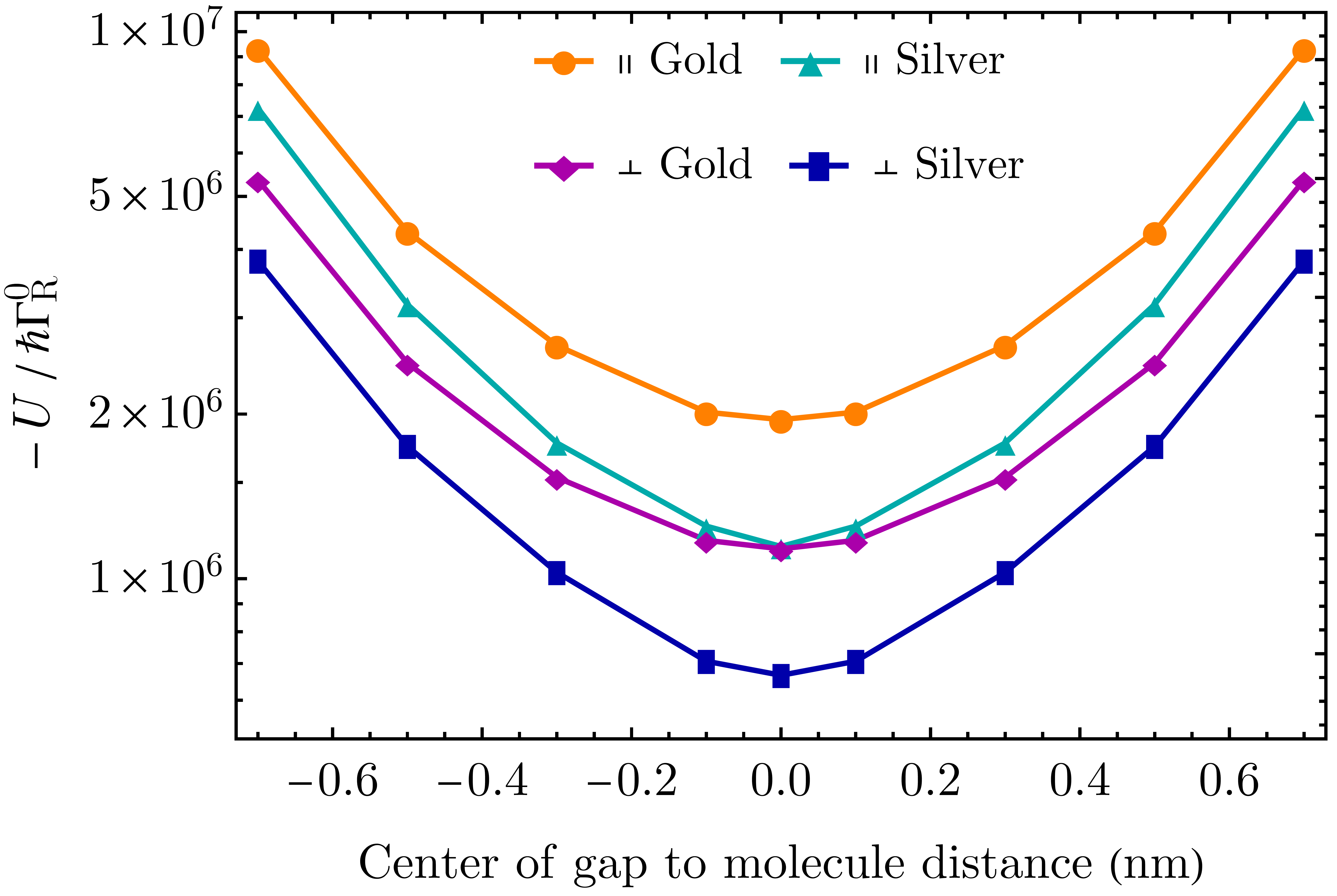}\label{fig:M_bisph_para_perp_gold_silver_10_10_2}}\subfloat[]{\includegraphics[scale=0.33]{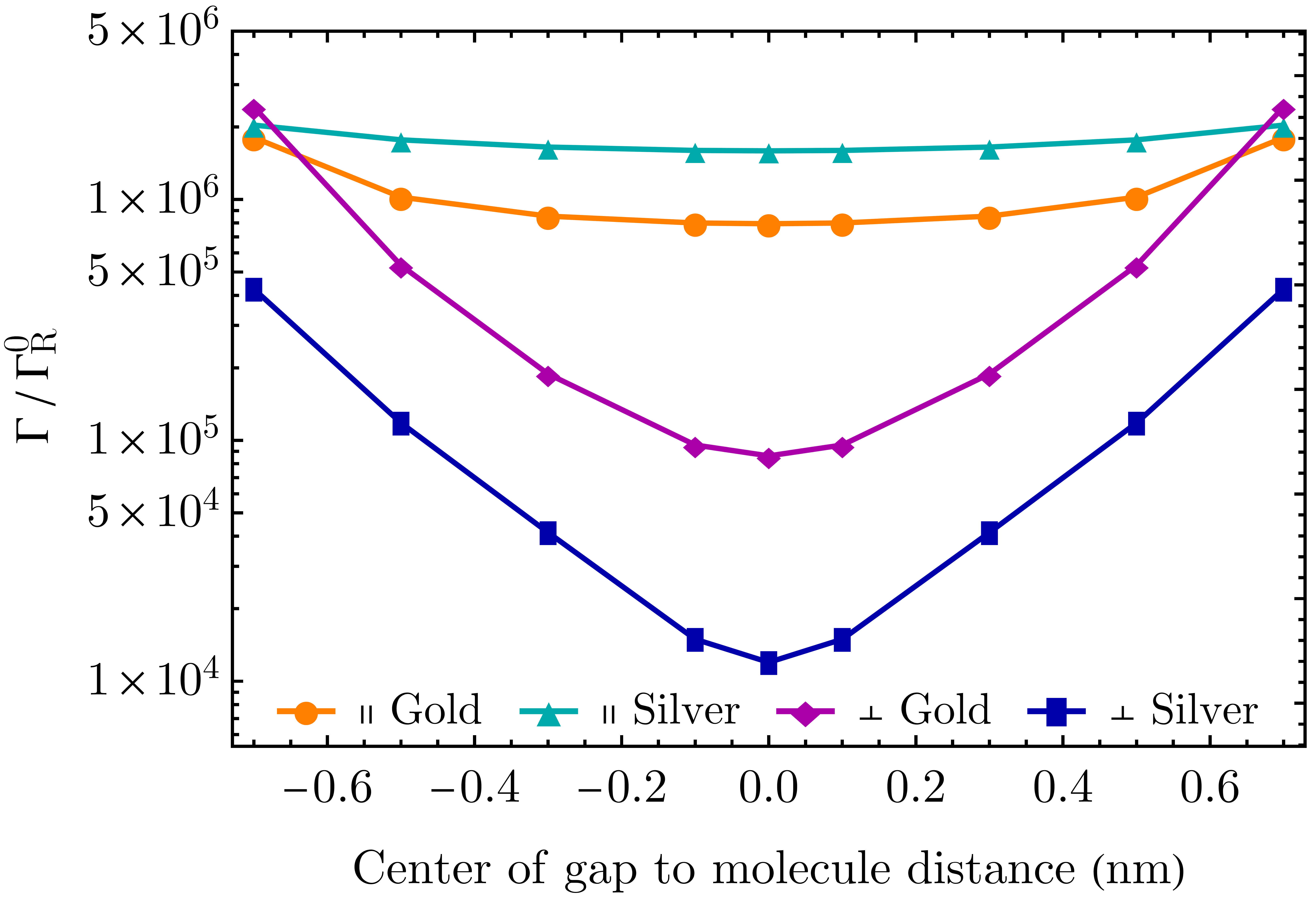}

\label{fig:Gamma_bisph_para_perp_gold_silver_10_10_2}}

\caption{(a) The CBR parameter (b) the magnitude of coupling $-U/\hbar$, and
(c) the total relaxation rate $\Gamma$ as functions of the position
of a molecular dipole in between the two nanospheres on the intersphere
axis, shown for molecules in parallel and perpendicular orientations
relative to this axis. The spheres are identical with $10$ nm radii
with a $2$ nm gap in between. The position of the molecule is shown
relative to the center of the gap. For the parallel configuration,
$\omega_{0}$ used in these calculations are $2.47$ eV (for gold)
and $3.07$ eV (for silver). For the perpendicular configurations,
they are taken as $2.64$ eV for gold and $3.41$ eV for silver. The
corresponding free molecule radiative relaxation rates are $6.20\times10^{7}$
$\text{s}^{-1}$ (gold, parallel), $1.19\times10^{8}$ $\text{s}^{-1}$
(silver, parallel), $7.53\times10^{7}$ $\text{s}^{-1}$ (gold, perpendicular),
and $1.64\times10^{8}$ $\text{s}^{-1}$ (silver, perpendicular).}
\label{fig:QSC_M_Gamma_bisph_para_perp_gold_silver_10_10_2}
\end{figure}

Next, we consider the effect of the gap size for the likes of the
configurations shown in Figure \ref{fig:QSC_M_Gamma_bisph_para_perp_gold_silver_10_10_2},
i.e., molecule in the gap in between nanosphere dimers, specifically
for the parallel molecular orientation relative to the intersphere
axis. In particular, in Figure \ref{fig:QSC_M_Gamma_bisph_para_gold_silver_10_10_dss},
the CBR parameter, the coupling magnitude, and the combined relaxation
rate are displayed for both gold and silver nanoparticle dimers. In
all calculations, the molecule is assumed to be positioned in the
middle of the gap. To note, the molecular frequency $\omega_{0}$
is taken to be in resonance with the higher plasmon peak for the configuration
with two $10$ nm metal nanoparticles with a $2$ nm gap. For this
reason, the calculated combined relaxation rate $\Gamma$ in Figure
\ref{fig:Gamma_bisph_para_gold_silver_10_10_dss} shows a clear maximum
for $2$ nm gap size for silver. This also explains the slight deviation
from the trend for gold. 
\begin{figure}[H]
\subfloat[]{\includegraphics[scale=0.33]{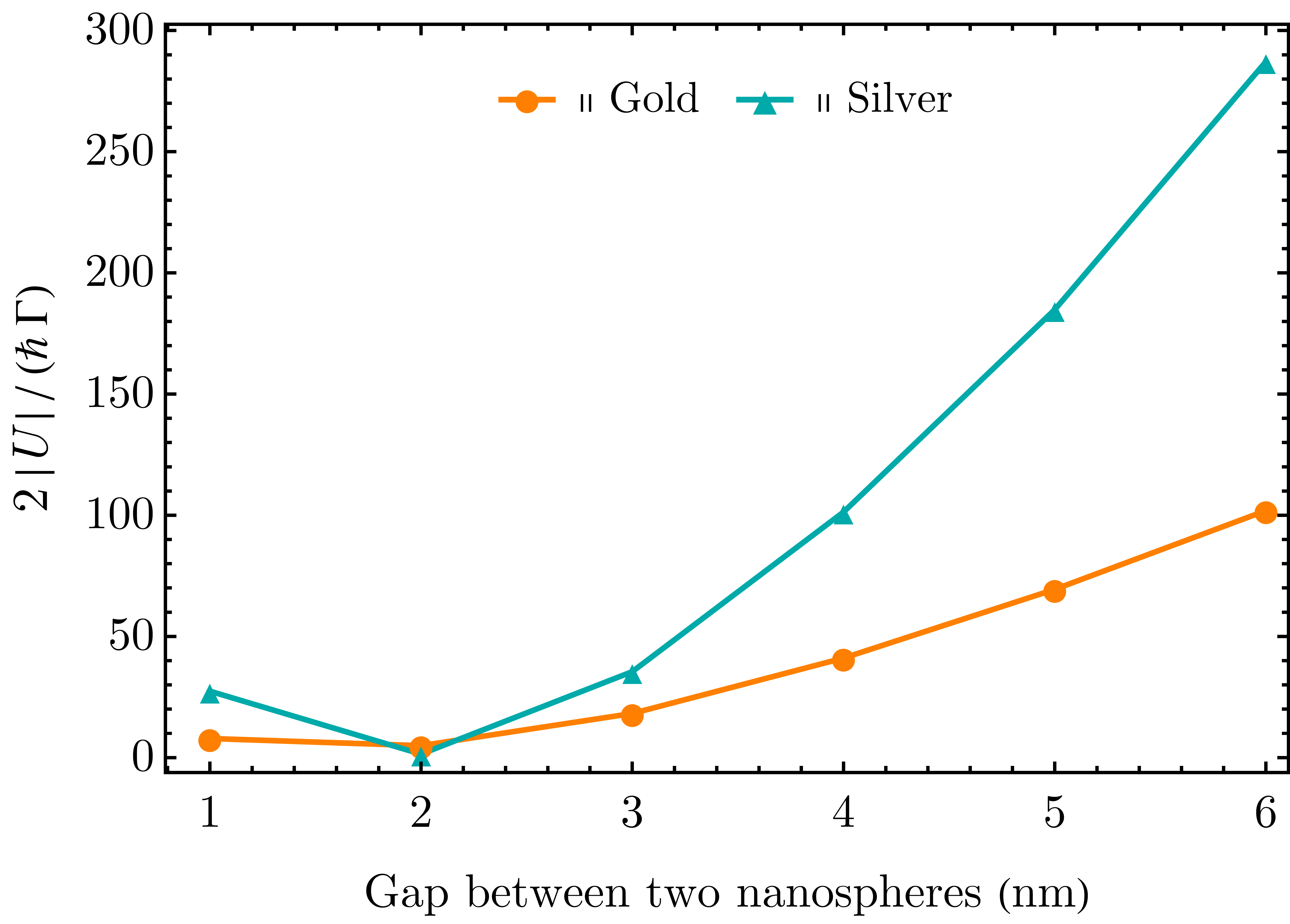}

\label{fig:QSC_bisph_para_gold_silver_10_10_dss}}\subfloat[]{\includegraphics[scale=0.33]{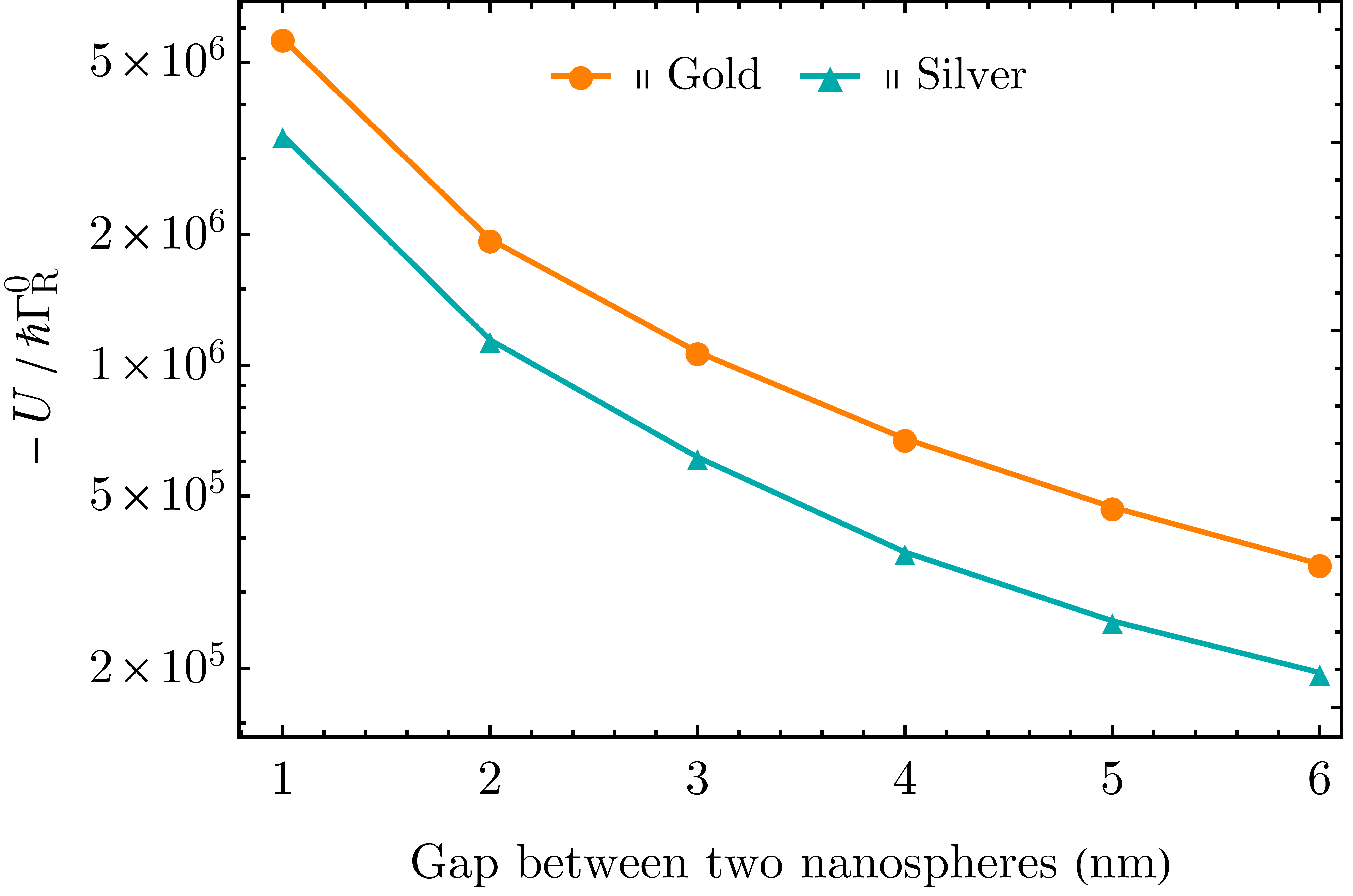}\label{fig:M_bisph_para_gold_silver_10_10_dss}}\subfloat[]{\includegraphics[scale=0.33]{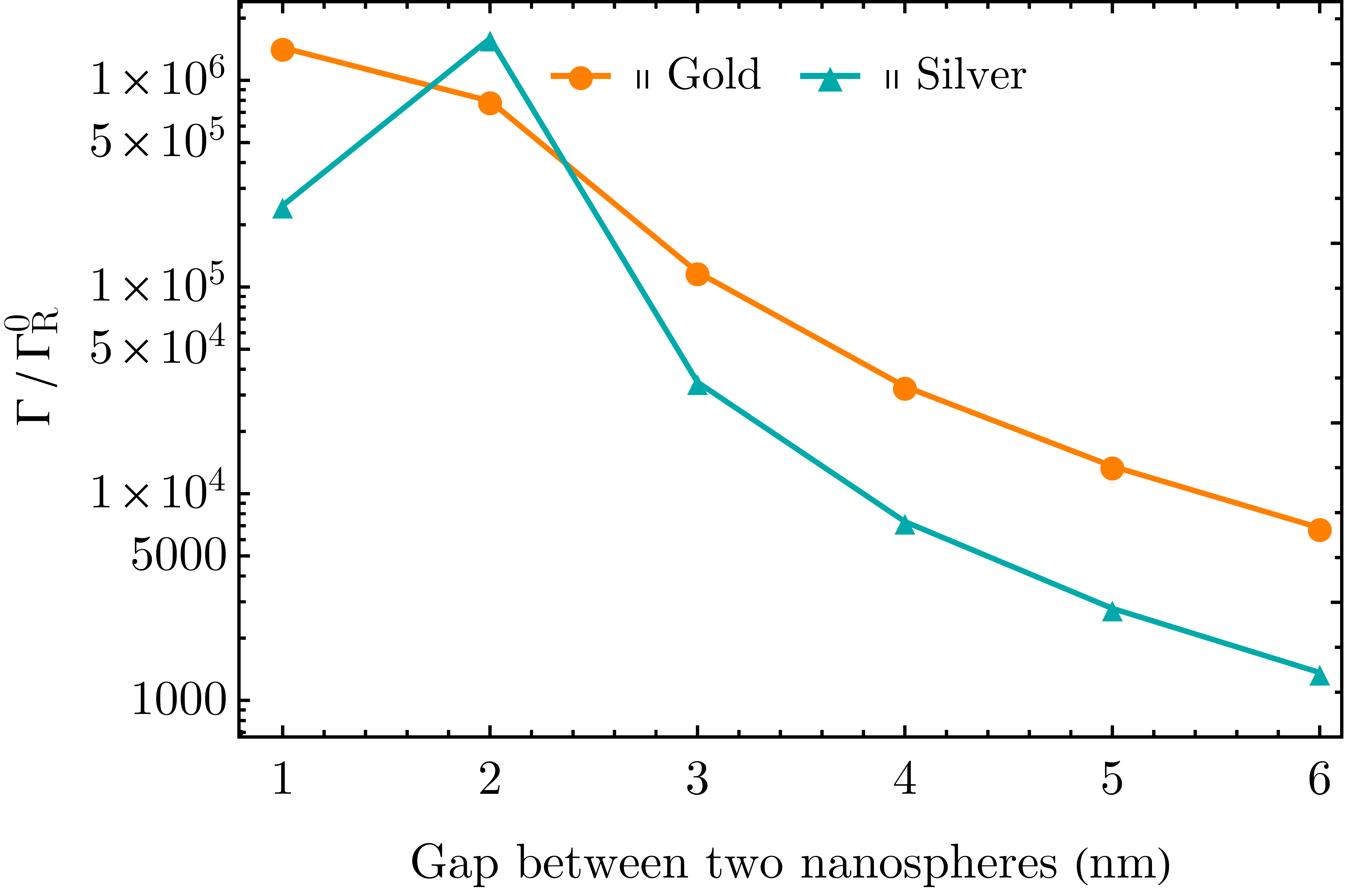}

\label{fig:Gamma_bisph_para_gold_silver_10_10_dss}}

\caption{Same as Figure \ref{fig:QSC_M_Gamma_bisph_para_perp_gold_silver_10_10_2},
where here $U$, $\Gamma$ and the CBR parameter, $2|U|/\hbar\Gamma$
are plotted against the gap size between the spherical gold and silver
particles, where the molecule is positioned at the center of the gap
and oriented parallel to the intersphere axis. The molecular frequencies
are taken the same as in Figure \ref{fig:QSC_M_Gamma_bisph_para_perp_gold_silver_10_10_2}
(in resonance with the brightest plasmon of the $2$ nm gap structure)
with the same choices for the free molecule radiative relaxation rates.}
\label{fig:QSC_M_Gamma_bisph_para_gold_silver_10_10_dss}
\end{figure}

The configuration considered here, with the molecule positioned in
the space between two nanoparticles, is often referred to as a plasmonic
cavity. The effective cavity volume can be determined from the calculated
$M=U/D_{12}$ using Eq. (\ref{eq:eff_volm}). Obviously this “effective
volume” can be defined also for configurations that cannot be perceived
as cavities, such as used in Figures \ref{fig:QSC_sphere_perp_para_silver_gold_10nm}-\ref{fig:QSC_U_Gamma_spheroid_major_perp_silver_aspect_ratio}
(dipole-sphere/spheroid systems), and it depends on the particular
geometry considered, including the dipole position and orientation.
It is therefore of limited value for molecules interacting with plasmonic
structures, but because of its prominence in many discussions we show
several examples of this parameter in the SI (Figures \ref{fig:SI_eff_vol_sph_sphd_perp_gold_silver_par_vol}-\ref{fig:SI_eff_vol_bisph_para_perp_gold_silver_gap_cube}).

Considering the results displayed in Figures \ref{fig:QSC_sphere_perp_para_silver_gold_10nm}-\ref{fig:QSC_M_Gamma_bisph_para_gold_silver_10_10_dss},
the following observations are notable:

(A) The interplay between coupling and broadening that determine “strong
coupling” according to Eq. \ref{eq:SC_condition} depends strongly
on geometry. While this fact is widely appreciated, the particulars
of these dependence are sometimes surprising as exemplified in points
(C) and (D) below.

(B) In the plasmonic structures studied, the intrinsic coupling-broadening
ratio (CBR) parameter is usually large, and by itself satisfies the
strong coupling criterion. Here, ``intrinsic'' implies that we consider
only spectral broadening effects that reflect (a) the molecular radiative
lifetime and fluorescence yield associated with the transition considered
and (b) the effect of the electromagnetic molecule-metal interaction
on the molecule radiative and non-radiative relaxation rates. This
intrinsic broadening disregards other sources of spectral broadening
such as congestion of many overlapping transitions and environmentally
induced thermal relaxation.

(C) While both the molecule-plasmon coupling and the intrinsic molecular
lifetime broadening (by ``intrinsic'' we mean broadening associated
only with the isolated molecule, the plasmonic nanostructure and their
interaction) strongly decrease with increasing molecule-metal surface
distance\footnote{For the metal-molecule distances considered in our calculations, the
intrinsic lifetime broadening is dominated by the molecule-metal interaction.}, the CBR remains large and the corresponding criterion for “strong
coupling” persists at large distances. This observation by itself
is meaningless as other sources of broadening as well as limited resolution
will usually mask the Rabi structure. This does indicates, however,
that if other sources of broadening and relaxation are eliminated
(such as with zero-phonon transitions at cryogenic temperatures),
signature of strong coupling would be observed in single molecule
plasmonic cavities, in agreement with a recent observation\cite{pscherer_single-molecule_2021}.

(D) Because of the interplay between local electromagnetic coupling
and particle induced relaxation, “hotspots” characterized by a large
enhancement of the local electromagnetic fields do not necessarily
stand out with regard to the CBR. To illustrate this consider the
perpendicular and parallel orientation of a molecular dipole of $10$
D transition dipole moment positioned $1$ nm apart from a silver
nanosphere of $10$ nm radius. For the perpendicular orientation,
the coupling strength is larger in magnitude than the parallel orientation:
$-U_{\perp}/\hbar=1.272\times10^{14}$ $\text{s}^{-1}$ (perpendicular)
and $-U_{\parallel}/\hbar=8.935\times10^{13}$ $\text{s}^{-1}$ (parallel).
However, the associated nonradiative relaxation rate is also larger
for perpendicular orientation: $\Gamma_{\text{NR}}^{\perp}=6.437\times10^{12}$
$\text{s}^{-1}$ (perpendicular) and $\Gamma_{\text{NR}}^{\parallel}=2.091\times10^{11}$
$\text{s}^{-1}$ (parallel) (the radiative decay rate is smaller than
these number by $\sim1$ order of magnitude at this small distance
from metal surface, so does not have significant contribution to the
CBR). As a result, the value of CBR for parallel orientation is more
than twice the value for perpendicular case ($\text{CBR}^{\perp}=36.733$,
$\text{CBR}^{\parallel}=80.744$).

Finally, consider the self-consistent treatment of Eq. (\ref{eq:A_omega}).
We reiterate that the results shown in Figures \ref{fig:QSC_sphere_perp_para_silver_gold_10nm}-\ref{fig:QSC_M_Gamma_bisph_para_gold_silver_10_10_dss}
are based on a calculation of $U$, $\Gamma_{\text{R}}$, $\Gamma_{\text{NR}}$
computed in a model in which the molecular dipole drives the system
and do not reflect the actual lineshape. In particular, such a calculation
does not account for spectral lineshifts induced by the proximity
of the molecule to the metal nanostructure and for the relative peaks
intensities, even if Rabi splitting occurs. Therefore, the results
shown in Figures \ref{fig:QSC_sphere_perp_para_silver_gold_10nm}-\ref{fig:QSC_M_Gamma_bisph_para_gold_silver_10_10_dss}
only indicate the fulfillment of the strong coupling criterion (Eq.
(\ref{eq:SC_condition})) but not the observability of an actual Rabi
splitting. Figures \ref{fig:alter_treatment_SC_sphere_gold}-\ref{fig:alter_treatment_SC_two_sphere_para_gold}
show results based on the self-consistent calculation, Eqs. (\ref{eq:EOM_damped_HO})-(\ref{eq:A_omega}),
which yields the actual absorption lineshape for the process in which
a dipolar emitter, coupled to the metal nanostructure, is driven by
an incident external field. 

Figure \ref{fig:alter_treatment_SC_sphere_gold} shows this lineshape,
calculated for a dipole emitter near a spherical gold nanosphere (radius
$10$ nm), oriented perpendicular to the sphere surface and parallel
to the incident field. The transition frequency $\omega_{0}$ is taken
$2.62$ eV, in resonance with the dipolar plasmon frequency of the
gold nanosphere. As before, the relaxation rate $\Gamma^{0}$ of the
free molecule is taken to be twice the radiative relaxation rate,
corresponding to an emission yield of $0.5$ for the free molecule.
Figure \ref{fig:SC_Abs_Lnshp_sphere_perp_gold_2D_10D_a0_10_r1_0.5}
depicts the absorption lineshape for a dipole situated at $0.5$ nm
away from the sphere surface for different values of the molecular
transition dipole moment in the range $2-10$ D. Figure \ref{fig:SC_Abs_Lnshp_sphere_perp_gold_10D_a0_10_r1_0.5_5}
displays the absorption lineshape for an emitter with transition dipole
moment $10$ D placed at different distances ($0.5-5$ nm) from the
sphere surface. Similar results for silver are shown in Figure \ref{subsec:SI_self_cons_silver}.
Figure \ref{fig:alter_treatment_SC_two_sphere_para_gold} shows similar
results for a molecular dipole in the middle of the gap between two
identical gold spheres (radii $10$ nm) oriented parallel to the intersphere
axis, where the incident electric field is parallel to this axis as
well. For a bispherical structure, the plasmonic response for an incident
plane field is characterized by a double-peak structure which results
from the interacting dipolar plasmonic responses of the individual
spheres and depends on the inter-sphere distance. The emitter transition
frequency is taken to be in resonance with the lower frequency and
more intense one of these peaks (see Figure \ref{fig:SI_rad_decay_bisph_uni_field_gold_para_perp}).
The results shown in Figure \ref{fig:Abs_Lnshp_bisph_para_gold_2_10D_10_10_1_0.5}
are for an intersphere gap size $1$ nm and different values of the
molecular transition dipole moment, while Figure \ref{fig:SC_Abs_Lnshp_bisph_para_gold_10D_10_10_gap_vary}
shows the lineshapes for a transition dipole $10$ D at varying gap
sizes.

\begin{figure}[H]
\subfloat[]{\includegraphics[scale=0.5]{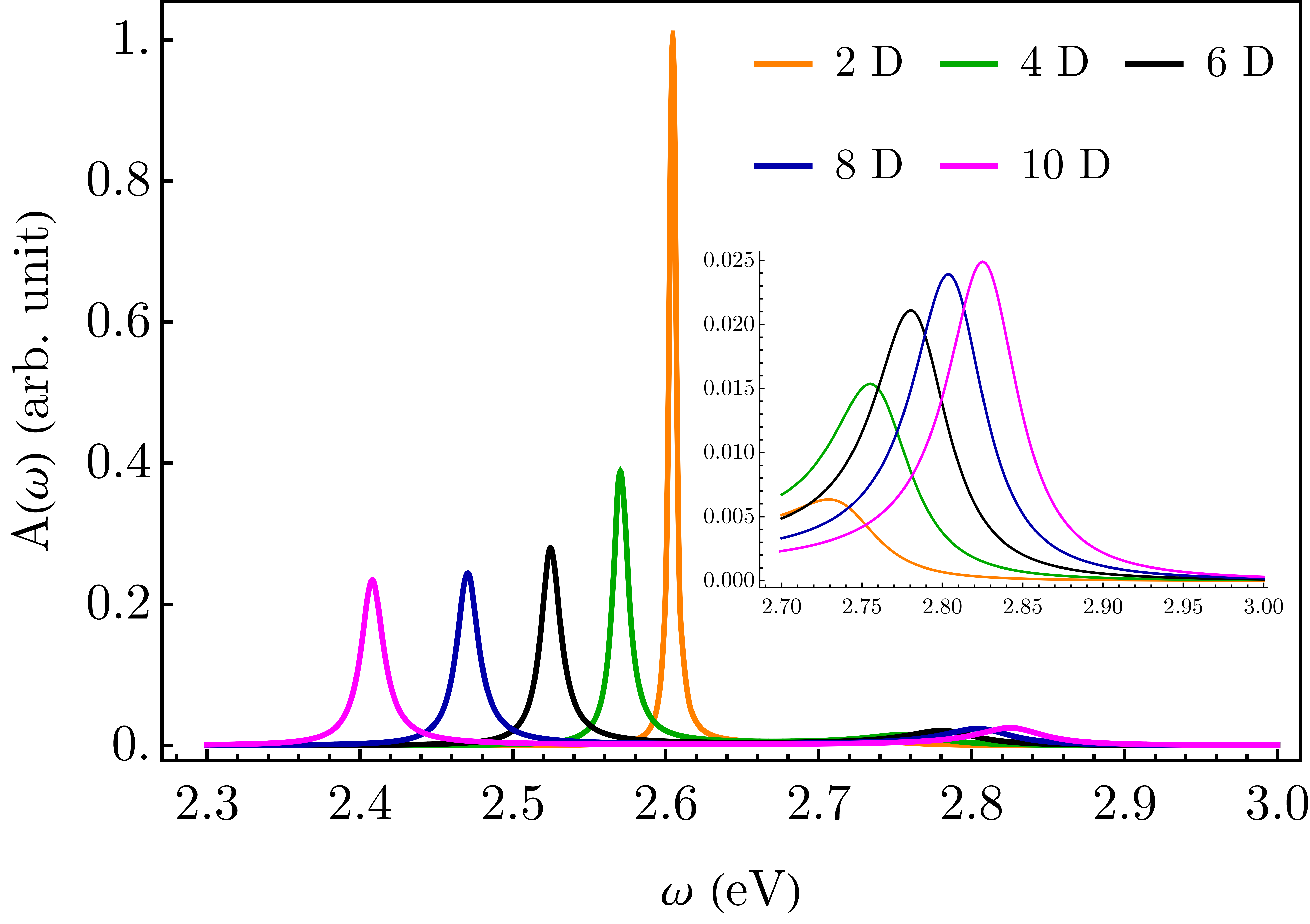}

\label{fig:SC_Abs_Lnshp_sphere_perp_gold_2D_10D_a0_10_r1_0.5}}\subfloat[]{\includegraphics[scale=0.5]{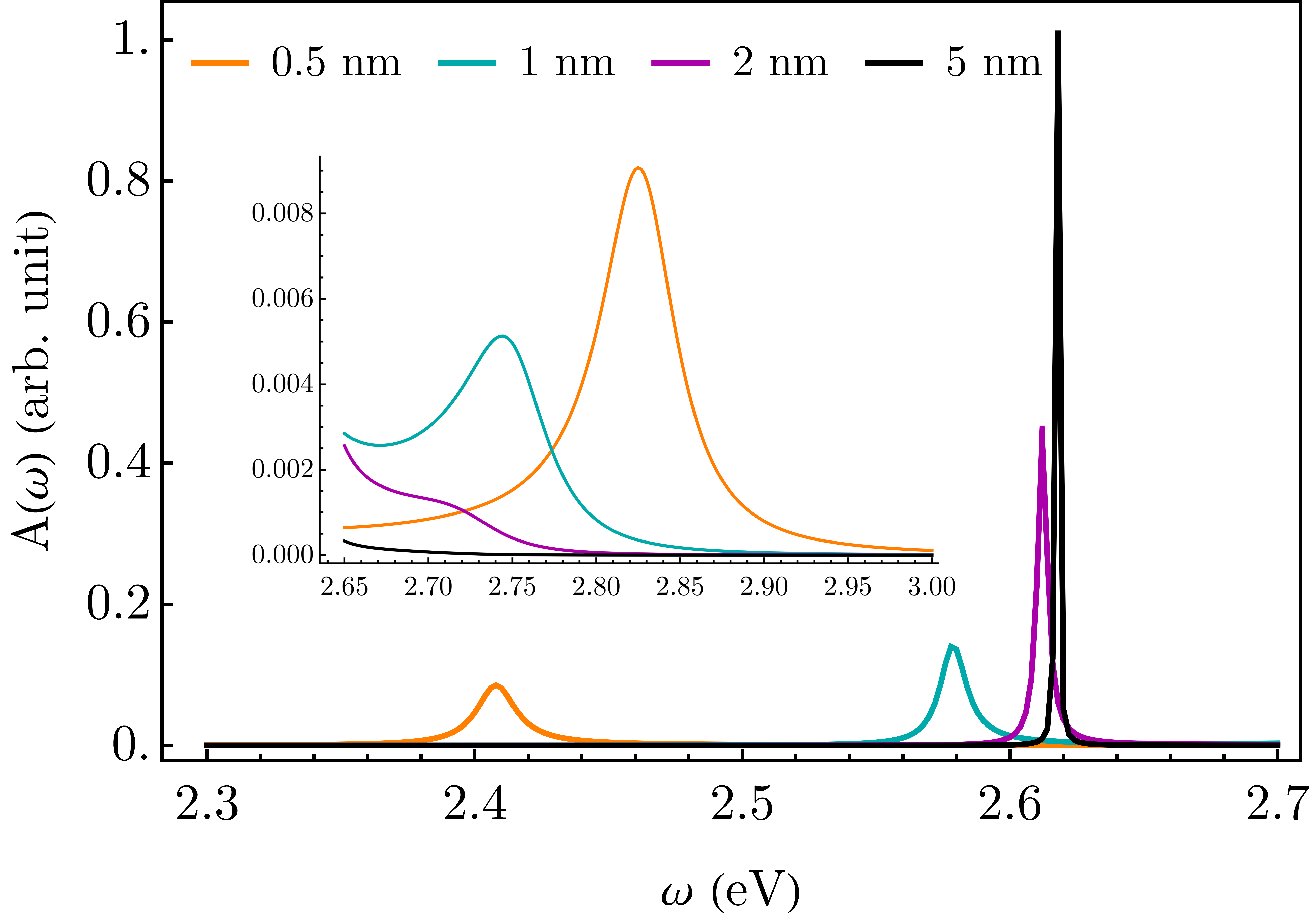}

\label{fig:SC_Abs_Lnshp_sphere_perp_gold_10D_a0_10_r1_0.5_5}}

\caption{The absorption lineshape, $A(\omega)$ (Eq. (\ref{eq:A_omega})),
displayed as a function of the driving frequency $\omega$ for an
emitter near a spherical gold nanoparticle of radius $10$ nm, perpendicular
to the sphere surface. The emitter transition frequency is taken $2.62$
eV, in resonance with the dipolar plasmon of a small gold particle,
and the relaxation rate of the free molecule is taken to be twice
the corresponding radiative relaxation rate (the latter are $2.95\times10^{6}$
$\text{s}^{-1}$, $1.18\times10^{7}$ $\text{s}^{-1}$, $2.66\times10^{7}$
$\text{s}^{-1}$, $4.72\times10^{7}$ $\text{s}^{-1}$ and $7.38\times10^{7}$
$\text{s}^{-1}$ for $D_{12}=2,4,6,8,10$ D respectively). (a) An
emitter with varying dipole moment placed at a distance $0.5$ nm
from the sphere surface. The inset shows a closeup of the small higher
energy peak structure that is dominated by the metal plasmon response.
(b) An emitter with a transition dipole moment $10$ D is placed at
different distances from the sphere surface. The inset displays the
small higher energy peak structures (not shown in the main figure
as they are practically invisible).}
\label{fig:alter_treatment_SC_sphere_gold}
\end{figure}

Consider first the case of an emitter near a spherical gold nanoparticle
(Figure \ref{fig:alter_treatment_SC_sphere_gold}). Keeping in mind
that in this model calculation the incident field is taken to be coupled
only to the emitter, it is the emitter absorption peak that is seen
for weak coupling (orange line in panel (a) and black line in panel
(b)). The main effect of strong coupling upon increasing the molecular
transition dipole is seen to be a strong red shift of the molecule-dominated
peak. The particle's optical signature, resulting the molecule-particle
interaction is seen on the high energy side of the spectrum, is small,
leading us to conclude that a pronounced Rabi splitting is not expected
in this configuration. In contrast, the optical response of a system
comprising a molecule positioned between two gold spheres does show
a clear evidence of such splitting as seen in Figure \ref{fig:alter_treatment_SC_two_sphere_para_gold}.
This observation is consistent with that of Ref. (\citen{heintz_few-molecule_2021})
that for a dimer-like structure of gold nanoparticles, a strong coupling
situation is likely to result when the interparticle gap is less than
$2$ nm.

\begin{figure}[H]
\subfloat[]{\includegraphics[scale=0.5]{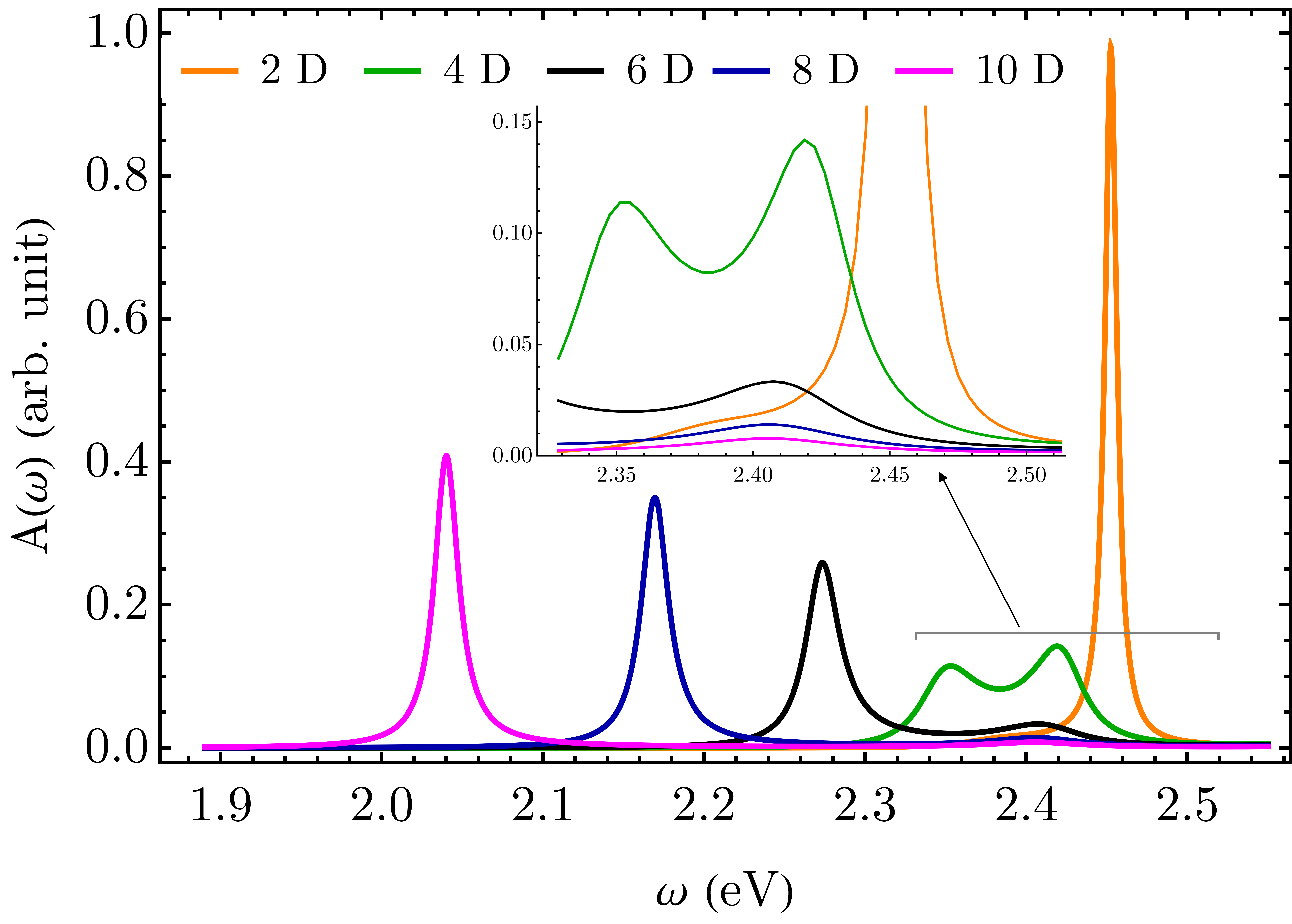}

\label{fig:Abs_Lnshp_bisph_para_gold_2_10D_10_10_1_0.5}}\subfloat[]{\includegraphics[scale=0.5]{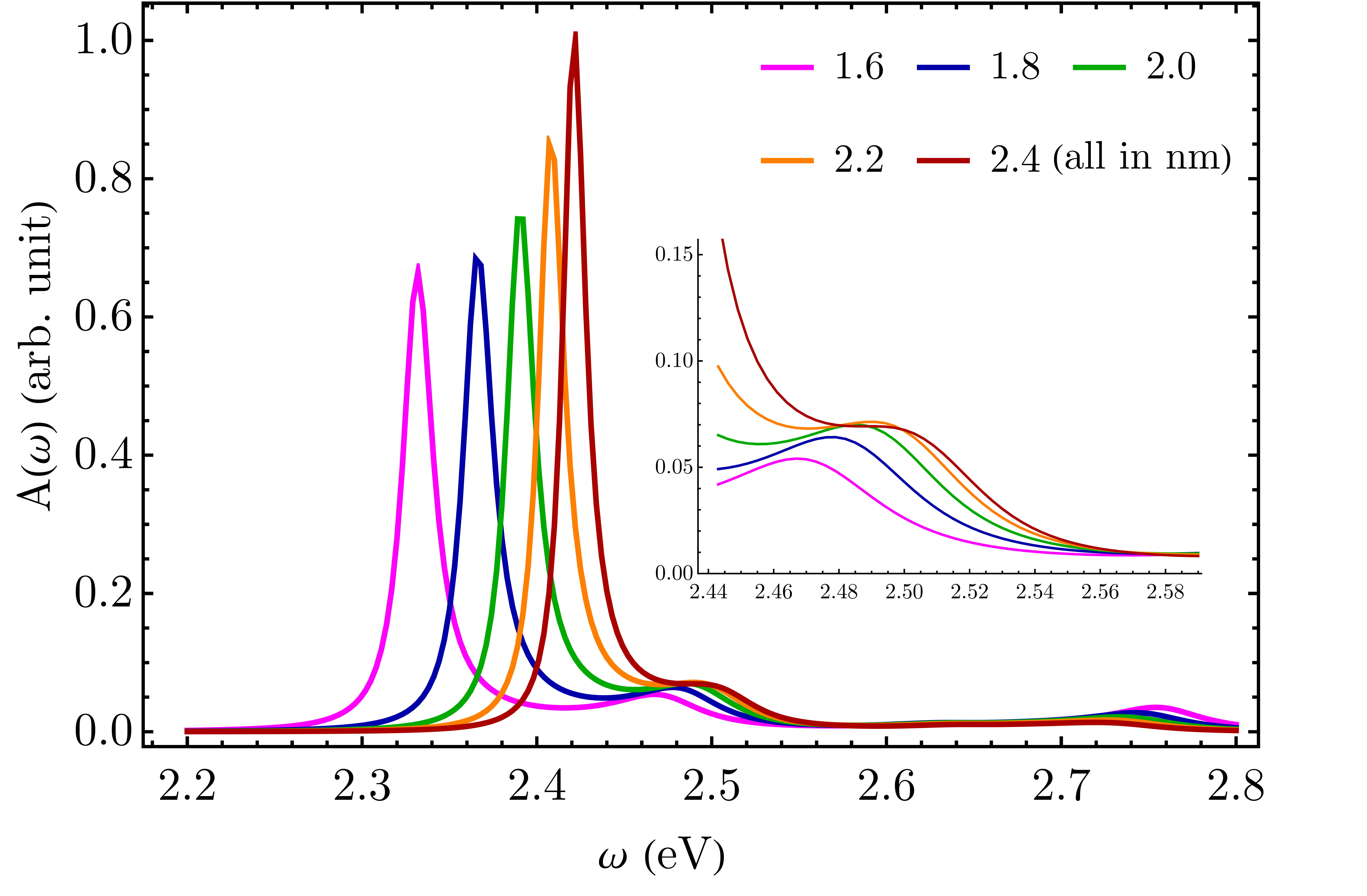}

\label{fig:SC_Abs_Lnshp_bisph_para_gold_10D_10_10_gap_vary}}

\caption{Same as Figure \ref{fig:alter_treatment_SC_sphere_gold} for an emitter
placed at the middle of the gap between two identical gold nanospheres
of $10$ nm radii oriented parallel to the bispherical axis and the
incident field. As in Figure \ref{fig:QSC_M_Gamma_bisph_para_perp_gold_silver_10_10_2}
(for gold, parallel configuration), the molecular transition frequency
is chosen to be in resonance with the brightest plasmon peak of the
corresponding gold bispherical structure with gap size $2$ nm ($2.47$
eV). (a) The intersphere gap size is $1$ nm and result are shown
for different emitter transition dipole moments in the range $2-10$
D. (b) The emitter transition dipole moment is $10$ D and the lineshape
is shown for gap sizes ranging from $1.6-2.4$ nm. The inset shows
a closeup for the higher energy peaks arising from the metal-molecule
coupling. Because the molecular transition frequency used here is
different from that used in Figure \ref{fig:alter_treatment_SC_sphere_gold},
the corresponding radiative relaxation rates are slightly different:
$2.48\times10^{6}$ $\text{s}^{-1}$, $9.92\times10^{6}$ $\text{s}^{-1}$,
$2.23\times10^{7}$ $\text{s}^{-1}$, $3.97\times10^{7}$ $\text{s}^{-1}$
and $6.20\times10^{7}$ $\text{s}^{-1}$ for $2,4,6,8,10$ D respectively.}
\label{fig:alter_treatment_SC_two_sphere_para_gold}
\end{figure}

While “strong coupling” between a single molecule and a metal-plasmon
excitation, manifested in the split-peak structure of the absorption
lineshape, is seen in the calculated spectra in Figure \ref{fig:alter_treatment_SC_two_sphere_para_gold},
these results also indicate that the simpler model calculation of
the CBR shown in Figures \ref{fig:QSC_sphere_perp_para_silver_gold_10nm}-\ref{fig:QSC_M_Gamma_bisph_para_gold_silver_10_10_dss}
is not by itself sufficient to predict the observability of this spectral
feature. The CBR, Eq. (\ref{eq:SC_condition}), does provide an estimate
of the relationship between the molecule-metal nanostructure coupling
and the associated broadening, but does not necessarily lead to an
observed Rabi splitting which is usually assumed to be a prime manifestation
of such strong coupling. We have identified the strong spectral shift,
another manifestation of this coupling, as the main reason for this
apparent discrepancy. This may also lead to apparent counter-intuitive
observations: for example, comparing the lineshapes in Figure \ref{fig:Abs_Lnshp_bisph_para_gold_2_10D_10_10_1_0.5},
a clear Rabi splitting is seen for an emitter with transition dipole
$4$ D, while a single peak is seen for a $2$ D emitter. In the stronger
coupling case of an emitter with transition dipole moment $\ge6$
D, a larger Rabi splitting is indicated by the calculation, but the
dominance of the lower energy peak may appear as a single peak in
a realistic observations. It should be noted that this shift will
also affect the distance dependence of the radiative and radiationless
relaxation rates of a molecule approaching a metal surface and may
potentially play a role in the observed ``quenching of quench'',
where the emission yield of an emission appears to increase, rather
than decrease, as the molecule-surface distance decreases (see Ref.
(\citen{kongsuwan_suppressed_2018} ) and references therein). We
emphasize however that no such trend in the emission yield $Y$, Eq.
(\ref{eq:q_yield}), was seen in our calculations.

\section{Conclusions \label{sec:conclusion}}

Strong radiation matter coupling is often characterized in the literature
as a relative concept, by comparing the absolute magnitudes of the
coupling $U$ and the broadening $\Gamma$, and the strong coupling
criterion $2|U|/\hbar\Gamma$, an indication that Rabi splitting may
be observed when the molecular and plasmon optical transition come
into resonance, is often used. While usually observed when an optical
mode interacts with many molecules, it has been noted, as outlined
in the introduction, that this criterion can be satisfied even for
single molecules interacting with metallic plasmonic structures. A
necessary condition for observing this phenomenon in such setting
is that the the coupling-broadening ratio (CBR), $2|U|/\hbar\Gamma$,
satisfies the strong coupling criterion when $\Gamma$ and $U$ are
derived from the interaction between a single molecular transition
when the molecule interacts only with the plasmonic structure, in
the absence of other sources of broadening. Using simple analytically
solvable model structures, representing the molecule as a point dipole
and treating radiation-matter coupling in the long wavelength approximation
whereas all distances are assumed small relative to the radiation
wavelength, we have found that this condition is easily satisfied.
Surprisingly, we observe that this remains true when the molecule-metal
distance increases so that the absolute coupling itself becomes small.
This suggest that Rabi splitting can be observed in systems involving
a single molecule interacting with small metal particles if other
sources of broadening can be suppressed, as was recently observed
using the low temperature molecular zero-phonon transition\cite{pscherer_single-molecule_2021}.

It should be kept in mind that even if the strong coupling criterion,
Eq. (\ref{eq:SC_condition}), is satisfied, such observation by itself
is not an indication of strong radiation-matter coupling when this
coupling is compared to other molecular energetic parameters. Furthermore,
by considering the results of a self-consistent calculation that addresses
directly the absorption lineshape in a system of interacting a molecule
and the plasmonic structure, we find that the intrinsic CBR satisfying
the strong coupling criterion does not guarantee the observation of
Rabi splitting in the absorption spectrum because the coupling induced
spectral shift together with the strong frequency dependence of the
imaginary part of plasmonic response can result in a spectrum dominated
by a single, albeit shifted, molecular peak.
\begin{acknowledgement}
M.S. acknowledges the financial support by the Air Force Office of
Scientific Research under Grant No. FA9550-19-1-0009. A.N. acknowledges
the support of the U.S. Department of Energy, Office of Science, Basic
Energy Sciences, Chemical Sciences, Geosciences, and Biosciences Division.
\end{acknowledgement}
 \clearpage

\bibliography{bibliography/library}

\end{document}


For completeness, here we provide details of the calculations described
in Section \ref{sec:response_function_relaxation} of the main text.
Those done with prolate spheroid follow the procedure of Ref. (\citen{Gersten1981}),
while those associated with a bispherical nanostructure follow Ref.
(\citen{aravind_interaction_1981}).

\section{A molecular dipole near a prolate spheroidal nanoparticle (arbitrary
orientation) \label{SI:sol_BC_spheroid_gen}}

\subsection{Prolate spheroidal coordinate system \label{SI:spheroid_coord_sys}}

This orthogonal coordinate system is obtained by rotating elliptic
coordinates along the major axis of the ellipse. With the foci of
the spheroid at $x=0,\,y=0,\,z=\pm a/2$, and distances $r_{1},\,r_{2}$
from the foci to the point $(x,y,z)$ respectively, the prolate spheroidal
coordinates $(\xi,\eta,\phi)$ are defined as:
\[
\xi=\frac{r_{1}+r_{2}}{a}\,,\,\,\eta=\frac{r_{1}-r_{2}}{a}\,,\,\,\phi=\arctan\Big(\frac{y}{x}\Big)
\]
where
\begin{align*}
r_{1} & =\sqrt{x^{2}+y^{2}+(z+\frac{a}{2})^{2}}\\
r_{2} & =\sqrt{x^{2}+y^{2}+(z-\frac{a}{2})^{2}}
\end{align*}
and
\[
1\le\xi<\infty,\,-1\le\eta\le1,\:0\le\phi<2\pi
\]
Cartesian coordinates are related to prolate spheroidal coordinates
by following relations:
\begin{align*}
x & =\frac{a}{2}\sqrt{(\xi^{2}-1)(1-\eta^{2})}\cos\phi
\end{align*}
\[
y=\frac{a}{2}\sqrt{(\xi^{2}-1)(1-\eta^{2})}\sin\phi
\]
\[
z=\frac{a\xi\eta}{2}
\]
The scale factors in this system are: 
\[
h_{\xi}=\frac{a}{2}\sqrt{\frac{\xi^{2}-\eta^{2}}{\xi^{2}-1}}\quad h_{\eta}=\frac{a}{2}\sqrt{\frac{\xi^{2}-\eta^{2}}{1-\eta^{2}}}
\]

\[
h_{\phi}=\frac{a}{2}\sqrt{(1-\eta^{2})(\xi^{2}-1)}
\]
The aspect ratio of semimajor (long) axis and semiminor (short) axis
of a spheroid, $p:q$ is given by: $\frac{p}{q}=\frac{\xi}{\sqrt{\xi^{2}-1}}$.
Increasing $a$ keeping $\xi$ constant gives larger spheroids with
same aspect ratio. Volume of a spheroid is given by: $V_{\text{sphd}}=\frac{4}{3}\pi(a/2)^{3}\xi(\xi^{2}-1)$

\subsection{The electric potential and field}

An oscillating dipole $\textbf{D}$ is located at $\textbf{r}_{1}(\xi_{1},\eta_{1},0)$
near a prolate spheroidal metal nanoparticle surface described by
$\xi_{0}$. The general solution for the electric potential inside
the nanoparticle is given by
\begin{align}
\Phi^{\text{in}}(\xi,\eta,\phi) & =\sum_{l=0}^{\infty}\sum_{m=0}^{l}A_{l,m}P_{l}^{m}(\xi)P_{l}^{m}(\eta)\cos m\phi\label{eq:phi_in_spheroid_SI}
\end{align}
Outside the nanoparticle,
\begin{align}
\Phi^{\text{out}}(\xi,\eta,\phi) & =\sum_{l=0}^{\infty}\sum_{m=0}^{l}B_{l,m}Q_{l}^{m}(\xi)P_{l}^{m}(\eta)\cos m\phi\nonumber \\
 & +\frac{1}{4\pi\varepsilon_{0}\varepsilon_{s}}\frac{{\bf D}\cdot({\bf r}-{\bf r}_{1})}{|{\bf r}-{\bf r}_{1}|^{3}}
\end{align}
We have
\begin{align}
\frac{({\bf r}-{\bf r}_{1})}{|{\bf r}-{\bf r}_{1}|^{3}} & =\boldsymbol{\nabla}_{1}\frac{1}{|{\bf r}-{\bf r}_{1}|}\label{eq:dipole_potential_general}
\end{align}
and
\begin{align}
 & \frac{1}{|{\bf r}-{\bf r}_{1}|}\nonumber \\
 & =\sum_{l=0}^{\infty}\sum_{m=0}^{l}K_{l,m}\cos m\phi P_{l}^{m}(\eta_{1})P_{l}^{m}(\eta)\nonumber \\
 & \times\begin{cases}
P_{l}^{m}(\xi_{1})Q_{l}^{m}(\xi); & \xi>\xi_{1}\\
P_{l}^{m}(\xi)Q_{l}^{m}(\xi_{1}); & \xi<\xi_{1}
\end{cases}\label{eq:green's_function_spheroid}
\end{align}
where
\begin{align*}
K_{l,m} & =(-1)^{m}\frac{2}{a}(2l+1)(2-\delta_{m0})\left[\frac{(l-m)!}{(l+m)!}\right]^{2}
\end{align*}
For $\xi<\xi_{1}$ this implies
\begin{align}
 & \Phi^{\text{out}}(\xi,\eta,\phi)\nonumber \\
 & =\sum_{l=0}^{\infty}\sum_{m=0}^{l}\Bigg[B_{l,m}Q_{l}^{m}(\xi)+\frac{H_{l,m}(\xi_{1},\eta_{1})}{\varepsilon_{s}}P_{l}^{m}(\xi)\Bigg]\nonumber \\
 & \times P_{l}^{m}(\eta)\cos m\phi\label{eq:phi_out_spheroid_SI}
\end{align}
where
\begin{align*}
 & H_{l,m}(\xi_{1},\eta_{1})\\
 & =\frac{K_{l,m}}{2\pi\varepsilon_{0}a}\Bigg[\sqrt{\frac{\xi_{1}^{2}-1}{\xi_{1}^{2}-\eta_{1}^{2}}}Q_{l}^{m^{\prime}}(\xi_{1})P_{l}^{m}(\eta_{1})D_{\xi}\\
 & +\sqrt{\frac{1-\eta_{1}^{2}}{\xi_{1}^{2}-\eta_{1}^{2}}}Q_{l}^{m}(\xi_{1})P_{l}^{m^{\prime}}(\eta_{1})D_{\eta}\Bigg]
\end{align*}
We use the following boundary conditions:
\begin{equation}
\Phi^{\text{in}}(\xi_{0},\eta,\phi)=\Phi^{\text{out}}(\xi_{0},\eta,\phi)\label{eq:BC_1_spheroid}
\end{equation}
and 

\begin{equation}
\varepsilon_{d}\frac{\partial\Phi^{\text{in}}(\xi,\eta,\phi)}{\partial\xi}\Bigg|_{\xi=\xi_{0}}=\varepsilon_{s}\frac{\partial\Phi^{\text{out}}(\xi,\eta,\phi)}{\partial\xi}\Bigg|_{\xi=\xi_{0}}\label{eq:BC_2_spheroid}
\end{equation}
So,
\begin{align}
A_{l,m} & =B_{l,m}\frac{Q_{l}^{m}(\xi_{0})}{P_{l}^{m}(\xi_{0})}+\frac{H_{l,m}(\xi_{1},\eta_{1})}{\varepsilon_{s}}\label{eq:BC_1_spheroid_explicit}
\end{align}

\begin{align}
A_{l,m} & =\frac{\varepsilon_{s}}{\varepsilon_{d}}\frac{Q_{l}^{m^{\prime}}(\xi_{0})}{P_{l}^{m^{\prime}}(\xi_{0})}B_{l,m}+\frac{1}{\varepsilon_{d}}H_{l,m}(\xi_{1},\eta_{1})\label{eq:BC_2_spheroid_explicit}
\end{align}
Solving \ref{eq:BC_1_spheroid_explicit} and \ref{eq:BC_2_spheroid_explicit}
we have

\begin{align}
A_{l,m} & =H_{l,m}(\xi_{1},\eta_{1})\nonumber \\
 & \times\frac{P_{l}^{m^{\prime}}(\xi_{0})Q_{l}^{m}(\xi_{0})-P_{l}^{m}(\xi_{0})Q_{l}^{m^{\prime}}(\xi_{0})}{\varepsilon_{d}P_{l}^{m^{\prime}}(\xi_{0})Q_{l}^{m}(\xi_{0})-\varepsilon_{s}P_{l}^{m}(\xi_{0})Q_{l}^{m^{\prime}}(\xi_{0})}\label{eq:soln_Alm_spehroid_general}
\end{align}
and 

\begin{align}
B_{l,m} & =-H_{l,m}(\xi_{1},\eta_{1})\frac{\varepsilon_{d}-\varepsilon_{s}}{\varepsilon_{s}}\nonumber \\
 & \times\frac{P_{l}^{m}(\xi_{0})P_{l}^{m^{\prime}}(\xi_{0})}{\varepsilon_{d}P_{l}^{m^{\prime}}(\xi_{0})Q_{l}^{m}(\xi_{0})-\varepsilon_{s}P_{l}^{m}(\xi_{0})Q_{l}^{m^{\prime}}(\xi_{0})}\label{eq:soln_Blm_spehroid_general}
\end{align}
Now that we have $A_{l,m}$ and $B_{l,m}$ we find the electric field
and hence the component of ${\bf \underline{G}}$ in the dipole direction:

\begin{align}
 & G_{\text{sphd-D}}({\bf r}_{1})\nonumber \\
 & =-\frac{\textbf{D}}{|\textbf{D}|^{2}}\cdot\boldsymbol{\nabla}\Phi^{\text{out}}(\xi_{1},\eta_{1},\phi_{1})\nonumber \\
 & =\sum_{l=0}^{\infty}\sum_{m=0}^{l}\frac{K_{l,m}}{\pi\varepsilon_{0}a^{2}|\textbf{D}|^{2}}\frac{\varepsilon_{d}-\varepsilon_{s}}{\varepsilon_{s}}\nonumber \\
 & \times\frac{P_{l}^{m}(\xi_{0})P_{l}^{m^{\prime}}(\xi_{0})}{\varepsilon_{d}P_{l}^{m^{\prime}}(\xi_{0})Q_{l}^{m}(\xi_{0})-\varepsilon_{s}P_{l}^{m}(\xi_{0})Q_{l}^{m^{\prime}}(\xi_{0})}\nonumber \\
 & \Bigg[\sqrt{\frac{\xi_{1}^{2}-1}{\xi_{1}^{2}-\eta_{1}^{2}}}Q_{l}^{m^{\prime}}(\xi_{1})P_{l}^{m}(\eta_{1})D_{\xi}\nonumber \\
 & \qquad\qquad+\sqrt{\frac{1-\eta_{1}^{2}}{\xi_{1}^{2}-\eta_{1}^{2}}}Q_{l}^{m}(\xi_{1})P_{l}^{m^{\prime}}(\eta_{1})D_{\eta}\Bigg]^{2}\label{eq:U_spheroid_general}
\end{align}
\textbf{Special case 1: Molecule on major (=}\textbf{\textit{maj}}\textbf{)
axis, dipole perpendicular to nanoparticle surface:}

$\eta_{1}=1$, $\textbf{D}=D_{\xi}\boldsymbol{\widehat{\xi}}$

\begin{align}
 & G_{\text{sphd}}^{\text{maj,}\perp}(\xi_{1})\nonumber \\
 & =\frac{2(\varepsilon_{d}-\varepsilon_{s})}{\pi\varepsilon_{0}\varepsilon_{s}a^{3}}\sum_{l=0}^{\infty}(2l+1)\nonumber \\
 & \times\frac{P_{l}(\xi_{0})P_{l}^{\prime}(\xi_{0})Q_{l}^{\prime}(\xi_{1})^{2}}{\varepsilon_{d}P_{l}^{\prime}(\xi_{0})Q_{l}(\xi_{0})-\varepsilon_{s}P_{l}(\xi_{0})Q_{l}^{\prime}(\xi_{0})}\label{eq:V_spheroid_major_perp}
\end{align}
\textbf{Special case 2: Molecule on major axis, dipole parallel to
nanoparticle surface:}

$\eta_{1}=1$, $\textbf{D}=D_{\eta}\boldsymbol{\widehat{\eta}}$

\begin{align}
 & G_{\text{sphd}}^{\text{maj,}\parallel}(\xi_{1})\nonumber \\
 & =\frac{(\varepsilon_{d}-\varepsilon_{s})}{\pi\varepsilon_{0}\varepsilon_{s}a^{3}(1-\xi_{1}^{2})}\sum_{l=1}^{\infty}(2l+1)\nonumber \\
 & \times\frac{P_{l}^{1}(\xi_{0})P_{l}^{1^{\prime}}(\xi_{0})Q_{l}^{1}(\xi_{1})^{2}}{\varepsilon_{d}P_{l}^{1^{\prime}}(\xi_{0})Q_{l}^{1}(\xi_{0})-\varepsilon_{s}P_{l}^{1}(\xi_{0})Q_{l}^{1^{\prime}}(\xi_{0})}\label{eq:V_spheroid_major_para}
\end{align}
\textbf{Special case 3: Molecule on minor (=}\textbf{\textit{min}}\textbf{)
axis, dipole perpendicular to nanoparticle surface:}

$\eta_{1}=0$, $\textbf{D}=D_{\xi}\boldsymbol{\widehat{\xi}}$

\begin{align}
 & G_{\text{sphd}}^{\text{min,}\perp}(\xi_{1})\nonumber \\
 & =\frac{(\varepsilon_{d}-\varepsilon_{s})}{\pi\varepsilon_{0}\varepsilon_{s}a^{2}}\frac{(\xi_{1}^{2}-1)}{\xi_{1}^{2}}\sum_{l=0}^{\infty}\sum_{m=0}^{l}K_{l,m}\nonumber \\
 & \times\frac{P_{l}^{m}(\xi_{0})P_{l}^{m^{\prime}}(\xi_{0})Q_{l}^{m^{\prime}}(\xi_{1})^{2}P_{l}^{m}(0)^{2}}{\varepsilon_{d}P_{l}^{m^{\prime}}(\xi_{0})Q_{l}^{m}(\xi_{0})-\varepsilon_{s}P_{l}^{m}(\xi_{0})Q_{l}^{m^{\prime}}(\xi_{0})}\label{eq:V_spheroid_minior_perp}
\end{align}
\textbf{Special case 4: Molecule on minor axis, dipole parallel to
nanoparticle surface:}

$\eta_{1}=0$, $\textbf{D}=D_{\eta}\boldsymbol{\widehat{\eta}}$
\begin{align}
 & G_{\text{sphd}}^{\text{min,}\parallel}(\xi_{1})\nonumber \\
 & =\frac{(\varepsilon_{d}-\varepsilon_{s})P_{l}^{m}}{\pi\varepsilon_{0}\varepsilon_{s}a^{2}\xi_{1}^{2}}\sum_{l=0}^{\infty}\sum_{m=0}^{l}K_{l,m}\nonumber \\
 & \times\frac{(\xi_{0})P_{l}^{m^{\prime}}(\xi_{0})Q_{l}^{m}(\xi_{1})^{2}P_{l}^{m^{\prime}}(0)^{2}}{\varepsilon_{d}P_{l}^{m^{\prime}}(\xi_{0})Q_{l}^{m}(\xi_{0})-\varepsilon_{s}P_{l}^{m}(\xi_{0})Q_{l}^{m^{\prime}}(\xi_{0})}\label{eq:V_spheroid_minor_para}
\end{align}

\subsection{The nonradiative decay rate}

The nonradiative decay rate is given by the following volume integral
of the amplitude of electric field inside the nanoparticle squared
\cite{Gersten1981}: 

\begin{align}
\Gamma_{\text{NR,sphd}} & =\frac{\sigma}{2\hbar\omega}\int_{V}d\textbf{r}|\textbf{E}_{\text{in}}|^{2}\nonumber \\
 & =\frac{\varepsilon_{0}\text{Im}[\varepsilon_{d}(\omega)]}{2\hbar}\oint_{S}\Phi^{\text{in}^{*}}\boldsymbol{\nabla}\Phi^{\text{in}}\cdot d\textbf{S}\label{eq:NR_decay_rate_spheroid_general}
\end{align}

\begin{align}
 & \oint_{S}\Phi^{\text{in}^{*}}\boldsymbol{\nabla}\Phi^{\text{in}}\cdot d\textbf{S}\nonumber \\
 & =\frac{a}{2}\left(\xi_{0}^{2}-1\right)\oint_{S}\Phi^{\text{in}^{*}}\frac{\partial\Phi^{\text{in}}}{\partial\xi}d\eta\,d\phi\nonumber \\
 & =\pi a\left(\xi_{0}^{2}-1\right)\sum_{l,m}\big|A_{l,m}\big|^{2}P_{l}^{m}(\xi_{0})P_{l}^{m^{\prime}}(\xi_{0})\nonumber \\
 & \times\frac{(1+\delta_{m0})(l+m)!}{(2l+1)(l-m)!}
\end{align}

Hence,

\begin{align}
 & \Gamma_{\text{NR,sphd}}\nonumber \\
 & =\frac{\pi a\varepsilon_{0}\text{Im}[\varepsilon_{d}(\omega)]\left(\xi_{0}^{2}-1\right)}{2\hbar}\nonumber \\
 & \times\sum_{l,m}\big|A_{l,m}\big|^{2}P_{l}^{m}(\xi_{0})P_{l}^{m^{\prime}}(\xi_{0})\frac{(1+\delta_{m0})(l+m)!}{(2l+1)(l-m)!}\label{eq:NR_decay_rate_spheroid_explicit}
\end{align}

\textbf{Special case 1: Molecule on major axis, dipole perpendicular
to nanoparticle surface:}

\begin{align}
 & \Gamma_{\text{NR,sphd}}^{\text{maj,}\perp}\nonumber \\
 & =\frac{D_{\xi}^{2}\text{Im}[\varepsilon_{d}(\omega)]}{\pi\varepsilon_{0}a^{3}\hbar\left(\xi_{0}^{2}-1\right)}\nonumber \\
 & \times\sum_{l=0}^{\infty}\frac{(2l+1)P_{l}(\xi_{0})P_{l}^{\prime}(\xi_{0})Q_{l}^{\prime}(\xi_{1})^{2}}{\Big|\varepsilon_{d}P_{l}^{\prime}(\xi_{0})Q_{l}(\xi_{0})-\varepsilon_{s}P_{l}(\xi_{0})Q_{l}^{\prime}(\xi_{0})\Big|^{2}}\label{eq:NR_decay_rate_spheroid_major_perp}
\end{align}

\textbf{Special case 2: Molecule on major axis, dipole parallel to
nanoparticle surface:}

\begin{align}
 & \Gamma_{\text{NR,sphd}}^{\text{maj,}\parallel}\nonumber \\
 & =\frac{D_{\eta}^{2}\text{Im}[\varepsilon_{d}(\omega)]}{2\pi\varepsilon_{0}a^{3}\hbar(\xi_{1}^{2}-1)\left(\xi_{0}^{2}-1\right)}\nonumber \\
 & \times\sum_{l=1}^{\infty}\frac{l(l+1)(2l+1)P_{l}^{1}(\xi_{0})P_{l}^{1^{\prime}}(\xi_{0})Q_{l}^{1}(\xi_{1})^{2}}{\Big|\varepsilon_{d}P_{l}^{1^{\prime}}(\xi_{0})Q_{l}^{1}(\xi_{0})-\varepsilon_{s}P_{l}^{1}(\xi_{0})Q_{l}^{1^{\prime}}(\xi_{0})\Big|^{2}}\label{eq:NR_decay_rate_spheroid_major_para}
\end{align}

\textbf{Special case 3: Molecule on minor axis, dipole perpendicular
to nanoparticle surface:}

\begin{align}
 & \Gamma_{\text{NR,sphd}}^{\text{min,}\perp}\nonumber \\
 & =\frac{D_{\xi}^{2}\text{Im}[\varepsilon_{d}(\omega)](\xi_{1}^{2}-1)}{\pi\varepsilon_{0}a^{3}\hbar\left(\xi_{0}^{2}-1\right)\xi_{1}^{2}}\nonumber \\
 & \times\sum_{l,m}(2l+1)(2-\delta_{m0})\frac{(l-m)!}{(l+m)!}\nonumber \\
 & \times\frac{P_{l}^{m}(\xi_{0})P_{l}^{m^{\prime}}(\xi_{0})Q_{l}^{m^{\prime}}(\xi_{1})^{2}P_{l}^{m}(0)^{2}}{\Big|\varepsilon_{d}P_{l}^{m^{\prime}}(\xi_{0})Q_{l}^{m}(\xi_{0})-\varepsilon_{s}P_{l}^{m}(\xi_{0})Q_{l}^{m^{\prime}}(\xi_{0})\Big|^{2}}\label{eq:NR_decay_rate_spheroid_minor_perp}
\end{align}

\textbf{Special case 4: Molecule on minor axis, dipole parallel to
nanoparticle surface:}

\begin{align}
 & \Gamma_{\text{NR,sphd}}^{\text{min,}\parallel}\nonumber \\
 & =\frac{D_{\eta}^{2}\text{Im}[\varepsilon_{d}(\omega)]}{\pi\varepsilon_{0}a^{3}\hbar\left(\xi_{0}^{2}-1\right)\xi_{1}^{2}}\nonumber \\
 & \times\sum_{l,m}(2l+1)(2-\delta_{m0})\frac{(l-m)!}{(l+m)!}\nonumber \\
 & \times\frac{P_{l}^{m}(\xi_{0})P_{l}^{m^{\prime}}(\xi_{0})Q_{l}^{m}(\xi_{1})^{2}P_{l}^{m^{\prime}}(0)^{2}}{\Big|\varepsilon_{d}P_{l}^{m^{\prime}}(\xi_{0})Q_{l}^{m}(\xi_{0})-\varepsilon_{s}P_{l}^{m}(\xi_{0})Q_{l}^{m^{\prime}}(\xi_{0})\Big|^{2}}\label{eq:NR_decay_rate_spheroid_minor_para}
\end{align}

\subsection{The radiative decay rate}

The radiative decay rate can be calculated by evaluating the effective
dipole in the system and then using

\begin{align}
\Gamma_{\text{R}}(\omega) & =\frac{\sqrt{\varepsilon_{s}}\omega^{3}}{3\hbar c^{3}4\pi\varepsilon_{0}}\Big|{\bf D}_{\text{tot}}(\omega)\Big|^{2}\label{eq:rad_decay_expression}
\end{align}
The needed total dipole can be evaluated from the asymptotic behavior
of the calculated electrostatic potential at large distance from the
molecule-particle system. This asymptotic behavior corresponds to
the net dipole moment of the system.

In this limit, 

\begin{align*}
 & \Phi^{\text{out}}\rightarrow\sum_{m=0}^{1}\left[B_{1m}+J_{1,m}(\xi_{1},\eta_{1})\right]\\
 & \quad\quad\quad\quad\quad\quad\quad\times\cos m\phi Q_{1}^{m}(\xi)P_{1}^{m}(\eta)
\end{align*}

Where

\begin{align*}
 & J_{l,m}(\xi_{1},\eta_{1})\\
 & =\frac{2}{a}\frac{K_{l,m}}{4\pi\varepsilon_{0}\varepsilon_{s}}\Bigg[\sqrt{\frac{\xi_{1}^{2}-1}{\xi_{1}^{2}-\eta_{1}^{2}}}P_{l}^{m^{\prime}}(\xi_{1})P_{l}^{m}(\eta_{1})D_{\xi}\\
 & +\sqrt{\frac{1-\eta_{1}^{2}}{\xi_{1}^{2}-\eta_{1}^{2}}}P_{l}^{m}(\xi_{1})P_{l}^{m^{\prime}}(\eta_{1})D_{\eta}\Bigg]
\end{align*}

\textbf{Special case 1: Molecule on major axis, dipole perpendicular
to nanoparticle surface:}

\begin{align}
 & D_{\text{tot}}^{\text{maj,}\perp}\nonumber \\
 & =D_{\xi}\left[1-\frac{(\varepsilon_{d}-\varepsilon_{s})\xi_{0}Q_{1}^{\prime}(\xi_{1})}{\varepsilon_{s}\left[\varepsilon_{d}Q_{1}(\xi_{0})-\varepsilon_{s}\xi_{0}Q_{1}^{\prime}(\xi_{0})\right]}\right]\label{eq:dipole_net_spheroid_major_perp}
\end{align}

\textbf{Special case 2: Molecule on major axis, dipole parallel to
nanoparticle surface:}

\begin{align}
 & D_{\text{tot}}^{\text{maj,}\parallel}\nonumber \\
 & =D_{\eta}\left[1-\frac{(\varepsilon_{d}-\varepsilon_{s})P_{1}^{1}(\xi_{0})P_{1}^{1^{\prime}}(\xi_{0})Q_{1}^{1}(\xi_{1})}{\varepsilon_{s}P_{1}^{1}(\xi_{1})\left[\varepsilon_{d}P_{1}^{1^{\prime}}(\xi_{0})Q_{1}^{1}(\xi_{0})-\varepsilon_{s}P_{1}^{1}(\xi_{0})Q_{1}^{1^{\prime}}(\xi_{0})\right]}\right]\label{eq:dipole_net_spheroid_major_para}
\end{align}

\textbf{Special case 3: Molecule on minor axis, dipole perpendicular
to nanoparticle surface:}

\begin{align}
 & D_{\text{tot}}^{\text{min,}\perp}\nonumber \\
 & =D_{\xi}\left[1-\frac{(\varepsilon_{d}-\varepsilon_{s})P_{1}^{1}(\xi_{0})P_{1}^{1^{\prime}}(\xi_{0})Q_{1}^{1^{\prime}}(\xi_{1})}{\varepsilon_{s}P_{1}^{1^{\prime}}(\xi_{1})\left[\varepsilon_{d}P_{1}^{1^{\prime}}(\xi_{0})Q_{1}^{1}(\xi_{0})-\varepsilon_{s}P_{1}^{1}(\xi_{0})Q_{1}^{1^{\prime}}(\xi_{0})\right]}\right]\label{eq:dipole_net_spheroid_minor_perp}
\end{align}

\textbf{Special case 4: Molecule on minor axis, dipole parallel to
nanoparticle surface:}

\begin{align}
 & D_{\text{tot}}^{\text{min,}\parallel}\nonumber \\
 & =D_{\eta}\left[1-\frac{(\varepsilon_{d}-\varepsilon_{s})P_{1}(\xi_{0})Q_{1}(\xi_{1})}{\varepsilon_{s}P_{1}(\xi_{1})\left[\varepsilon_{d}Q_{1}(\xi_{0})-\varepsilon_{s}\xi_{0}Q_{1}^{\prime}(\xi_{0})\right]}\right]\label{eq:dipole_net_spheroid_minor_para}
\end{align}

\section{The spherical limit \label{SI:sol_BC_sphere_gen}}

Here we derive expressions for electric field and decay rates for
a system of an oscillating dipole $\textbf{D}$ located at $\textbf{r}_{1}$
near a spherical metal nanoparticle of radius $a$ by solving the
boundary conditions. 

\subsection{The electric potential and field}

As the natural symmetry of the system dictates, we will use spherical
coordinates. Spherical harmonics are defined by\cite{JacksonElectrodynamics1999}

\[
Y_{l,m}(\theta,\phi)=\sqrt{\frac{2l+1}{4\pi}\frac{(l-m)!}{(l+m)!}}P_{l}^{m}(\cos\theta)e^{im\phi}
\]

Inside the nanosphere, the electric potential in general is

\begin{align}
\Phi^{\text{in}}(r,\theta,\phi) & =\sum_{l=0}^{\infty}\sum_{m=-l}^{l}A_{l,m}r^{l}Y_{l,m}(\theta,\phi)
\end{align}

where $Y_{l,m}(\theta,\phi)$ are the spherical harmonics.

Outside, it is

\begin{align}
\Phi^{\text{out}}(r,\theta,\phi) & =\sum_{l=0}^{\infty}\sum_{m=-l}^{l}\frac{B_{l,m}}{r^{l+1}}Y_{l,m}(\theta,\phi)\nonumber \\
 & \quad\quad+\frac{1}{4\pi\varepsilon_{0}\varepsilon_{s}}\frac{{\bf D}\cdot({\bf r}-{\bf r}_{1})}{|{\bf r}-{\bf r}_{1}|^{3}}
\end{align}
and

\begin{align}
 & \frac{1}{|{\bf r}-{\bf r}_{1}|}\nonumber \\
 & =4\pi\sum_{l=0}^{\infty}\sum_{m=-l}^{l}\frac{1}{2l+1}\frac{r_{<}^{l}}{r_{>}^{l+1}}Y_{l,m}^{*}(\theta_{1},\phi_{1})Y_{l,m}(\theta,\phi)
\end{align}

For the region $r<r_{1}$,

\begin{align}
 & \frac{1}{4\pi\varepsilon_{0}\varepsilon_{s}}\frac{{\bf D}\cdot({\bf r}-{\bf r}_{1})}{|{\bf r}-{\bf r}_{1}|^{3}}\nonumber \\
 & ={\bf D}\cdot\frac{1}{\varepsilon_{0}\varepsilon_{s}}\sum_{l=0}^{\infty}\sum_{m=-l}^{l}\frac{r^{l}}{2l+1}\nonumber \\
 & \times Y_{l,m}(\theta,\phi)\mathcal{\boldsymbol{Q}}_{l,m}^{*}(r_{1},\theta_{1},\phi_{1})
\end{align}

where

\begin{align*}
 & \mathcal{\boldsymbol{Q}}_{l,m}^{*}(r_{1},\theta_{1},\phi_{1})\\
 & =\Bigg[-(l+1)\frac{Y_{l,m}^{*}(\theta_{1},\phi_{1})}{r_{1}^{l+2}}\boldsymbol{\widehat{r}}+\\
 & \frac{1}{r_{1}^{l+2}}\frac{\partial Y_{l,m}^{*}(\theta_{1},\phi_{1})}{\partial\theta_{1}}\boldsymbol{\widehat{\theta}}+\frac{1}{r_{1}^{l+2}\sin\theta_{1}}\frac{\partial Y_{l,m}^{*}(\theta_{1},\phi_{1})}{\partial\phi_{1}}\boldsymbol{\widehat{\phi}}\Bigg]
\end{align*}

The molecular dipole is in general in an arbitrary orientation $\textbf{D}=D_{r}\boldsymbol{\widehat{r}}+D_{\theta}\boldsymbol{\widehat{\theta}}+D_{\phi}\boldsymbol{\widehat{\phi}}$.
So,

\begin{align}
 & \frac{1}{4\pi\varepsilon_{0}\varepsilon_{s}}\frac{{\bf D}\cdot({\bf r}-{\bf r}_{1})}{|{\bf r}-{\bf r}_{1}|^{3}}\nonumber \\
 & =\frac{1}{\varepsilon_{0}\varepsilon_{s}}\sum_{l=0}^{\infty}\sum_{m=-l}^{l}\frac{r^{l}}{2l+1}Y_{l,m}(\theta,\phi)\mathcal{P}_{l,m}^{*}(r_{1},\theta_{1},\phi_{1})
\end{align}

Where

\begin{align*}
 & \mathcal{P}_{l,m}^{*}(r_{1},\theta_{1},\phi_{1})\\
 & =\Bigg[-D_{r}(l+1)\frac{Y_{l,m}^{*}(\theta_{1},\phi_{1})}{r_{1}^{l+2}}\\
 & +\frac{D_{\theta}}{r_{1}^{l+2}}\frac{\partial Y_{l,m}^{*}(\theta_{1},\phi_{1})}{\partial\theta_{1}}+\frac{D_{\phi}}{r_{1}^{l+2}\sin\theta_{1}}\frac{\partial Y_{l,m}^{*}(\theta_{1},\phi_{1})}{\partial\phi_{1}}\Bigg]
\end{align*}

We have two boundary conditions as follows: 

\begin{align}
A_{l,m} & =\frac{B_{l,m}}{a^{2l+1}}+\frac{1}{\varepsilon_{0}\varepsilon_{s}}\frac{\mathcal{P}_{l,m}^{*}(r_{1},\theta_{1},\phi_{1})}{2l+1}\label{eq:BC_Sphere_1}
\end{align}

and 

\begin{align}
A_{l,m} & =-\frac{\varepsilon_{s}}{\varepsilon_{d}}\frac{(l+1)}{l}\frac{B_{l,m}}{a^{2l+1}}+\frac{1}{\varepsilon_{0}\varepsilon_{d}}\frac{\mathcal{P}_{l,m}^{*}(r_{1},\theta_{1},\phi_{1})}{2l+1}\label{eq:BC_Sphere_2}
\end{align}

Solving \ref{eq:BC_Sphere_1} and \ref{eq:BC_Sphere_2}, we have:

\begin{align}
A_{l,m} & =\frac{\mathcal{P}_{l,m}^{*}(r_{1},\theta_{1},\phi_{1})}{\varepsilon_{0}\left[l\varepsilon_{d}+(l+1)\varepsilon_{s}\right]}
\end{align}

and 

\begin{align}
B_{l,m} & =\frac{l}{(2l+1)}\frac{\varepsilon_{s}-\varepsilon_{d}}{\varepsilon_{0}\varepsilon_{s}\left[l\varepsilon_{d}+(l+1)\varepsilon_{s}\right]}\nonumber \\
 & \quad\quad\quad\quad\times a^{2l+1}\mathcal{P}_{l,m}^{*}(r_{1},\theta_{1},\phi_{1})
\end{align}

So we have the potential inside and outside the nanoparticle. We obtain
the component of ${\bf \underline{G}}$ in the direction of the dipole:

\begin{align}
 & G_{\text{sph}}({\bf r}_{1})\nonumber \\
 & =\frac{1}{|\textbf{D}|^{2}}\sum_{l=0}^{\infty}\sum_{m=-l}^{l}\frac{l}{(2l+1)}\frac{\epsilon_{s}-\epsilon_{d}}{\epsilon_{0}\epsilon_{s}\left[l\epsilon_{d}+(l+1)\epsilon_{s}\right]}a^{2l+1}\nonumber \\
 & \times|\mathcal{P}_{l,m}(r_{1},\theta_{1},\phi_{1})|^{2}
\end{align}

\textbf{Special case 1: Dipole perpendicular to sphere surface }

$\textbf{D}=D_{r}\boldsymbol{\widehat{r}}$:

\begin{align}
 & G_{\text{sph}}^{\perp}(r_{1})\nonumber \\
 & =\sum_{l=0}^{\infty}\frac{l(l+1)^{2}}{(2l+1)}\frac{\epsilon_{s}-\epsilon_{d}}{\epsilon_{0}\epsilon_{s}\left[l\epsilon_{d}+(l+1)\epsilon_{s}\right]}\frac{a^{2l+1}}{r_{1}^{2l+4}}\nonumber \\
 & \times\sum_{m=-l}^{l}|Y_{l,m}^{*}(\theta_{1},\phi_{1})|^{2}\nonumber \\
 & =\frac{1}{4\pi\epsilon_{0}\epsilon_{s}a^{3}}\sum_{l=0}^{\infty}\frac{(\epsilon_{s}-\epsilon_{d})l(l+1)^{2}}{\left[l\epsilon_{d}+(l+1)\epsilon_{s}\right]}\Big(\frac{a}{r_{1}}\Big)^{2l+4}
\end{align}

In the last step we have used following property of Spherical Harmonics\cite{JacksonElectrodynamics1999}:

\[
\sum_{m=-l}^{l}|Y_{l,,m}(\theta,\phi)|^{2}=\frac{2l+1}{4\pi}
\]

\textbf{Special case 2: Dipole parallel to sphere surface}

$\textbf{D}=D_{\theta}\boldsymbol{\widehat{\theta}}$:

\begin{align}
 & G_{\text{sph}}^{\parallel}(r_{1})\nonumber \\
 & =\sum_{l=0}^{\infty}\frac{l}{(2l+1)}\frac{\epsilon_{s}-\epsilon_{d}}{\epsilon_{0}\epsilon_{s}\left[l\epsilon_{d}+(l+1)\epsilon_{s}\right]}\frac{a^{2l+1}}{r_{1}^{2l+4}}\nonumber \\
 & \times\sum_{m=-l}^{l}|\frac{\partial Y_{l,m}^{*}(\theta_{1},\phi_{1})}{\partial\theta_{1}}|^{2}\nonumber \\
 & =\frac{1}{8\pi\epsilon_{0}\epsilon_{s}a^{3}}\sum_{l=0}^{\infty}\frac{(\epsilon_{s}-\epsilon_{d})l^{2}(l+1)}{\left[l\epsilon_{d}+(l+1)\epsilon_{s}\right]}\Big(\frac{a}{r_{1}}\Big)^{2l+4}
\end{align}

where
\[
\sum_{m=-l}^{l}|\frac{\partial Y_{l,,m}(\theta,\phi)}{\partial\theta}|^{2}=\frac{l(l+1)(2l+1)}{8\pi}
\]

\subsection{The nonradiative decay rate}

The electric field inside the sphere is given by: 

\begin{align*}
 & \textbf{E}_{\text{in}}(r,\theta,\phi)\\
 & =-\sum_{l=0}^{\infty}\sum_{m=-l}^{l}A_{l,m}\Bigg[lr^{l-1}Y_{l,m}(\theta,\phi)\boldsymbol{\widehat{r}}\\
 & +r^{l-1}\frac{\partial Y_{l,m}(\theta,\phi)}{\partial\theta}\boldsymbol{\widehat{\theta}}+\frac{r^{l-1}}{\sin\theta}\frac{\partial Y_{l,m}(\theta,\phi)}{\partial\phi}\boldsymbol{\widehat{\phi}}\Bigg]
\end{align*}

As before,

\begin{align}
 & \oint_{S}\Phi^{\text{in}^{*}}\boldsymbol{\nabla}\Phi^{\text{in}}\cdot d\textbf{S}\nonumber \\
 & =\oint_{S}r^{2}\sin\theta\Phi^{\text{in}^{*}}\frac{\partial\Phi^{\text{in}}}{\partial r}d\theta\,d\phi\nonumber \\
 & =\sum_{l=0}^{\infty}\sum_{m=-l}^{l}l|A_{l,m}|^{2}a^{2l+1}
\end{align}

\textbf{Special case 1: Dipole perpendicular to sphere surface }
\begin{align}
 & \Gamma_{\text{NR,sph}}^{\perp}\nonumber \\
 & =\frac{\varepsilon_{0}\text{Im}[\varepsilon_{d}(\omega)]D_{r}^{2}}{2\hbar\varepsilon_{0}^{2}a^{3}}\sum_{l=0}^{\infty}\frac{l(l+1)^{2}}{\Big|l\varepsilon_{d}+(l+1)\varepsilon_{s}\Big|^{2}}\Big(\frac{a}{r_{1}}\Big)^{2l+4}\nonumber \\
 & \times\sum_{m=-l}^{l}|Y_{l,m}^{*}(\theta_{1},\phi_{1})|^{2}\nonumber \\
 & =\frac{\text{Im}[\varepsilon_{d}(\omega)]D_{r}^{2}}{8\pi\varepsilon_{0}\hbar a^{3}}\sum_{l=0}^{\infty}\frac{l(l+1)^{2}(2l+1)}{\Big|l\varepsilon_{d}+(l+1)\varepsilon_{s}\Big|^{2}}\Big(\frac{a}{r_{1}}\Big)^{2l+4}
\end{align}

\textbf{Special case 2: Dipole parallel to sphere surface}
\begin{align}
 & \Gamma_{\text{NR,sph}}^{\parallel}\nonumber \\
 & =\frac{\text{Im}[\varepsilon_{d}(\omega)]D_{\theta}^{2}}{2\hbar\varepsilon_{0}a^{3}}\sum_{l=0}^{\infty}\frac{l}{\Big|l\varepsilon_{d}+(l+1)\varepsilon_{s}\Big|^{2}}\Big(\frac{a}{r_{1}}\Big)^{2l+4}\nonumber \\
 & \times\sum_{m=-l}^{l}\Bigg|\frac{\partial Y_{l,m}^{*}(\theta_{1},\phi_{1})}{\partial\theta_{1}}\Bigg|^{2}\nonumber \\
 & =\frac{\text{Im}[\varepsilon_{d}(\omega)]D_{\theta}^{2}}{16\pi\varepsilon_{0}\hbar a^{3}}\sum_{l=0}^{\infty}\frac{l^{2}(l+1)(2l+1)}{\Big|l\varepsilon_{d}+(l+1)\varepsilon_{s}\Big|^{2}}\Big(\frac{a}{r_{1}}\Big)^{2l+4}
\end{align}

\subsection{The radiative decay rate}

In far field,
\begin{align}
 & \Phi^{\text{out}}(r>>r_{1},\theta,\phi)\nonumber \\
 & \rightarrow\sum_{m=-1}^{1}\Bigg[\frac{3\varepsilon_{0}\varepsilon_{s}B_{1,m}}{\mathcal{R}_{1,m}^{*}(r_{1},\theta_{1},\phi_{1})}+1\Bigg]\nonumber \\
 & \times\frac{\mathcal{R}_{1,m}^{*}(r_{1},\theta_{1},\phi_{1})}{3\varepsilon_{0}\varepsilon_{s}}\frac{Y_{1,m}(\theta,\phi)}{r^{2}}
\end{align}

where 

\begin{align*}
 & \mathcal{R}_{l,m}^{*}(r_{1},\theta_{1},\phi_{1})\\
 & =\Bigg[D_{r}lr_{1}^{l-1}Y_{l,m}^{*}(\theta_{1},\phi_{1})+D_{\theta}r_{1}^{l-1}\frac{\partial Y_{l,m}^{*}(\theta_{1},\phi_{1})}{\partial\theta_{1}}\\
 & \quad\quad\quad\quad\quad\quad\quad\quad\quad\quad+\frac{D_{\phi}r_{1}^{l-1}}{\sin\theta_{1}}\frac{\partial Y_{l,m}^{*}(\theta_{1},\phi_{1})}{\partial\phi_{1}}\Bigg]
\end{align*}

\textbf{Special case 1: Dipole perpendicular to sphere surface }

\begin{align*}
 & \Phi^{\text{out}}\\
 & \rightarrow\Bigg[1-\frac{2(\varepsilon_{s}-\varepsilon_{d})}{(\varepsilon_{d}+2\varepsilon_{s})}\frac{a^{3}}{r_{1}^{3}}\Bigg]\frac{D_{r}}{4\pi\varepsilon_{0}\varepsilon_{s}}\frac{\cos\theta}{r^{2}}\\
 & =\frac{D_{\text{tot,sph}}^{\perp}}{4\pi\varepsilon_{0}\varepsilon_{s}}\frac{\cos\theta}{r^{2}}
\end{align*}

where 
\begin{equation}
D_{\text{tot,sph}}^{\perp}=\Bigg[1-\frac{2(\varepsilon_{s}-\varepsilon_{d})}{(\varepsilon_{d}+2\varepsilon_{s})}\frac{a^{3}}{r_{1}^{3}}\Bigg]D_{r}
\end{equation}

\textbf{Special case 2: Dipole parallel to sphere surface}

\begin{align*}
 & \Phi^{\text{out}}\\
 & \rightarrow\Bigg[1+\frac{\varepsilon_{s}-\varepsilon_{d}}{\left[\varepsilon_{d}+2)\varepsilon_{s}\right]}\frac{a^{3}}{r_{1}^{3}}\Bigg]\frac{D_{\theta}}{4\pi\varepsilon_{0}\varepsilon_{s}}\frac{\sin\theta}{r^{2}}\cos\phi\\
 & =\frac{D_{\text{tot,sph}}^{\parallel}}{4\pi\varepsilon_{0}\varepsilon_{s}}\frac{\sin\theta}{r^{2}}\cos\phi
\end{align*}

where 
\begin{equation}
D_{\text{tot,sph}}^{\parallel}=\Bigg[1+\frac{\varepsilon_{s}-\varepsilon_{d}}{\left[\varepsilon_{d}+2\varepsilon_{s}\right]}\frac{a^{3}}{r_{1}^{3}}\Bigg]D_{\theta}
\end{equation}

\section{A molecule in the gap of a bispherical dimer \label{SI:sol_BC_bisph_para}}

\subsection{Bispherical coordinate system \label{SI:coord_sys_bisph}}

We describe the orthonormal set of coordinates \cite{MorsePhilipMcCordandFeshbach1946}$(\mu,\eta,\phi)$,
which are transformed into Cartesian coordinates $(x,y,z)$ by:

\begin{align*}
x & =\frac{a\sin\eta\cos\phi}{\cosh\mu-\cos\eta}\\
y & =\frac{a\sin\eta\sin\phi}{\cosh\mu-\cos\eta}\\
z & =\frac{a\sinh\mu}{\cosh\mu-\cos\eta}
\end{align*}

with scaling factors 

\begin{align*}
h_{\mu}=h_{\eta} & =\frac{a}{\cosh\mu-\cos\eta}\\
h_{\phi} & =\frac{a\sin\eta}{\cosh\mu-\cos\eta}
\end{align*}

with $-\infty<\mu<\infty$, $0\leq\eta<\pi$ and $0<\phi<2\pi$. The
two poles ($\mu=\pm\infty$) are at $x=y=0,z=\pm a$. The surface
with $\mu$ constant, say $\mu_{0}$, is a sphere of radius $a|\csch\mu_{0}|$
with center at $x=y=0$, $z=a\coth\mu_{0}$. 

We limit ourselves to geometries in which the molecule is placed between
the spheres, on the axis connecting their centers, with the mlecular
dipole perpendicular or parallel to this axis.

\subsection{Parallel configuration \label{SI:subsec:Parallel-configuration}}

Consider first the case where the molecule is on the axis connecting
the spheres centers with the molecular dipole parallel to this axis
(subsequently\textit{ parallel configuration}). The general solution
to the Laplace equation in this coordinate system is:

\begin{align}
 & \tilde{\Phi}(\mu,\eta,\phi)\nonumber \\
 & =\sqrt{\cosh\mu-\cos\eta}\nonumber \\
 & \times\sum_{n=0}^{\infty}\sum_{m=0}^{n}\Bigg[A_{n}e^{-(n+1/2)\mu}+B_{n}e^{(n+1/2)\mu}\Bigg]\nonumber \\
 & \times\Bigg[E_{m}\cos(m\phi)+F_{m}\sin(m\phi)\Bigg]P_{n}^{m}(\cos\eta)\label{eq:phi_general_bisph}
\end{align}

As usual, we will represent the molecule by a point dipole $\textbf{D}$,
located at $\textbf{r}_{1}(\mu_{1},\eta_{1},\phi_{1})$ (we may as
well set $\phi_{1}=0$) The Green's function $1/|\textbf{r}-\textbf{r}_{1}|$
reads:

\begin{align}
 & \frac{1}{|{\bf r}-{\bf r}_{1}|}\nonumber \\
 & =\frac{1}{a}\sqrt{(\cosh\mu-\cos\eta)(\cosh\mu_{1}-\cos\eta_{1})}\nonumber \\
 & \times\sum_{n=0}^{\infty}\sum_{m=0}^{n}(2-\delta_{m0})\frac{(n-m)!}{(n+m)!}\cos[m(\phi-\phi_{1})]\nonumber \\
 & \times P_{n}^{m}(\cos\eta_{1})P_{n}^{m}(\cos\eta)e^{-(n+1/2)|\mu-\mu_{1}|}\label{eq:green's_function_bisph_general}
\end{align}

In this particular orientation, the molecular transition dipole is
given by $\textbf{D}=D\boldsymbol{\widehat{\mu}}$. This position
choice corresponds to $x_{1}=y_{1}=0$ in Cartesian coordinates and
$\eta_{1}=\pi$ in bispherical coordinates. The resulting system has
an azimuthal symmetry and we have the simpler expressions as follows:

\begin{align*}
 & \tilde{\Phi}(\mu,\eta,\phi)=\sqrt{\cosh\mu-\cos\eta}\\
 & \times\sum_{n=0}^{\infty}\Bigg[A_{n}e^{-(n+1/2)\mu}+B_{n}e^{(n+1/2)\mu}\Bigg]P_{n}(\cos\eta)
\end{align*}

and

\begin{align}
 & \frac{1}{|{\bf r}-{\bf r}_{1}|}\nonumber \\
 & =\frac{1}{a}\sqrt{(\cosh\mu-\cos\eta)(\cosh\mu_{1}-\cos\eta_{1})}\nonumber \\
 & \times\sum_{n=0}^{\infty}e^{-(n+1/2)|\mu-\mu_{1}|}P_{n}(\cos\eta_{1})P_{n}(\cos\eta)\label{eq:green's_function_bisph_parallel}
\end{align}

Inside the sphere (on the ``right'') $\mu>\mu_{R}>0$

\begin{align}
 & \Phi_{\text{R}}^{\text{in}}(\mu,\eta)\nonumber \\
 & =\sqrt{\cosh\mu-\cos\eta}\sum_{n=0}^{\infty}A_{R,n}^{\text{in}}e^{-(n+1/2)\mu}P_{n}(\cos\eta)\label{eq:phi_in_R_bisph_parallel}
\end{align}

and inside the sphere (on the ``left'') $\ensuremath{\mu<\mu_{L}<0}$ 

\begin{align}
 & \Phi_{\text{L}}^{\text{in}}(\mu,\eta)\nonumber \\
 & =\sqrt{\cosh\mu-\cos\eta}\sum_{n=0}^{\infty}B_{L,n}^{\text{in}}e^{(n+1/2)\mu}P_{n}(\cos\eta)\label{eq:phi_in_L_bisph_parallel}
\end{align}

The potential outside both nanoparticles is:

\begin{align}
 & \Phi^{\text{out}}(\mu,\eta)=\sqrt{\cosh\mu-\cos\eta}\nonumber \\
 & \times\sum_{n=0}^{\infty}\Bigg[C_{n}^{\text{out}}e^{-(n+1/2)\mu}+D_{n}^{\text{out}}e^{(n+1/2)\mu}\Bigg]\nonumber \\
 & \times P_{n}(\cos\eta)+\frac{D}{4\pi\varepsilon_{0}\varepsilon_{s}}\boldsymbol{\widehat{\mu}}\cdot\boldsymbol{\nabla}_{1}\frac{1}{|{\bf r}-{\bf r}_{1}|}\label{eq:phi_out_bisph_parallel}
\end{align}

\begin{align*}
 & \boldsymbol{\widehat{\mu}}\cdot\boldsymbol{\nabla}_{1}\frac{1}{|{\bf r}-{\bf r}_{1}|}=\frac{1}{h_{\mu}}\frac{\partial}{\partial\mu_{1}}\frac{1}{|{\bf r}-{\bf r}_{1}|}\\
 & =\frac{\sqrt{\cosh\mu_{1}+1}}{a^{2}}\sqrt{(\cosh\mu-\cos\eta)}\\
 & \times\sum_{n=0}^{\infty}P_{n}(\cos\eta)(-1)^{n}\\
 & \times\Bigg[\frac{\sinh\mu_{1}}{2}+(-1)^{\text{H}(\mu-\mu_{1})+1}\Big(n+\frac{1}{2}\Big)\\
 & \times(\cosh\mu_{1}+1)\Bigg]e^{-(n+1/2)|\mu-\mu_{1}|}
\end{align*}

Where $\text{H}(\mu-\mu_{1})$ is the Heaviside function.

Following are the boundary conditions we apply: 

\begin{align}
\Phi_{\text{R}}^{\text{in}}(\mu_{R},\eta) & =\Phi^{\text{out}}(\mu_{\text{R}},\eta)\label{eq:BC_1_bisph_general}\\
\Phi_{\text{L}}^{\text{in}}(\mu_{L},\eta) & =\Phi^{\text{out}}(\mu_{\text{L}},\eta)\label{eq:BC_2_bisph_general}\\
\varepsilon_{R}\frac{\partial\Phi_{\text{R}}^{\text{in}}(\mu,\eta)}{\partial\mu}\Big|_{\mu=\mu_{\text{R}}} & =\varepsilon_{s}\frac{\partial\Phi^{\text{out}}(\mu,\eta)}{\partial\mu}\Big|_{\mu=\mu_{\text{R}}}\label{eq:BC_3_bisph_general}\\
\varepsilon_{L}\frac{\partial\Phi_{\text{L}}^{\text{in}}(\mu,\eta)}{\partial\mu}\Big|_{\mu=\mu_{\text{L}}} & =\varepsilon_{s}\frac{\partial\Phi^{\text{out}}(\mu,\eta)}{\partial\mu}\Big|_{\mu=\mu_{\text{L}}}\label{eq:BC_4_bisph_general}
\end{align}

Eqs. \ref{eq:BC_1_bisph_general}-\ref{eq:BC_4_bisph_general} when
written explicitly, gives following infinite linear system of equations
for the coefficients Eqs. \ref{eq:BC_1_bisph_parallel_explicit} through
\ref{eq:BC_4_bisph_parallel_explicit}.

\begin{algorithm*}
\begin{align}
A_{R,n}^{\text{in}}-C_{n}^{\text{out}}-D_{n}^{\text{out}}e^{(2n+1)\mu_{R}} & =\mathbb{P}_{n}(\mu_{1})\label{eq:BC_1_bisph_parallel_explicit}
\end{align}

\begin{align}
B_{L,n}^{\text{in}}-C_{n}^{\text{out}}e^{-(2n+1)\mu_{L}}-D_{n}^{\text{out}} & =\mathbb{Q}_{n}(\mu_{1})\label{eq:BC_2_bisph_parallel_explicit}
\end{align}

\begin{align}
 & \frac{\varepsilon_{R}}{\varepsilon_{s}}\mathbb{R}_{n}(\mu_{R})e^{-(n+1/2)\mu_{R}}A_{R,n}^{\text{in}}-\mathbb{R}_{n}(\mu_{R})e^{-(n+1/2)\mu_{R}}C_{n}^{\text{out}}-\mathbb{S}_{n}(\mu_{R})e^{(n+1/2)\mu_{R}}D_{n}^{\text{out}}\nonumber \\
 & +\frac{\varepsilon_{R}}{\varepsilon_{s}}\frac{n}{2}e^{-(n-1/2)\mu_{R}}A_{R,n-1}^{\text{in}}-\frac{n}{2}e^{-(n-1/2)\mu_{R}}C_{n-1}^{\text{out}}+\frac{n}{2}e^{(n-1/2)\mu_{R}}D_{n-1}^{\text{out}}\nonumber \\
 & +\frac{\varepsilon_{R}}{\varepsilon_{s}}\frac{n+1}{2}e^{-(n+3/2)\mu_{R}}A_{R,n+1}^{\text{in}}-\frac{n+1}{2}e^{-(n+3/2)\mu_{R}}C_{n+1}^{\text{out}}+\frac{n+1}{2}e^{(n+3/2)\mu_{R}}D_{n+1}^{\text{out}}\nonumber \\
 & =\mathbb{R}_{n}(\mu_{R})\mathbb{P}_{n}(\mu_{1})e^{-(n+1/2)\mu_{R}}+\frac{n}{2}\mathbb{P}_{n-1}(\mu_{1})e^{-(n-1/2)\mu_{R}}+\frac{n+1}{2}\mathbb{P}_{n+1}(\mu_{1})e^{-(n+3/2)\mu_{R}}\label{eq:BC_3_bisph_parallel_explicit}
\end{align}

\begin{flushleft}

and

\end{flushleft}

\begin{align}
 & -\frac{\varepsilon_{L}}{\varepsilon_{s}}\mathbb{S}_{n}(\mu_{L})e^{(n+1/2)\mu_{L}}B_{L,n}^{\text{in}}+\mathbb{R}_{n}(\mu_{L})e^{-(n+1/2)\mu_{L}}C_{n}^{\text{out}}+\mathbb{S}_{n}(\mu_{L})e^{(n+1/2)\mu_{L}}D_{n}^{\text{out}}\nonumber \\
 & +\frac{\varepsilon_{L}}{\varepsilon_{s}}\frac{n}{2}e^{(n-1/2)\mu_{L}}B_{L,n-1}^{\text{in}}+\frac{n}{2}e^{-(n-1/2)\mu_{L}}C_{n-1}^{\text{out}}-\frac{n}{2}e^{(n-1/2)\mu_{L}}D_{n-1}^{\text{out}}\nonumber \\
 & +\frac{\varepsilon_{L}}{\varepsilon_{s}}\frac{n+1}{2}e^{(n+3/2)\mu_{L}}B_{L,n+1}^{\text{in}}+\frac{n+1}{2}e^{-(n+3/2)\mu_{L}}C_{n+1}^{\text{out}}-\frac{n+1}{2}e^{(n+3/2)\mu_{L}}D_{n+1}^{\text{out}}\nonumber \\
 & =-\mathbb{S}_{n}(\mu_{L})\mathbb{Q}_{n}(\mu_{1})e^{(n+1/2)\mu_{L}}+\frac{n}{2}\mathbb{Q}_{n-1}(\mu_{1})e^{(n-1/2)\mu_{L}}+\frac{n+1}{2}\mathbb{Q}_{n+1}(\mu_{1})e^{(n+3/2)\mu_{L}}\label{eq:BC_4_bisph_parallel_explicit}
\end{align}

\begin{flushleft}

where

\end{flushleft}

\begin{align*}
\mathbb{P}_{n}(\mu_{1})=(-1)^{n}\mathbb{K} & \sqrt{\cosh\mu_{1}+1}\Bigg[\frac{\sinh\mu_{1}}{2}+\Big(n+\frac{1}{2}\Big)(\cosh\mu_{1}+1)\Bigg]e^{(n+1/2)\mu_{1}}\\
\mathbb{Q}_{n}(\mu_{1})=(-1)^{n} & \mathbb{K}\sqrt{\cosh\mu_{1}+1}\Bigg[\frac{\sinh\mu_{1}}{2}-\Big(n+\frac{1}{2}\Big)(\cosh\mu_{1}+1)\Bigg]e^{-(n+1/2)\mu_{1}}\\
\mathbb{R}_{n}(\mu) & =\Bigg[\frac{\sinh\mu}{2}-\Big(n+\frac{1}{2}\Big)\cosh\mu\Bigg]\\
\mathbb{S}_{n}(\mu) & =\Bigg[\frac{\sinh\mu}{2}+\Big(n+\frac{1}{2}\Big)\cosh\mu\Bigg]
\end{align*}

\begin{flushleft}

with $\mathbb{K}=D/(4\pi\varepsilon_{s}\varepsilon_{0}a^{2})$

\end{flushleft}
\end{algorithm*}

Once we have the coefficients calculated numerically, it is straightforward
to find the required component of ${\bf \underline{G}}$ and the decay
rates.

\begin{align*}
 & G_{\text{bisph}}^{\parallel}(r_{1})\\
 & =\frac{\sinh\mu_{1}\sqrt{\cosh\mu_{1}+1}}{2aD}\\
 & \times\sum_{n=0}^{\infty}(-1)^{n+1}\Bigg[C_{n}^{\text{out}}e^{-(n+1/2)\mu_{1}}+D_{n}^{\text{out}}e^{(n+1/2)\mu_{1}}\Bigg]\\
 & +\frac{(\cosh\mu_{1}+1)^{3/2}}{aD}\sum_{n=0}^{\infty}(-1)^{n+1}\Big(n+\frac{1}{2}\Big)\\
 & \times\Bigg[-C_{n}^{\text{out}}e^{-(n+1/2)\mu_{1}}+D_{n}^{\text{out}}e^{(n+1/2)\mu_{1}}\Bigg]
\end{align*}

In addition, we evaluate the non-radiative and radiative decay rates
as shown in following subsections.

\subsubsection{The nonradiative decay rate\label{subsec:NonRad_Dec_Bisph_Para}}

The volume element in the bispherical coordinate system is given by:

\[
dV=\frac{a^{3}\sin\eta}{(\cosh\mu-\cos\eta)^{3}}d\mu\,d\eta\,d\phi
\]

\begin{align*}
 & \Gamma_{\text{NR}}\sim\int_{V_{\text{L}}+V_{\text{R}}}d\textbf{r}|\textbf{E}|^{2}\\
 & =\oint_{S_{\text{L}}+S_{\text{R}}}\Phi^{*}\boldsymbol{\nabla}\Phi\cdot d\textbf{S}
\end{align*}

\[
d\textbf{S}=\frac{a^{2}\sin\eta}{(\cosh\mu-\cos\eta)^{2}}d\eta\,d\phi\boldsymbol{\widehat{\mu}}
\]

\begin{align}
 & \oint_{S_{\text{R}}}\Phi_{\text{R}}^{\text{in}^{*}}\boldsymbol{\nabla}\Phi_{\text{R}}^{\text{in}}\cdot d\text{\textbf{S}}\nonumber \\
 & =-\pi a\sum_{n=0}^{\infty}\sum_{n^{\prime}=0}^{\infty}\sinh\mu_{\text{R}}A_{R,n}^{\text{in}^{*}}A_{R,n^{\prime}}^{\text{in}}\nonumber \\
 & \times e^{-(n+n^{\prime}+1)\mu_{\text{R}}}\int_{-1}^{1}\frac{P_{n}(x)P_{n^{\prime}}(x)}{(\cosh\mu_{\text{R}}-x)}dx\nonumber \\
 & +2\pi a\sum_{n=0}^{\infty}\sum_{n^{\prime}=0}^{\infty}\left(n^{\prime}+\frac{1}{2}\right)A_{R,n}^{\text{in}^{*}}A_{R,n^{\prime}}^{\text{in}}\nonumber \\
 & \times e^{-(n+n^{\prime}+1)\mu_{\text{R}}}\int_{-1}^{1}P_{n}(x)P_{n^{\prime}}(x)dx\label{eq:NR_decay_rate_bisph_parallel_right}
\end{align}

where $x=\cos\eta$.

The second integral in Eq. \ref{eq:NR_decay_rate_bisph_parallel_right}
is evaluated from simple orthonormality property of Legendre polynomials:

\[
\int_{-1}^{1}P_{n}(x)P_{n^{\prime}}(x)dx=\frac{2}{2n+1}\delta_{nn^{\prime}}
\]

but the first integral can be obtained from a corollary of Neumann's
formula for the Legendre Functions\cite{Macrobert1952}:

\[
\frac{1}{2}\int_{-1}^{1}\frac{P_{n}(x)P_{n^{\prime}}(x)}{z-x}dx=P_{n^{\prime}}(z)Q_{n}(z),\quad n^{\prime}\le n
\]

where $|z|>1$ ($\cosh\mu\ge1$).

So,

\begin{align*}
 & \oint_{S_{\text{R}}}\Phi_{\text{R}}^{\text{in}^{*}}\boldsymbol{\nabla}\Phi_{\text{R}}^{\text{in}}\cdot d\text{\textbf{S}}\\
 & =-\pi a\sum_{n=0}^{\infty}\sum_{n^{\prime}=0}^{n}\sinh\mu_{\text{R}}A_{R,n}^{\text{in}^{*}}A_{R,n^{\prime}}^{\text{in}}\\
 & \times e^{-(n+n^{\prime}+1)\mu_{\text{R}}}\int_{-1}^{1}\frac{P_{n}(x)P_{n^{\prime}}(x)}{(\cosh\mu_{\text{R}}-x)}dx\\
 & -\pi a\sum_{n=0}^{\infty}\sum_{n^{\prime}=n+1}^{\infty}\sinh\mu_{\text{R}}A_{R,n}^{\text{in}^{*}}A_{R,n^{\prime}}^{\text{in}}\\
 & \times e^{-(n+n^{\prime}+1)\mu_{\text{R}}}\int_{-1}^{1}\frac{P_{n}(x)P_{n^{\prime}}(x)}{(\cosh\mu_{\text{R}}-x)}dx\\
 & +2\pi a\sum_{n=0}^{\infty}\left|A_{R,n}^{\text{in}}\right|^{2}e^{-(2n+1)\mu_{\text{R}}}
\end{align*}

\begin{align}
 & =-2\pi a\sum_{n=0}^{\infty}\sum_{n^{\prime}=0}^{n}\sinh\mu_{\text{R}}A_{R,n}^{\text{in}^{*}}A_{R,n^{\prime}}^{\text{in}}\nonumber \\
 & \times e^{-(n+n^{\prime}+1)\mu_{\text{R}}}P_{n^{\prime}}(\cosh\mu_{\text{R}})Q_{n}(\cosh\mu_{\text{R}})\nonumber \\
 & -2\pi a\sum_{n=0}^{\infty}\sum_{n^{\prime}=n+1}^{\infty}\sinh\mu_{\text{R}}A_{R,n}^{\text{in}^{*}}A_{R,n^{\prime}}^{\text{in}}\nonumber \\
 & \times e^{-(n+n^{\prime}+1)\mu_{\text{R}}}P_{n}(\cosh\mu_{\text{R}})Q_{n^{\prime}}(\cosh\mu_{\text{R}})\nonumber \\
 & +2\pi a\sum_{n=0}^{\infty}\left|A_{R,n}^{\text{in}}\right|^{2}e^{-(2n+1)\mu_{\text{R}}}\label{eq:NR_decay_rate_bisph_parallel_right_explicit}
\end{align}

For $\mu_{L}$,

\begin{align*}
 & \oint_{S_{\text{L}}}\Phi_{\text{L}}^{\text{in}^{*}}\boldsymbol{\nabla}\Phi_{\text{L}}^{\text{in}}\cdot d\textbf{S}\\
 & =\pi a\sum_{n=0}^{\infty}\sum_{n^{\prime}=0}^{\infty}\sinh\mu_{\text{L}}B_{L,n}^{\text{in}^{*}}B_{L,n^{\prime}}^{\text{in}}\\
 & \times e^{(n+n^{\prime}+1)\mu_{\text{L}}}\int_{-1}^{1}\frac{P_{n}(x)P_{n^{\prime}}(x)}{(\cosh\mu_{\text{L}}-x)}dx\\
 & +2\pi a\sum_{n=0}^{\infty}\sum_{n^{\prime}=0}^{\infty}\left(n^{\prime}+\frac{1}{2}\right)B_{L,n}^{\text{in}^{*}}B_{L,n^{\prime}}^{\text{in}}\\
 & \times e^{(n+n^{\prime}+1)\mu_{\text{L}}}\int_{-1}^{1}P_{n}(x)P_{n^{\prime}}(x)dx
\end{align*}

\begin{align}
 & =2\pi a\sum_{n=0}^{\infty}\sum_{n^{\prime}=0}^{n}\sinh\mu_{\text{L}}B_{L,n}^{\text{in}^{*}}B_{L,n^{\prime}}^{\text{in}}\nonumber \\
 & \times e^{(n+n^{\prime}+1)\mu_{\text{L}}}P_{n^{\prime}}(\cosh\mu_{\text{L}})Q_{n}(\cosh\mu_{\text{L}})\nonumber \\
 & +2\pi a\sum_{n=0}^{\infty}\sum_{n^{\prime}=n+1}^{\infty}\sinh\mu_{\text{L}}B_{L,n}^{\text{in}^{*}}B_{L,n^{\prime}}^{\text{in}}\nonumber \\
 & \times e^{(n+n^{\prime}+1)\mu_{\text{L}}}P_{n}(\cosh\mu_{\text{L}})Q_{n^{\prime}}(\cosh\mu_{\text{L}})\nonumber \\
 & +2\pi a\sum_{n=0}^{\infty}\left|B_{L,n}^{\text{in}}\right|^{2}e^{(2n+1)\mu_{\text{L}}}\label{eq:NR_decay_rate_bisph_parallel_left_explicit}
\end{align}

\subsubsection{The radiative decay rate \label{subsec:Rad_Dec_Bisph_Para}}

The total dipole moment of the composite system is given by ${\bf D}_{\text{tot}}(\omega)={\bf D}(\omega)+{\bf D}_{\text{ns}}(\omega)$
where ${\bf D}_{\text{ns}}(\omega)$ is the dipole induced on the
nanostructure given by the integral:

\begin{align}
{\bf D}_{\text{ns}}(\omega) & =\big(\varepsilon_{\text{ns}}(\omega)-\varepsilon_{s}\big)\varepsilon_{0}\int_{V_{\text{ns}}}{\bf E}({\bf r},\omega)d{\bf r}\label{eq:induced_dipole_ns}
\end{align}

\begin{align}
 & \int_{V_{\text{ns}}}{\bf E}({\bf r},\omega)d{\bf r}\nonumber \\
 & =-\oint_{S_{\text{L}}}\Phi_{\text{L}}^{\text{in}}({\bf r},\omega)d{\bf S}-\oint_{S_{\text{R}}}\Phi_{\text{R}}^{\text{in}}({\bf r},\omega)d{\bf S}\label{eq:rad_decay_vol_int_to_surf}
\end{align}

For bispherical coordinate, the unit vector $\boldsymbol{\widehat{\mu}}$
is given by:

\begin{align}
 & \boldsymbol{\widehat{\mu}}=\nonumber \\
 & \Bigg[-\boldsymbol{\widehat{i}}\frac{\sinh\mu\sin\eta\cos\phi}{(\cosh\mu-\cos\eta)}-\boldsymbol{\widehat{j}}\frac{\sinh\mu\sin\eta\sin\phi}{(\cosh\mu-\cos\eta)}\nonumber \\
 & +\boldsymbol{\widehat{k}}\Bigg\{\cosh\mu-\frac{\sinh^{2}\mu}{(\cosh\mu-\cos\eta)}\Bigg\}\Bigg]\label{eq:mu_unit_vector_bisph}
\end{align}
where $\boldsymbol{\widehat{i}},\boldsymbol{\widehat{j}},\boldsymbol{\widehat{k}}$
are unit vectors in Cartesian coordinates.

For this particular configuration, the integral corresponding to the
sphere $\mu_{\text{R}}$ is:

\begin{align*}
 & -\oint_{S_{\text{R}}}\Phi_{\text{R}}^{\text{in}}({\bf r},\omega)d{\bf S}\\
 & =\boldsymbol{\widehat{k}}2\pi a^{2}\sum_{n=0}^{\infty}A_{R,n}^{\text{in}}e^{-(n+1/2)\mu_{\text{R}}}\\
 & \times\int_{0}^{\pi}\Bigg\{\cosh\mu_{\text{R}}-\frac{\sinh^{2}\mu}{(\cosh\mu_{\text{R}}-\cos\eta)}\Bigg\}\\
 & \times\frac{P_{n}(x)\sin\eta}{(\cosh\mu_{\text{R}}-\cos\eta)^{3/2}}d\eta
\end{align*}

For the sphere $\mu_{\text{L}}$

\begin{align*}
 & -\oint_{S_{\text{L}}}\Phi_{\text{L}}^{\text{in}}({\bf r},\omega)d{\bf S}\\
 & =-\boldsymbol{\widehat{k}}2\pi a^{2}\sum_{n=0}^{\infty}B_{L,n}^{\text{in}}e^{(n+1/2)\mu_{\text{L}}}\\
 & \times\int_{0}^{\pi}\Bigg\{\cosh\mu_{\text{L}}-\frac{\sinh^{2}\mu_{\text{L}}}{(\cosh\mu_{\text{L}}-\cos\eta)}\Bigg\}\\
 & \times\frac{P_{n}(x)\sin\eta}{(\cosh\mu_{\text{L}}-\cos\eta)^{3/2}}dx
\end{align*}

These integrals are obtained numerically. 

\subsection{Perpendicular configuration\label{SI:sol_BC_bisph_perp}}

When the molecule lies on the line connecting centers of the two nanoparticles
orienting perpendicular to this line, molecular transition dipole
is given by $\textbf{D}=D\boldsymbol{\widehat{\eta}}$. This choice
of position also corresponds to $x_{1}=y_{1}=0$ in Cartesian coordinates
and $\eta_{1}=\pi$ in bispherical coordinates. The resulting composite
system does not have the azimuthal symmetry anymore. 

\begin{align*}
 & \frac{1}{|{\bf r}-{\bf r}_{1}|}\\
 & =\frac{1}{a}\sqrt{(\cosh\mu-\cos\eta)(\cosh\mu_{1}-\cos\eta_{1})}\\
 & \times\sum_{n=0}^{\infty}\sum_{m=0}^{n}(2-\delta_{m0})\frac{(n-m)!}{(n+m)!}\cos[m(\phi-\phi_{1})]\\
 & \times P_{n}^{m}(\cos\eta_{1})P_{n}^{m}(\cos\eta)e^{-(n+1/2)|\mu-\mu_{1}|}
\end{align*}

We have:

\begin{align*}
 & \boldsymbol{\widehat{\eta}}\cdot\boldsymbol{\nabla}_{1}\frac{1}{|{\bf r}-{\bf r}_{1}|}=\frac{1}{h_{\eta}}\frac{\partial}{\partial\eta_{1}}\frac{1}{|{\bf r}-{\bf r}_{1}|}\\
 & =\frac{\cosh\mu_{1}-\cos\eta_{1}}{a^{2}}\sqrt{\cosh\mu-\cos\eta}\\
 & \times\sum_{n=0}^{\infty}\sum_{m=0}^{n}\Bigg[\frac{\sin\eta_{1}P_{n}^{m}(\cos\eta_{1})}{2\sqrt{\cosh\mu_{1}-\cos\eta_{1}}}\\
 & -\sin\eta_{1}P_{n}^{m^{\prime}}(\cos\eta_{1})\sqrt{\cosh\mu_{1}-\cos\eta_{1}}\Bigg]\\
 & \times(2-\delta_{m0})\frac{(n-m)!}{(n+m)!}\cos[m(\phi-\phi_{1})]\\
 & \times P_{n}^{m}(\cos\eta)e^{-(n+1/2)|\mu-\mu_{1}|}
\end{align*}

Because $\sin\eta_{1}=0$ and in the limit $\eta_{1}\rightarrow\pi$,
$\sin\eta_{1}P_{n}^{m^{\prime}}(\cos\eta_{1})$ is only nonzero for
$m=1$ with the value $\frac{(-1)^{n}n(n+1)}{2}$. From this one finds:

\begin{align}
 & \Phi^{\text{out}}(\mu,\eta,\phi)=\sqrt{\cosh\mu-\cos\eta}\nonumber \\
 & \times\sum_{n=1}^{\infty}\Bigg[C_{n}^{\text{out}}e^{-(n+1/2)\mu}+D_{n}^{\text{out}}e^{(n+1/2)\mu}\nonumber \\
 & +(-1)^{n+1}\mathbb{K}(\cosh\mu_{1}+1)^{3/2}e^{-(n+1/2)|\mu-\mu_{1}|}\Bigg]\nonumber \\
 & \times\cos(\phi-\phi_{1})P_{n}^{1}(\cos\eta)\label{eq:phi_out_bisph_perp}
\end{align}

where $\mathbb{K}=D/(4\pi\varepsilon_{s}\varepsilon_{0}a^{2})$.

The potential inside the spheres must be finite. This implies that
for the sphere (on the ``right'') $\mu>\mu_{R}>0$ 

\begin{align}
 & \Phi_{\text{R}}^{\text{in}}(\mu,\eta,\phi)\nonumber \\
 & =\sqrt{\cosh\mu-\cos\eta}\sum_{n=1}^{\infty}A_{R,n}^{\text{in}}e^{-(n+1/2)\mu}\nonumber \\
 & \times P_{n}^{1}(\cos\eta)\cos(\phi-\phi_{1})\label{eq:phi_in_R_bisph_perp}
\end{align}

and for the sphere (on the ``left'') $\ensuremath{\mu<\mu_{L}<0}$ 

\begin{align}
 & \Phi_{\text{L}}^{\text{in}}(\mu,\eta,\phi)\nonumber \\
 & =\sqrt{\cosh\mu-\cos\eta}\sum_{n=1}^{\infty}B_{L,n}^{\text{in}}e^{(n+1/2)\mu}\nonumber \\
 & \times P_{n}^{1}(\cos\eta)\cos(\phi-\phi_{1})\label{eq:phi_in_L_bisph_perp}
\end{align}

From same boundary conditions as with the parallel orientation of
the dipole with the internuclear axis (Eqs. \ref{eq:BC_1_bisph_general}-\ref{eq:BC_4_bisph_general}),
we obtain the following infinite system of linear equations (Eqs.
\ref{eq:BC_1_bisph_perp_explicit}-\ref{eq:BC_4_bisph_perp_explicit})
and solve this system numerically by truncating to find the the coefficients. 

\begin{algorithm*}[p]
\begin{align}
A_{R,n}^{\text{in}}-C_{n}^{\text{out}}-D_{n}^{\text{out}}e^{(2n+1)\mu_{R}} & =\mathbb{T}_{n}(\mu_{1})e^{(n+1/2)\mu_{1}}\label{eq:BC_1_bisph_perp_explicit}
\end{align}

\begin{align}
B_{R,n}^{\text{in}}-C_{n}^{\text{out}}e^{-(2n+1)\mu_{L}}-D_{n}^{\text{out}} & =\mathbb{T}_{n}(\mu_{1})e^{-(n+1/2)\mu_{1}}\label{eq:BC_2_bisph_perp_explicit}
\end{align}

\begin{align}
 & \frac{\varepsilon_{R}}{\varepsilon_{s}}\mathbb{R}_{n}(\mu_{R})e^{-(n+1/2)\mu_{R}}A_{R,n}^{\text{in}}-\mathbb{R}_{n}(\mu_{R})e^{-(n+1/2)\mu_{R}}C_{n}^{\text{out}}-\mathbb{S}_{n}(\mu_{R})e^{(n+1/2)\mu_{R}}D_{n}^{\text{out}}\nonumber \\
 & +\frac{\varepsilon_{R}}{\varepsilon_{s}}\frac{n-1}{2}e^{-(n-1/2)\mu_{R}}A_{R,n-1}^{\text{in}}-\frac{n-1}{2}e^{-(n-1/2)\mu_{R}}C_{n-1}^{\text{out}}+\frac{n-1}{2}e^{(n-1/2)\mu_{R}}D_{n-1}^{\text{out}}\nonumber \\
 & +\frac{\varepsilon_{R}}{\varepsilon_{s}}\frac{n+2}{2}e^{-(n+3/2)\mu_{R}}A_{R,n+1}^{\text{in}}-\frac{n+2}{2}e^{-(n+3/2)\mu_{R}}C_{n+1}^{\text{out}}+\frac{n+2}{2}e^{(n+3/2)\mu_{R}}D_{n+1}^{\text{out}}\nonumber \\
 & =\mathbb{R}_{n}(\mu_{R})\mathbb{T}_{n}(\mu_{1})e^{-(n+1/2)(\mu_{R}-\mu_{1})}+\frac{n-1}{2}\mathbb{T}_{n-1}(\mu_{1})e^{-(n-1/2)(\mu_{R}-\mu_{1})}\nonumber \\
 & \quad\quad\quad\quad\quad\quad\quad\quad\quad\quad\quad\quad\quad\quad\quad\quad\quad\quad\quad+\frac{n+2}{2}\mathbb{T}_{n+1}(\mu_{1})e^{-(n+3/2)(\mu_{R}-\mu_{1})}\label{eq:BC_3_bisph_perp_explicit}
\end{align}

\begin{flushleft}

and

\end{flushleft}

\begin{align}
 & -\frac{\varepsilon_{L}}{\varepsilon_{s}}\mathbb{S}_{n}(\mu_{L})e^{(n+1/2)\mu_{L}}B_{L,n}^{\text{in}}+\mathbb{R}_{n}(\mu_{L})e^{-(n+1/2)\mu_{L}}C_{n}^{\text{out}}+\mathbb{S}_{n}(\mu_{L})e^{(n+1/2)\mu_{L}}D_{n}^{\text{out}}\nonumber \\
 & +\frac{\varepsilon_{L}}{\varepsilon_{s}}\frac{n-1}{2}e^{(n-1/2)\mu_{L}}B_{L,n-1}^{\text{in}}+\frac{n-1}{2}e^{-(n-1/2)\mu_{L}}C_{n-1}^{\text{out}}-\frac{n-1}{2}e^{(n-1/2)\mu_{L}}D_{n-1}^{\text{out}}\nonumber \\
 & +\frac{\varepsilon_{L}}{\varepsilon_{s}}\frac{n+2}{2}e^{(n+3/2)\mu_{L}}B_{L,n+1}^{\text{in}}+\frac{n+2}{2}e^{-(n+3/2)\mu_{L}}C_{n+1}^{\text{out}}-\frac{n+2}{2}e^{(n+3/2)\mu_{L}}D_{n+1}^{\text{out}}\nonumber \\
 & =-\mathbb{S}_{n}(\mu_{L})\mathbb{T}_{n}(\mu_{1})e^{(n+1/2)(\mu_{L}-\mu_{1})}+\frac{n-1}{2}\mathbb{T}_{n-1}(\mu_{1})e^{(n-1/2)(\mu_{L}-\mu_{1})}\nonumber \\
 & \quad\quad\quad\quad\quad\quad\quad\quad\quad\quad\quad\quad\quad\quad\quad\quad\quad\quad\quad+\frac{n+2}{2}\mathbb{T}_{n+1}(\mu_{1})e^{(n+3/2)(\mu_{L}-\mu_{1})}\label{eq:BC_4_bisph_perp_explicit}
\end{align}

\begin{flushleft}

where

\end{flushleft}

\begin{align*}
\mathbb{T}_{n}(\mu_{1}) & =(-1)^{n+1}\mathbb{K}(\cosh\mu_{1}+1)^{3/2}\\
\mathbb{R}_{n}(\mu) & =\Bigg[\frac{\sinh\mu}{2}-\Big(n+\frac{1}{2}\Big)\cosh\mu\Bigg]\\
\mathbb{S}_{n}(\mu) & =\Bigg[\frac{\sinh\mu}{2}+\Big(n+\frac{1}{2}\Big)\cosh\mu\Bigg]
\end{align*}
\end{algorithm*}

From the calculated coefficients, the required component of ${\bf \underline{G}}$
is obtained as
\begin{align*}
 & G_{\text{bisph}}^{\perp}(r_{1})\\
 & =\frac{(\cosh\mu_{1}+1)^{3/2}}{aD}\\
 & \times\sum_{n=1}^{\infty}(-1)^{n}\Bigg[C_{n}^{\text{out}}e^{-(n+1/2)\mu_{1}}+D_{n}^{\text{out}}e^{(n+1/2)\mu_{1}}\Bigg]\\
 & \times\sin\eta P_{n}^{1^{\prime}}(\cos\eta)
\end{align*}

using the limit $\lim_{x\to-1+}(1-x)^{1/2}P_{n}^{1^{\prime}}(x)=(-1)^{n}n(n+1)/2$.

\subsubsection{The nonradiative decay rate\label{subsec:NonRad_Dec_Bisph_Perp}}

\begin{align*}
 & \Phi_{\text{R}}^{\text{in}^{*}}\boldsymbol{\nabla}\Phi_{\text{R}}^{\text{in}}\cdot d\textbf{S}\\
 & =a\sum_{n=1}^{\infty}\sum_{n^{\prime}=1}^{\infty}A_{R,n}^{\text{in}^{*}}A_{R,n^{\prime}}^{\text{in}}e^{-(n^{\prime}+n+1)\mu}\\
 & \times\left[\frac{\sinh\mu}{2(\cosh\mu-\cos\eta)}-\left(n^{\prime}+\frac{1}{2}\right)\right]\\
 & \times\sin\eta P_{n}^{1}(\cos\eta)P_{n^{\prime}}^{1}(\cos\eta)\,\cos^{2}(\phi-\phi_{1})d\eta\,d\phi
\end{align*}

\begin{align*}
 & \oint_{S_{\text{R}}}\Phi_{\text{R}}^{\text{in}^{*}}\boldsymbol{\nabla}\Phi_{\text{R}}^{\text{in}}\cdot d\textbf{S}\\
 & =-\frac{\pi a}{2}\sum_{n=1}^{\infty}\sum_{n^{\prime}=1}^{\infty}\sinh\mu_{\text{R}}A_{R,n}^{\text{in}^{*}}A_{R,n^{\prime}}^{\text{in}}\\
 & \times e^{-(n+n^{\prime}+1)\mu_{\text{R}}}\int_{-1}^{1}\frac{P_{n}^{1}(x)P_{n^{\prime}}^{1}(x)}{(\cosh\mu_{\text{R}}-x)}dx\\
 & +\pi a\sum_{n=1}^{\infty}\sum_{n^{\prime}=1}^{\infty}\left(n^{\prime}+\frac{1}{2}\right)A_{R,n}^{\text{in}^{*}}A_{R,n^{\prime}}^{\text{in}}\\
 & \times e^{-(n+n^{\prime}+1)\mu_{\text{R}}}\int_{-1}^{1}P_{n}^{1}(x)P_{n^{\prime}}^{1}(x)dx
\end{align*}

where $x=\cos\eta$. 

The second integral is evaluated from simple orthonormality property
of associated Legendre polynomials:

\[
\int_{-1}^{1}P_{n}^{1}(x)P_{n^{\prime}}^{1}(x)dx=\frac{2n(n+1)}{2n+1}\delta_{nn^{\prime}}
\]

The first integral is obtained from a corollary of Neumann's formula
for the Legendre Functions\cite{Macrobert1952}:

\[
-\frac{1}{2}\int_{-1}^{1}\frac{P_{n}^{1}(x)P_{n^{\prime}}^{1}(x)}{z-x}dx=P_{n^{\prime}}^{1}(z)Q_{n}^{1}(z),\quad n^{\prime}\le n
\]

where $|z|>1$ ($\cosh\mu\ge1$).

So,

\begin{align*}
 & \oint_{S_{\text{R}}}\Phi_{\text{R}}^{\text{in}^{*}}\boldsymbol{\nabla}\Phi_{\text{R}}^{\text{in}}\cdot d\textbf{S}\\
 & =-\frac{\pi a}{2}\sum_{n=1}^{\infty}\sum_{n^{\prime}=1}^{n}\sinh\mu_{\text{R}}A_{R,n}^{\text{in}^{*}}A_{R,n^{\prime}}^{\text{in}}\\
 & \times e^{-(n+n^{\prime}+1)\mu_{\text{R}}}\int_{-1}^{1}\frac{P_{n}^{1}(x)P_{n^{\prime}}^{1}(x)}{(\cosh\mu_{\text{R}}-x)}dx\\
 & -\frac{\pi a}{2}\sum_{n=1}^{\infty}\sum_{n^{\prime}=n+1}^{\infty}\sinh\mu_{\text{R}}A_{R,n}^{\text{in}^{*}}A_{R,n^{\prime}}^{\text{in}}\\
 & \times e^{-(n+n^{\prime}+1)\mu_{\text{R}}}\int_{-1}^{1}\frac{P_{n}^{1}(x)P_{n^{\prime}}^{1}(x)}{(\cosh\mu_{\text{R}}-x)}dx\\
 & +\pi a\sum_{n=1}^{\infty}\left|A_{R,n}^{\text{in}}\right|^{2}n(n+1)e^{-(2n+1)\mu_{\text{R}}}
\end{align*}

\begin{align}
 & =\pi a\sum_{n=1}^{\infty}\sum_{n^{\prime}=1}^{n}\sinh\mu_{\text{R}}A_{R,n}^{\text{in}^{*}}A_{R,n^{\prime}}^{\text{in}}\nonumber \\
 & \times e^{-(n+n^{\prime}+1)\mu_{\text{R}}}P_{n^{\prime}}^{1}(\cosh\mu_{\text{R}})Q_{n}^{1}(\cosh\mu_{\text{R}})\nonumber \\
 & +\pi a\sum_{n=1}^{\infty}\sum_{n^{\prime}=n+1}^{\infty}\sinh\mu_{\text{R}}A_{R,n}^{\text{in}^{*}}A_{R,n^{\prime}}^{\text{in}}\nonumber \\
 & \times e^{-(n+n^{\prime}+1)\mu_{\text{R}}}P_{n}^{1}(\cosh\mu_{\text{R}})Q_{n^{\prime}}^{1}(\cosh\mu_{\text{R}})\nonumber \\
 & +\pi a\sum_{n=1}^{\infty}\left|A_{R,n}^{\text{in}}\right|^{2}n(n+1)e^{-(2n+1)\mu_{\text{R}}}\label{eq:NR_decay_rate_bisph_perp_right}
\end{align}

For $\mu_{L}$,

\begin{align*}
 & \oint_{S_{\text{L}}}\Phi_{\text{L}}^{\text{in}^{*}}\overrightarrow{\nabla}\Phi_{\text{L}}^{\text{in}}\cdot d\overrightarrow{S}\\
 & =\frac{\pi a}{2}\sum_{n=0}^{\infty}\sum_{n^{\prime}=1}^{\infty}\sinh\mu_{\text{L}}B_{L,n}^{\text{in}^{*}}B_{L,n^{\prime}}^{\text{in}}\\
 & \times e^{(n+n^{\prime}+1)\mu_{\text{L}}}\int_{-1}^{1}\frac{P_{n}^{1}(x)P_{n^{\prime}}^{1}(x)}{(\cosh\mu_{\text{L}}-x)}dx\\
 & +\pi a\sum_{n=1}^{\infty}\sum_{n^{\prime}=1}^{\infty}\left(n^{\prime}+\frac{1}{2}\right)B_{L,n}^{\text{in}^{*}}B_{L,n^{\prime}}^{\text{in}}\\
 & \times e^{(n+n^{\prime}+1)\mu_{\text{L}}}\int_{-1}^{1}P_{n}^{1}(x)P_{n^{\prime}}^{1}(x)dx
\end{align*}

\begin{align}
 & =-\pi a\sum_{n=1}^{\infty}\sum_{n^{\prime}=1}^{n}\sinh\mu_{\text{L}}B_{L,n}^{\text{in}^{*}}B_{L,n^{\prime}}^{\text{in}}\nonumber \\
 & \times e^{(n+n^{\prime}+1)\mu_{\text{L}}}P_{n^{\prime}}^{1}(\cosh\mu_{\text{L}})Q_{n}^{1}(\cosh\mu_{\text{L}})\nonumber \\
 & -\pi a\sum_{n=1}^{\infty}\sum_{n^{\prime}=n+1}^{\infty}\sinh\mu_{\text{L}}B_{L,n}^{\text{in}^{*}}B_{L,n^{\prime}}^{\text{in}}\nonumber \\
 & \times e^{(n+n^{\prime}+1)\mu_{\text{L}}}P_{n}^{1}(\cosh\mu_{\text{L}})Q_{n^{\prime}}^{1}(\cosh\mu_{\text{L}})\nonumber \\
 & +\pi a\sum_{n=1}^{\infty}\left|B_{L,n}^{\text{in}}\right|^{2}n(n+1)e^{(2n+1)\mu_{\text{L}}}\label{eq:NR_decay_rate_bisph_perp_left}
\end{align}

\subsubsection{The radiative decay rate \label{subsec:Rad_Dec_Bisph_Perp}}

As evaluated in Subsection \ref{subsec:Rad_Dec_Bisph_Para}, the total
dipole moment of the composite system, given by ${\bf D}_{\text{tot}}(\omega)$,
is obtained from Eq. \ref{eq:induced_dipole_ns}. For this particular
configuration, the integral corresponding to the sphere $\mu_{\text{R}}$
is:
\begin{align*}
 & -\oint_{S_{\text{R}}}\Phi_{\text{R}}^{\text{in}}({\bf r},\omega)d{\bf S}\\
 & =\boldsymbol{\widehat{i}}\pi a^{2}\sinh\mu_{\text{R}}\sum_{n=1}^{\infty}A_{R,n}^{\text{in}}e^{-(n+1/2)\mu_{\text{R}}}\\
 & \times\int_{0}^{\pi}\frac{P_{n}^{1}(\cos\eta)\sin^{2}\eta}{(\cosh\mu_{\text{R}}-\cos\eta)^{5/2}}d\eta
\end{align*}

For the sphere $\mu_{\text{L}}$

\begin{align*}
 & -\oint_{S_{\text{L}}}\Phi_{\text{L}}^{\text{in}}({\bf r},\omega)d{\bf S}\\
 & =-\boldsymbol{\widehat{i}}\pi a^{2}\sinh\mu_{\text{L}}\sum_{n=1}^{\infty}B_{L,n}^{\text{in}}e^{(n+1/2)\mu_{\text{L}}}\\
 & \times\int_{0}^{\pi}\frac{P_{n}^{1}(\cos\eta)\sin^{2}\eta}{(\cosh\mu_{\text{L}}-\cos\eta)^{5/2}}d\eta
\end{align*}

These integrals are obtained numerically. 

\section{Results for some additional configurations\label{SI_results_additional}}

\subsection{A molecule near a prolate spheroidal nanoparticle on the major axis
with the dipole perpendicular to the nanoparticle surface \label{subsec:spheroid_perp_results_gold_silver}}

For the perpendicular orientation of the transition dipole near gold
and silver spheroidal nanoparticles as shown in main text, the variations
of nonratiative and radiative relaxation rates as functions of molecule
to nanoparticle surface distance are shown here. 

\begin{figure}
\subfloat[]{\includegraphics[viewport=0bp 0bp 468bp 328.061bp,scale=0.5]{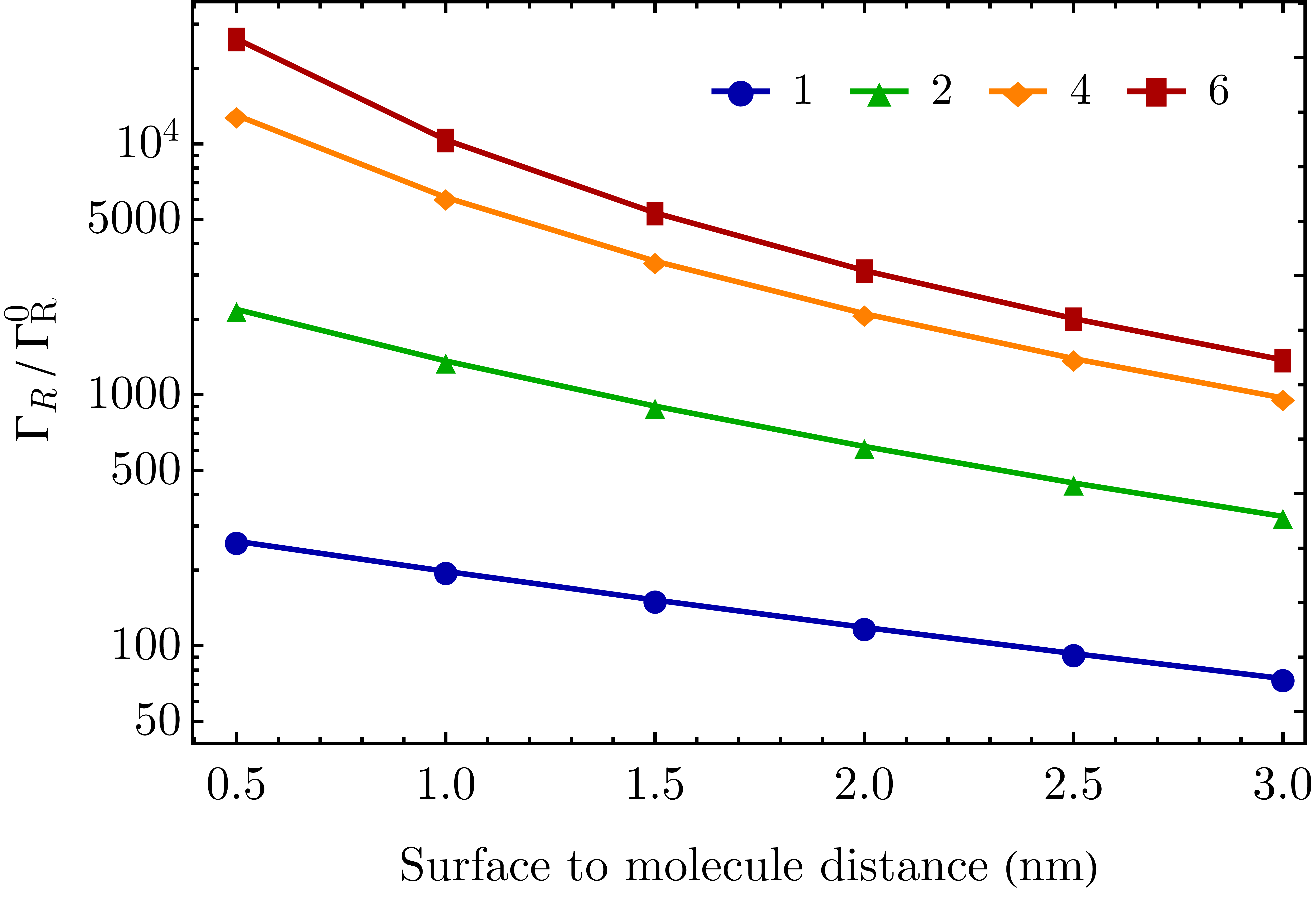}

\label{fig:SI_GamR_spheroid_major_perp_gold_aspect_ratio}}\subfloat[]{\includegraphics[viewport=0bp 0bp 468bp 328.314bp,scale=0.5]{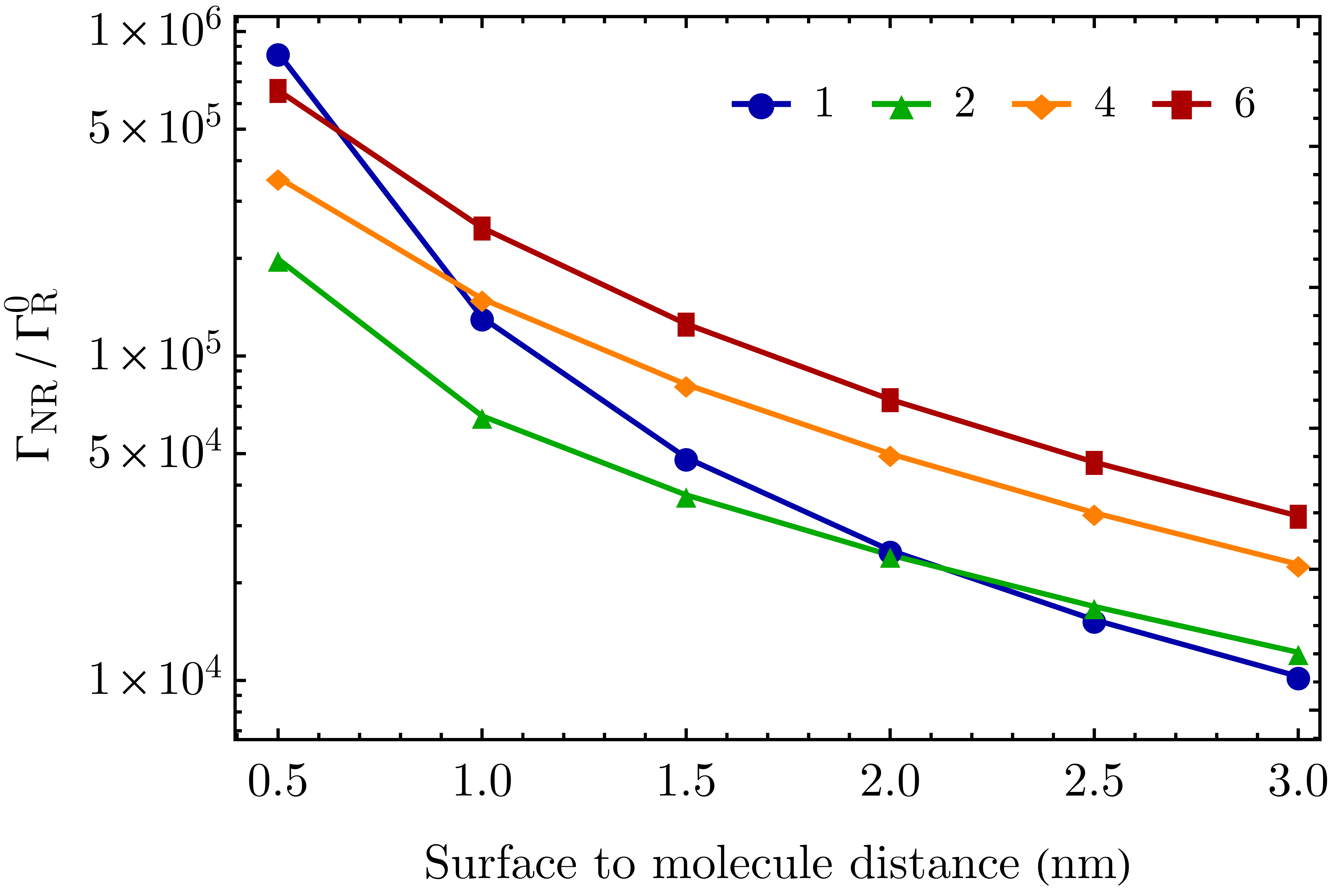}

\label{fig:SI_GamNR_spheroid_major_perp_gold_aspect_ratio}}

\caption{(a) Radiative ($\Gamma_{\text{R}}$) and (b) nonradiative ($\Gamma_{\text{NR}}$)
relaxation rates as functions of the molecule to spheroidal gold nanoparticle
surface distance as aspect ratio of the spheroids $p/q$ varies from
$1$ to $6$ keeping the particle volume constant (and all have the
volume of a sphere of radius $10$ nm). The molecule is perpendicular
to the surface of the nanospheroid on long axis For each aspect ratio
we take the molecular frequency equal to the frequency of the respective
longitudinal dipolar plasmon (excited along the particle long axis).}
\label{fig:SI_GamR_GamNR_spheroid_major_perp_gold_aspect_ratio}
\end{figure}

\begin{figure}
\subfloat[]{\includegraphics[viewport=0bp 0bp 468bp 317.705bp,scale=0.5]{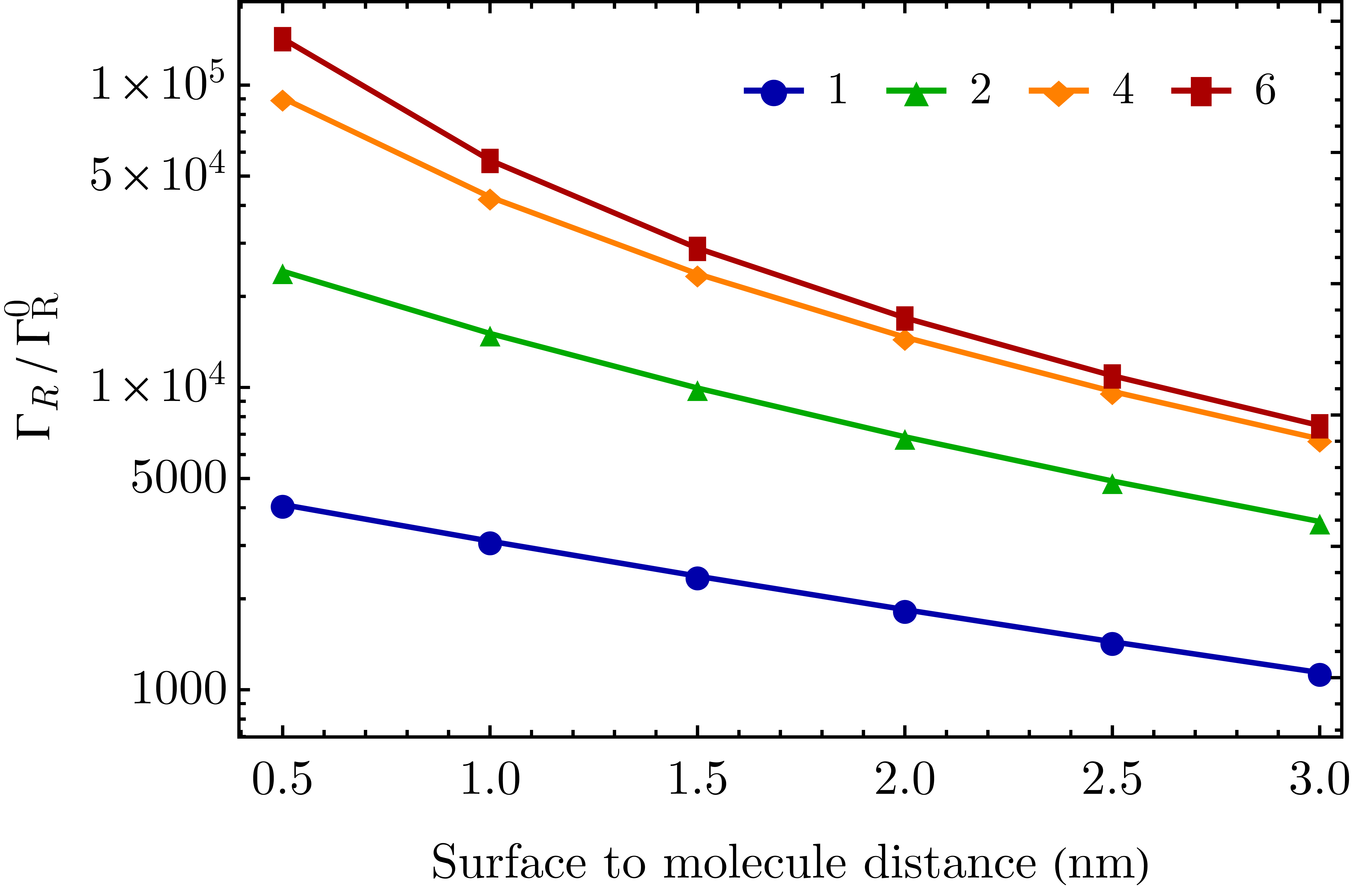}\label{fig:SI_GamR_spheroid_major_perp_silver_aspect_ratio}}\subfloat[]{\includegraphics[viewport=0bp 0bp 468bp 309.64bp,scale=0.5]{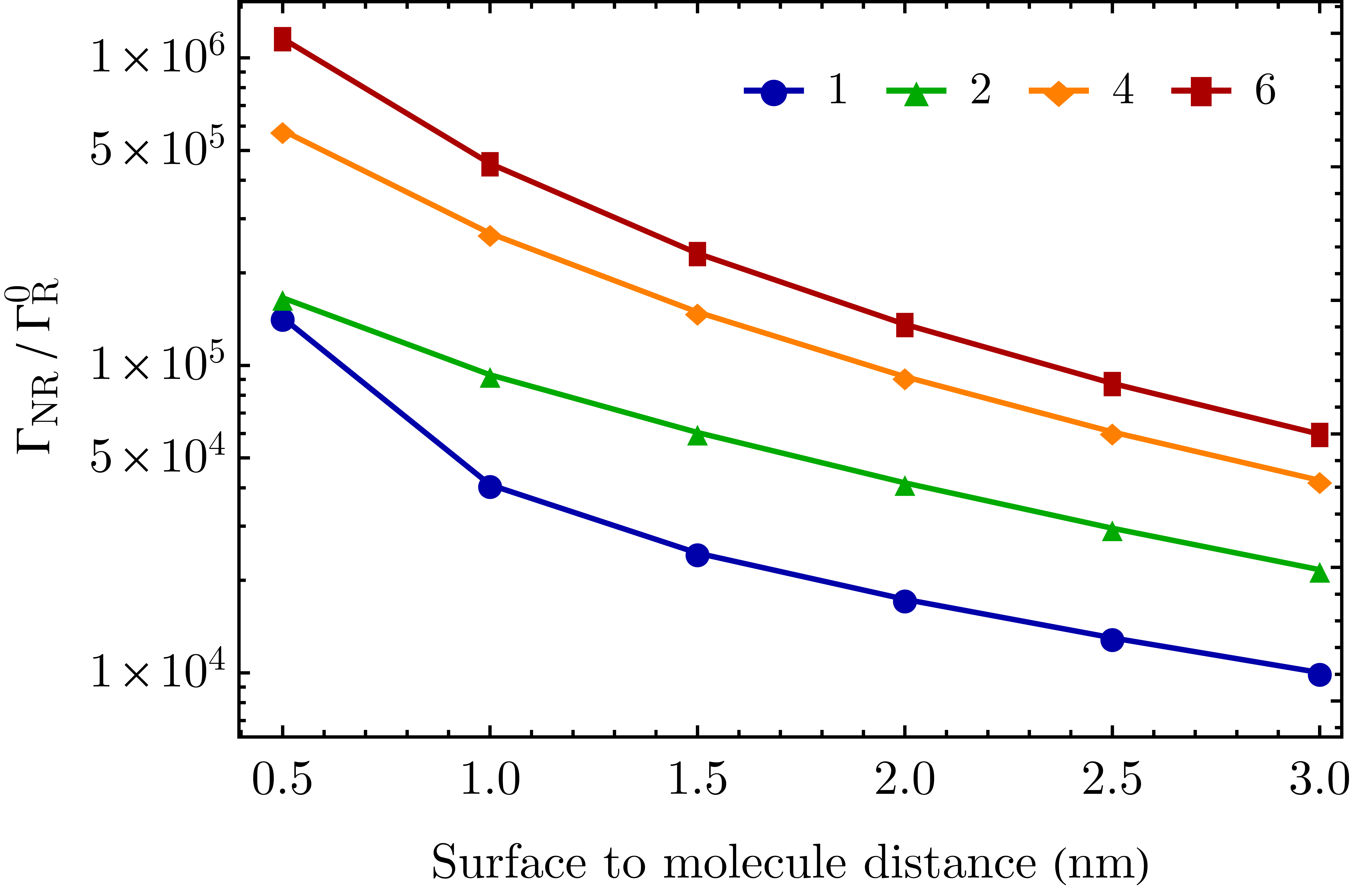}

\label{fig:SI_GamNR_spheroid_major_perp_silver_aspect_ratio}}\caption{(a) Radiative ($\Gamma_{\text{R}}$) and (b) nonradiative ($\Gamma_{\text{NR}}$)
relaxation rates as functions of the molecule to spheroidal silver
nanoparticle surface distance as aspect ratio of the spheroids $p/q$
varies from $1$ to $6$ keeping the particle volume constant (and
all have the volume of a sphere of radius $10$ nm). The molecule
is perpendicular to the surface of the nanospheroid on long axis For
each aspect ratio we take the molecular frequency equal to the frequency
of the respective longitudinal dipolar plasmon (excited along the
particle long axis).}
\label{fig:SI_GamR_GamNR_spheroid_major_perp_silver_aspect_ratio}
\end{figure}

\subsection{A molecule near a prolate spheroidal nanoparticle on the major axis
with the dipole parallel to the nanoparticle surface \label{subsec:spheroid_para_results_gold_silver}}

In addition to the perpendicular orientation of the transition dipole
near gold and silver spheroidal nanoparticles discussed in the main
text, here we show the results for the parallel orientation. The parameter
$M$, the total relaxation rate $\Gamma$ normalized by the radiative
relaxation rate of a free molecule and the strong coupling parameter
$2|U|/(\hbar\Gamma)$ are plotted against the distance of the molecule
from the nanoparticle surface. Four different aspect ratios ($=1,2,4,6$)
of the nanospheroid are considered. Their volumes are kept constant
to that of a $10$ nm nanosphere. For each aspect ratio, molecular
frequency is fixed to be the resonance frequency of the longitudinal
plasmon of the respective geometry. As usual, the transition dipole
is taken to be $10$ D. For each aspect ratio we take the molecular
frequency equal to the frequency of the respective transverse dipolar
plasmon (excited along the particle short axis).

\begin{figure}
\subfloat[]{\includegraphics[viewport=0bp 0bp 468bp 328bp,scale=0.33]{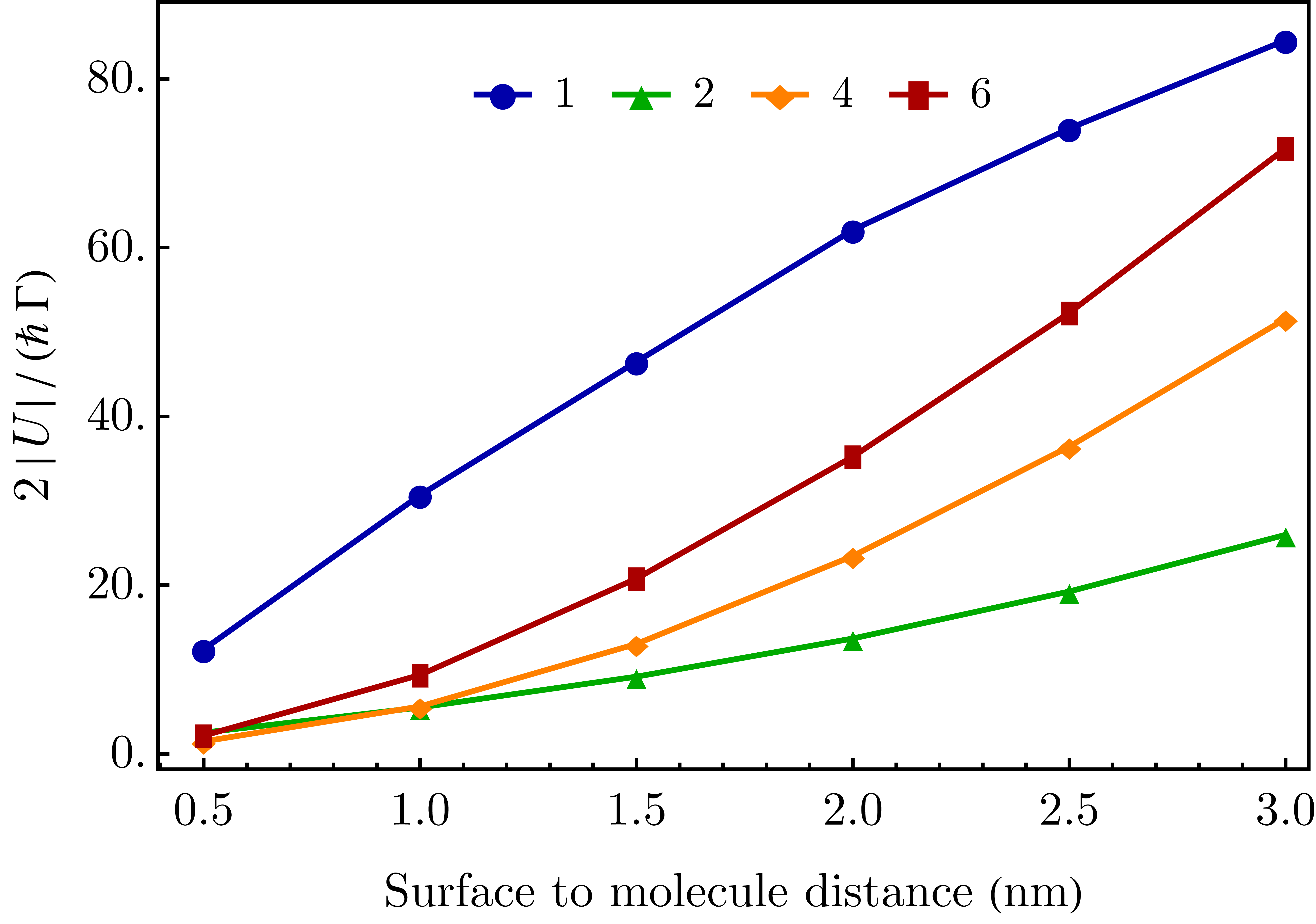}

\label{fig:SI_QSC_spheroid_major_para_gold_aspect_ratio}}\subfloat[]{\includegraphics[viewport=0bp 0bp 468bp 317.705bp,scale=0.33]{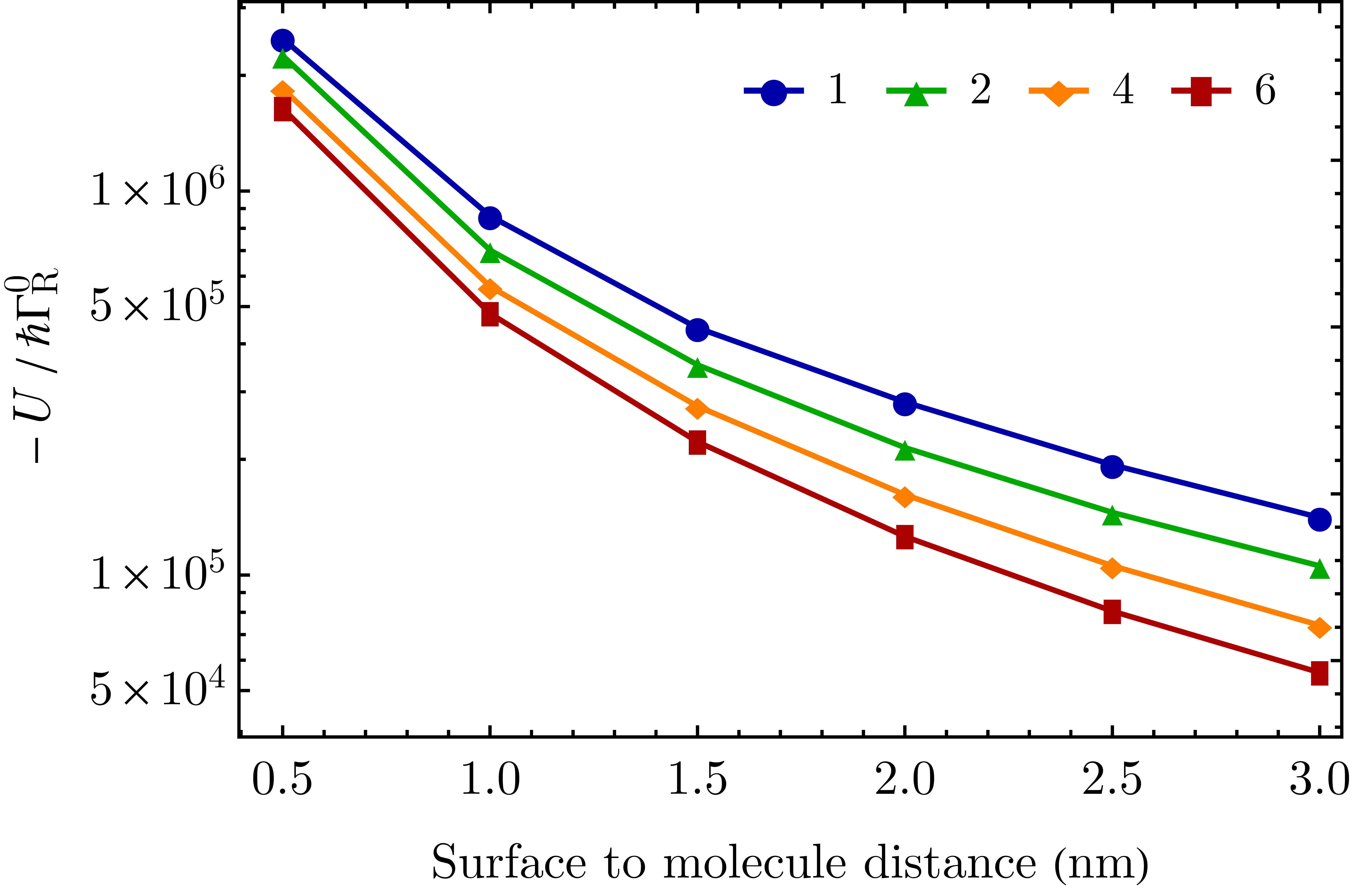}\label{fig:SI_U_spheroid_major_para_gold_aspect_ratio}}\subfloat[]{\includegraphics[viewport=0bp 0bp 468bp 309.22bp,scale=0.33]{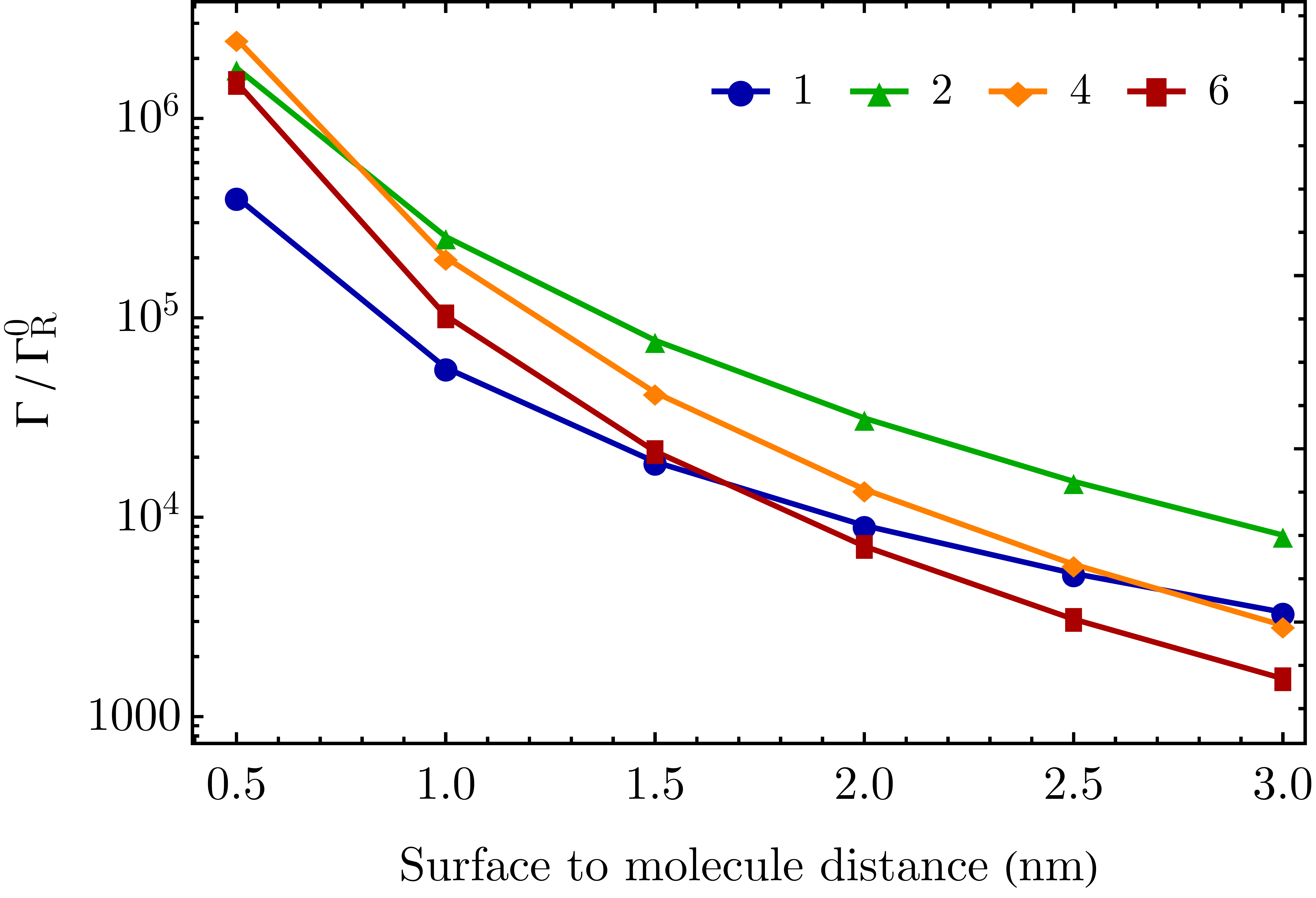}

\label{fig:SI_Gamma_spheroid_major_para_gold_aspect_ratio}}

\caption{(a) The CBR parameter (b) magnitude of coupling ($-U/\hbar$), and
(c) total relaxation rate $\Gamma$ as functions of the molecule to
spheroidal gold nanoparticle surface distance when a molecule is oriented
parallel to the surface of gold nanospheroid on the long axis as the
aspect ratio of the spheroids $p/q$ varies from $1$ to $6$. Molecular
frequency and the corresponding free molecule radiative decay rate
are chosen as: $2.62$ eV and $7.38\times10^{7}$ $\text{s}^{-1}$
for $p/q=1$, $2.68$ eV and $7.96\times10^{7}$ $\text{s}^{-1}$
for $p/q=2$, $2.72$ eV and $8.24\times10^{7}$ $\text{s}^{-1}$
for $p/q=4$ and $2.72$ eV and $8.32\times10^{7}$ $\text{s}^{-1}$
for $p/q=6$.}
\label{fig:SI_QSC_U_Gamma_spheroid_major_para_gold_aspect_ratio}
\end{figure}

\begin{figure}[H]
\subfloat[]{\includegraphics[scale=0.33]{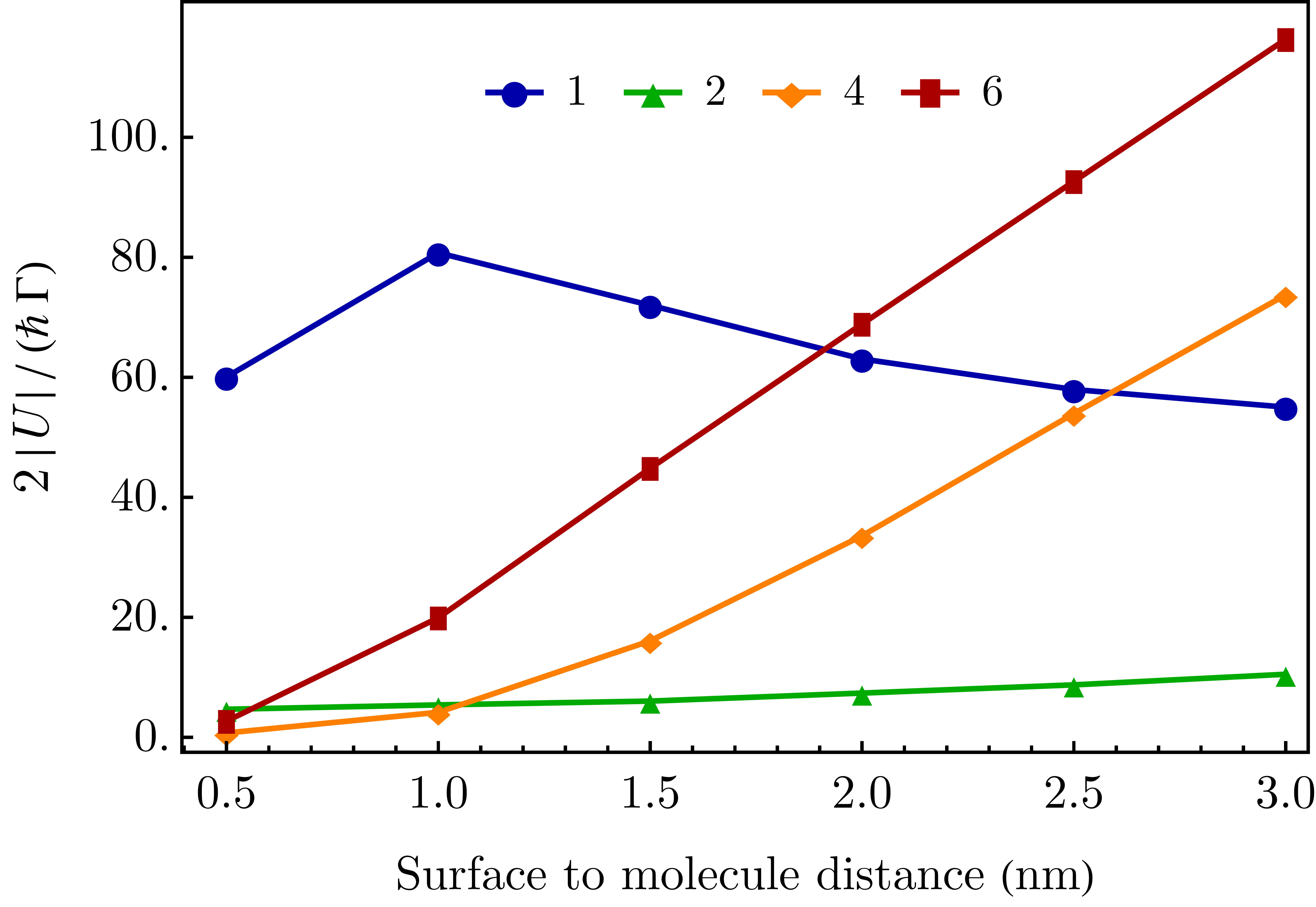}

\label{fig:SI_QSC_spheroid_major_para_silver_aspect_ratio}}\subfloat[]{\includegraphics[scale=0.33]{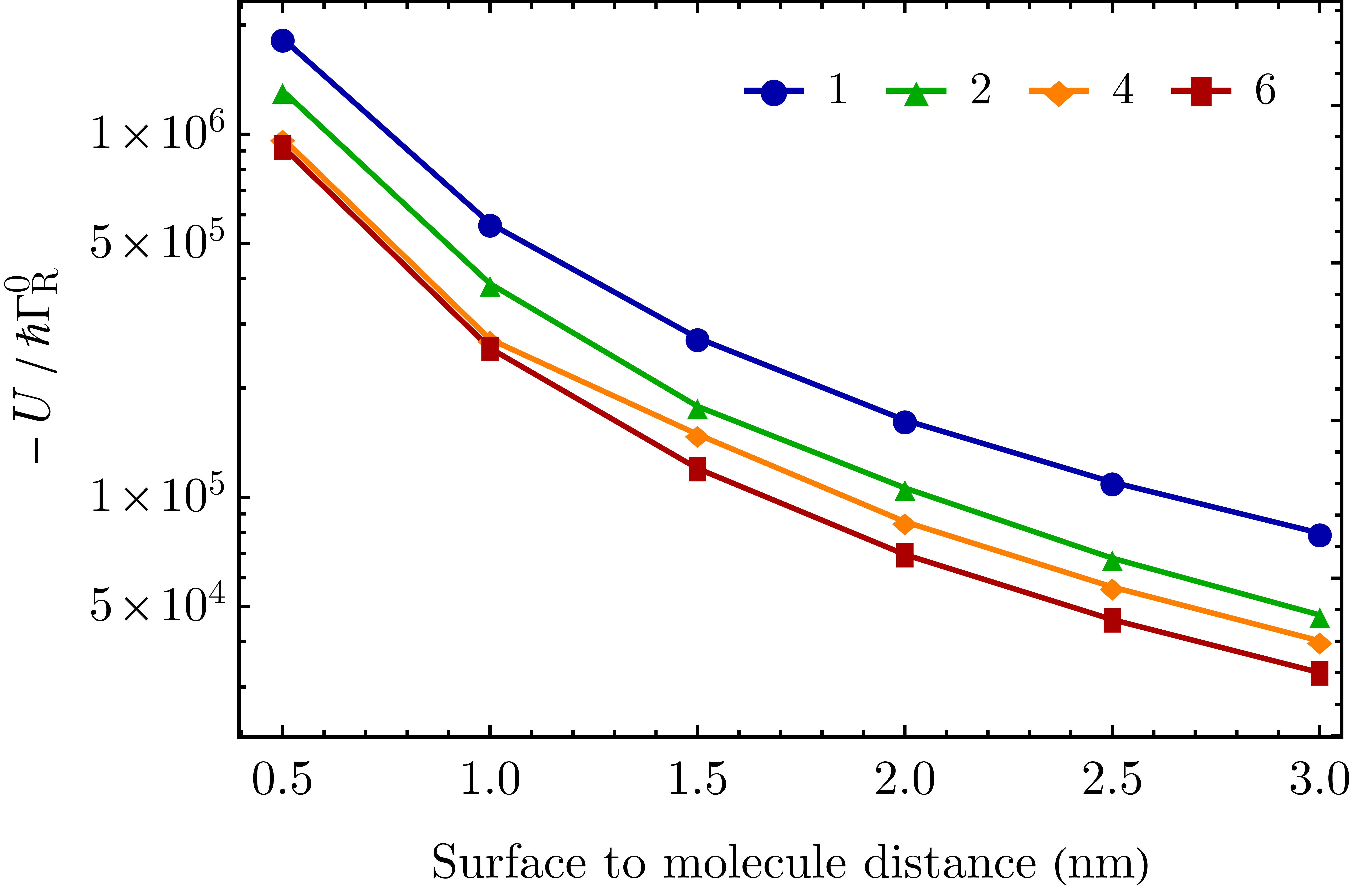}\label{fig:SI_U_spheroid_major_para_silver_aspect_ratio}}\subfloat[]{\includegraphics[scale=0.33]{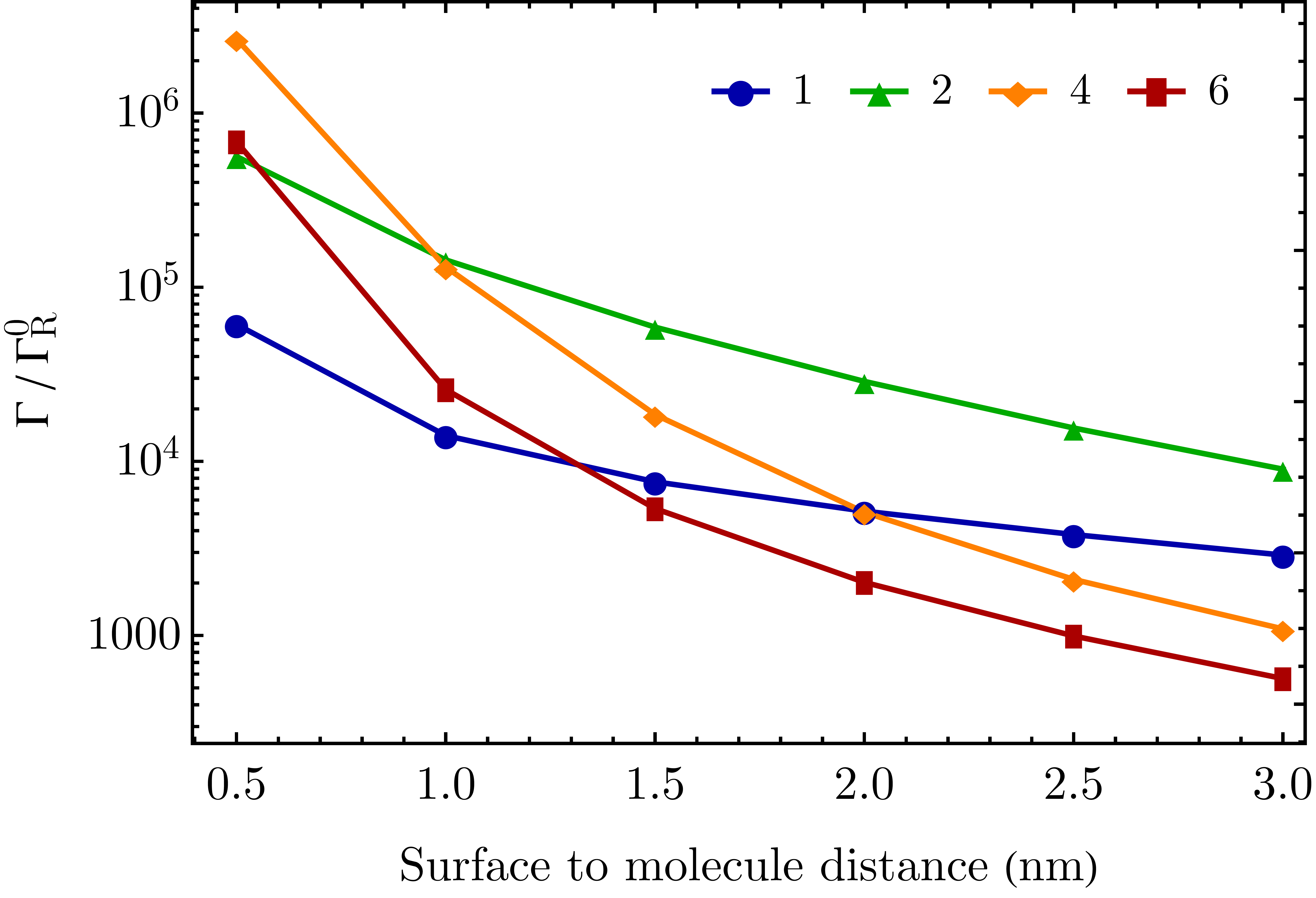}

\label{fig:SI_Gamma_spheroid_major_para_silver_aspect_ratio}}

\caption{(a) The CBR parameter (b) magnitude of coupling ($-U/\hbar$), and
(c) total relaxation rate $\Gamma$ as functions of the molecule to
spheroidal silver nanoparticle surface distance when a molecule is
oriented parallel to the surface of silver nanospheroid on the long
axis as the aspect ratio of the spheroids $p/q$ varies from $1$
to $6$. Molecular frequency and the corresponding free molecule radiative
decay rate are chosen as: $3.37$ eV and $1.57\times10^{8}$ $\text{s}^{-1}$
for $p/q=1$, $3.52$ eV and $1.79\times10^{8}$ $\text{s}^{-1}$
for $p/q=2$, $3.59$ eV and $1.90\times10^{8}$ $\text{s}^{-1}$
for $p/q=4$ and $3.61$ eV and $1.94\times10^{8}$ $\text{s}^{-1}$
for $p/q=6$.}
\label{fig:SI_QSC_U_Gamma_spheroid_major_para_silver_aspect_ratio}
\end{figure}

\subsection{Effective volume\label{subsec:effective_volume}}

\begin{figure}[H]
\subfloat[]{\includegraphics[scale=0.5]{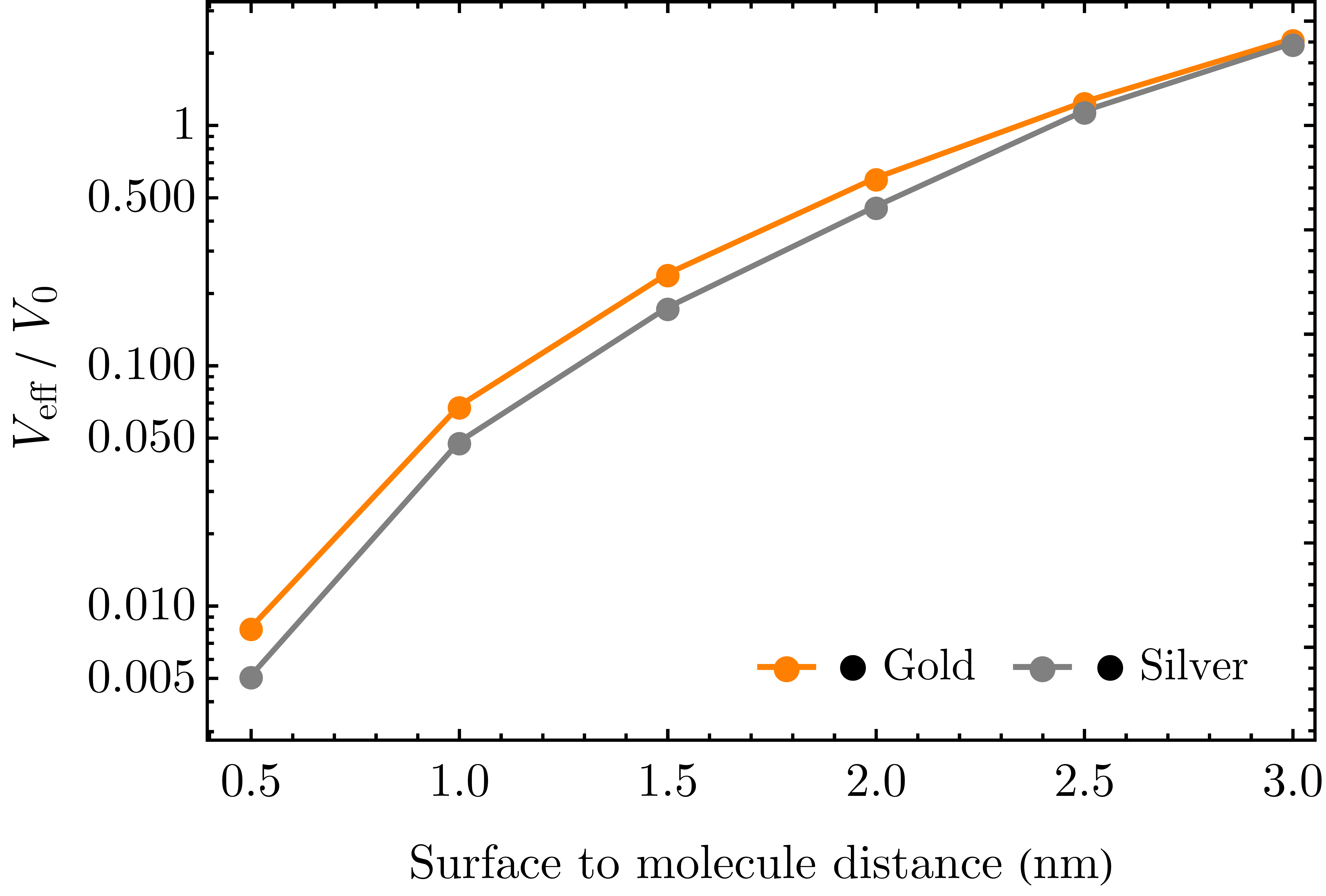}

\label{fig:SI_eff_vol_sph_perp_gold_silver_par_vol}}\subfloat[]{\includegraphics[scale=0.5]{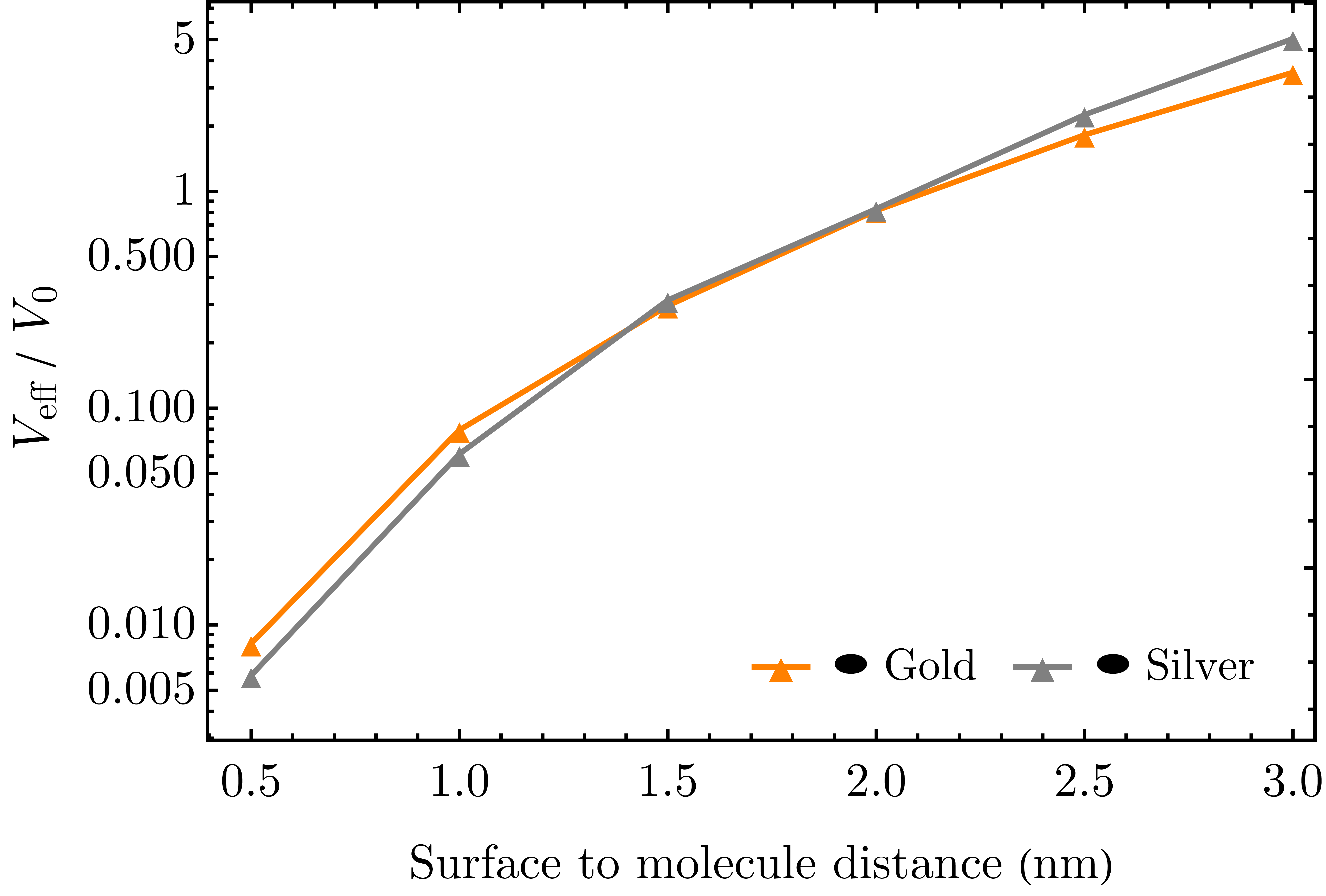}\label{fig:SI_eff_vol_sphd_perp_gold_silver_par_vol}}

\caption{Effective volume $V_{\text{eff}}$ of the resulting plasmonic cavity
for a molecular dipole near (a) spherical nanoparticles of $10$ nm
radius and (b) spheroidal nanoparticles of aspect ratio $2$ with
same volume as (a), plotted against molecule to surface distance.
In both cases, molecular dipole is in perpendicular orientation. $V_{\text{eff}}$
is normalized by the nanoparticle volume ($V_{0}$).}
\label{fig:SI_eff_vol_sph_sphd_perp_gold_silver_par_vol}
\end{figure}
\begin{figure}[H]
\subfloat[]{\includegraphics[scale=0.5]{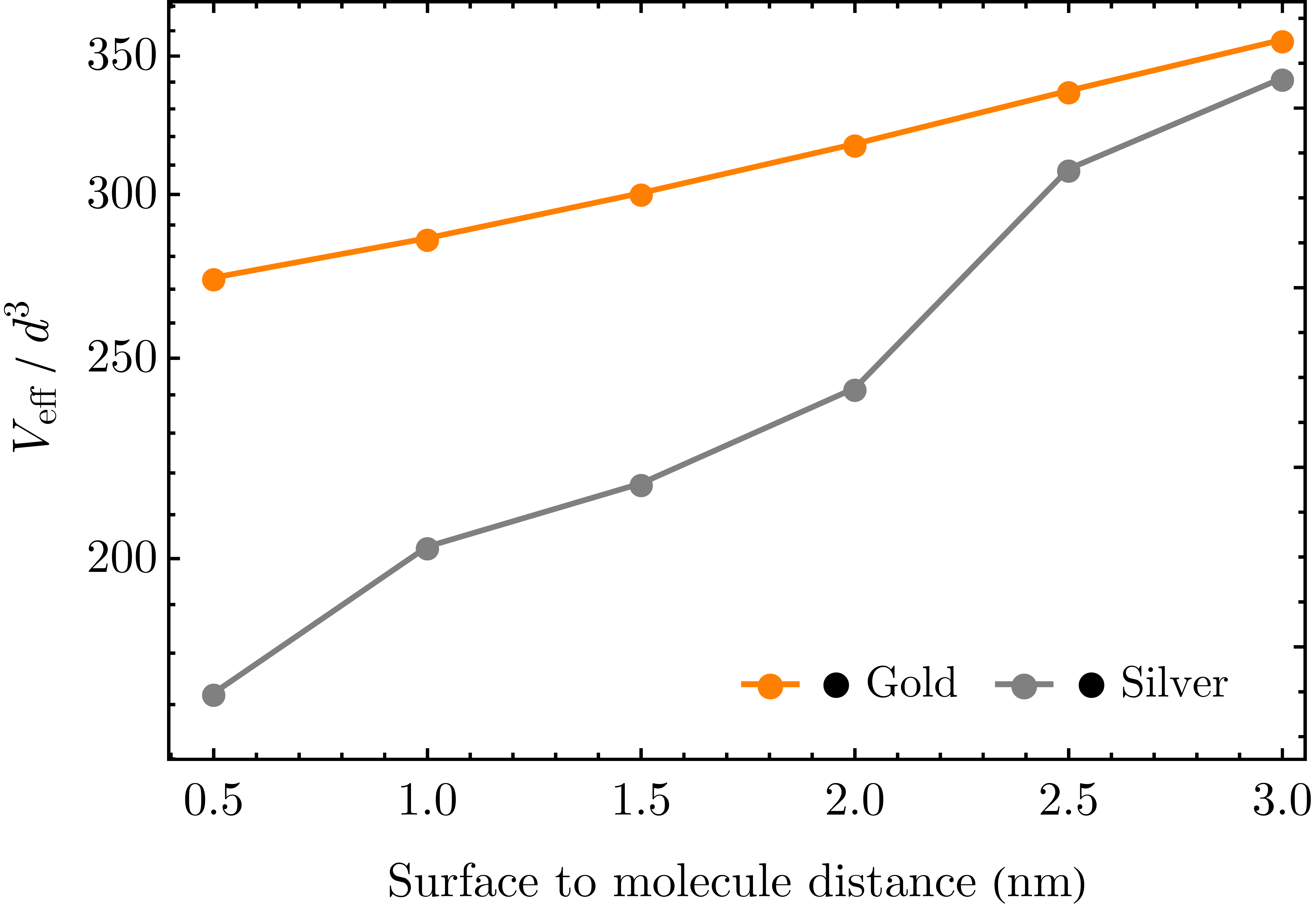}

\label{fig:SI_eff_vol_sph_perp_gold_silver_dist_cube}}\subfloat[]{\includegraphics[scale=0.5]{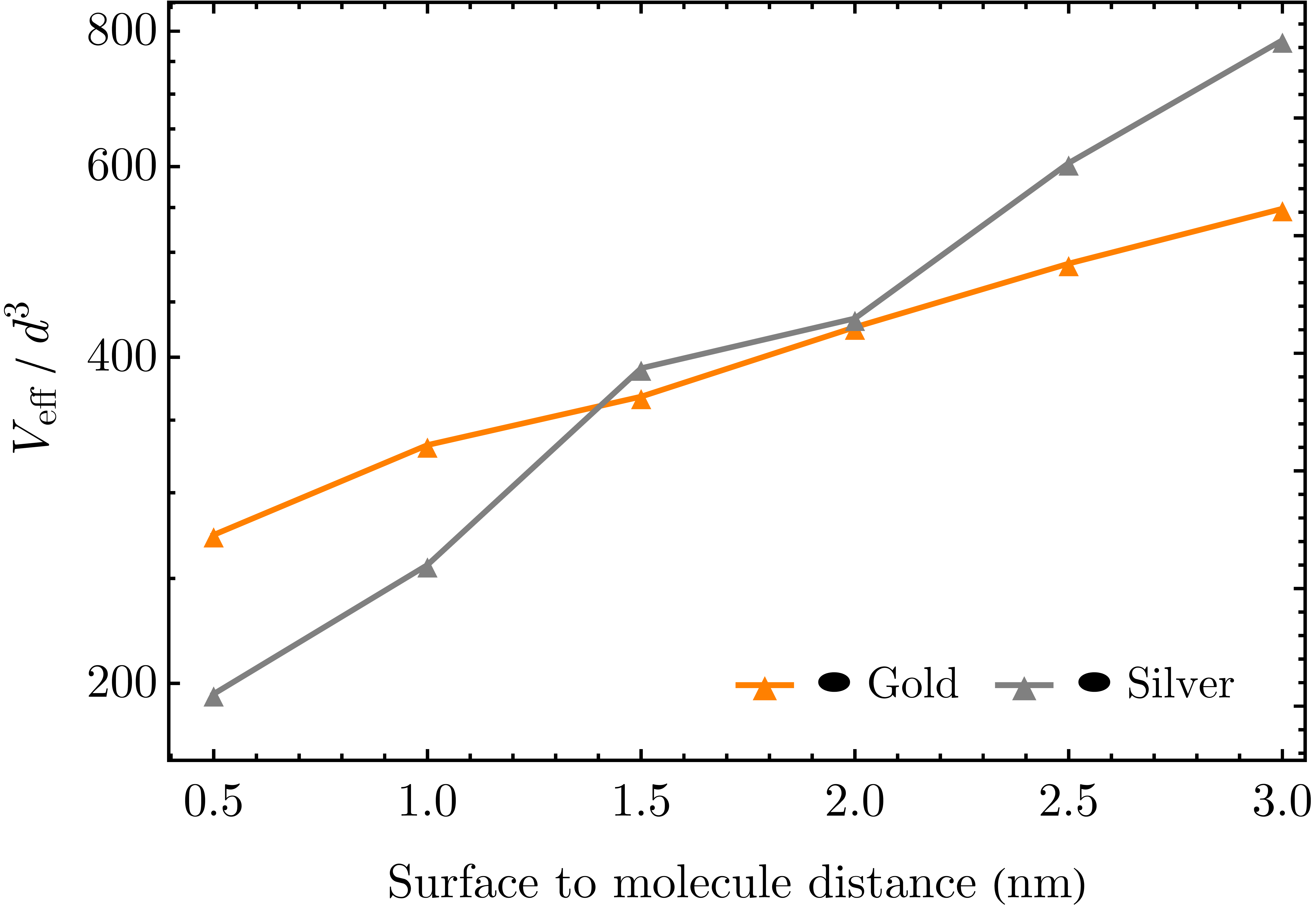}\label{fig:SI_eff_vol_sphd_perp_gold_silver_dist_cube}}

\caption{Same as Figure \ref{fig:SI_eff_vol_sph_sphd_perp_gold_silver_par_vol},
but now $V_{\text{eff}}$ is normalized by the cube of the respective
molecule to surface distance ($d^{3}$). }
\label{fig:SI_eff_vol_sph_sphd_perp_gold_silver_dist_cube}
\end{figure}
\begin{figure}[H]
\includegraphics[scale=0.65]{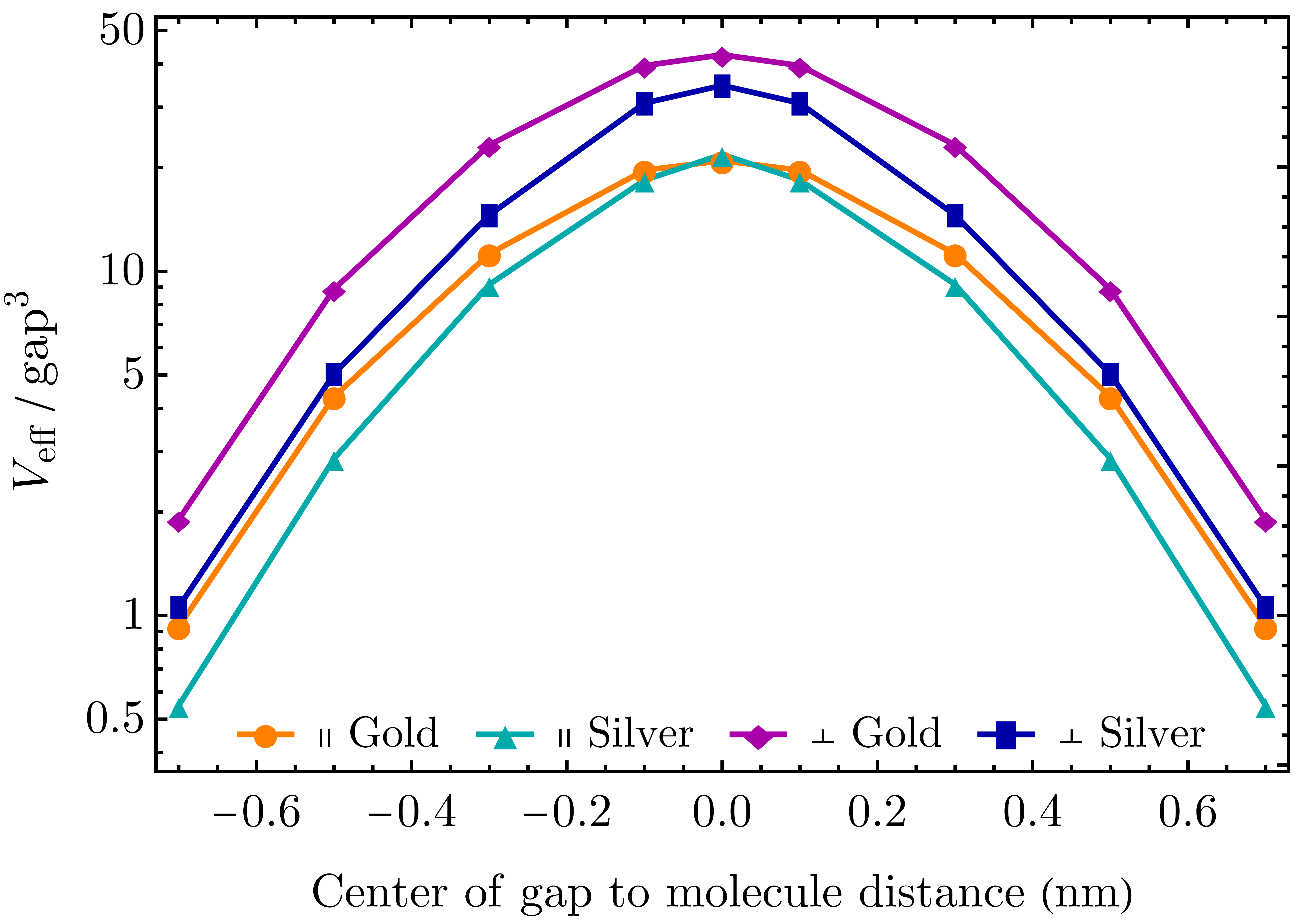}\caption{Effective volume $V_{\text{eff}}$ of the plasmonic cavity when a
molecular dipole is placed in a $2$ nm gap of a nanoparticle dimer
(each of $10$ nm radius) of silver and gold in both orienatations,
parallel and perpendicular, relative to the intersphere axis plotted
against the position of the molecule relative to the center of the
gap. Here $V_{\text{eff}}$ is normalized by the cube of the gap size.}
\label{fig:SI_eff_vol_bisph_para_perp_gold_silver_gap_cube}
\end{figure}

\subsection{Nanoparticle dimer in uniform electric field\label{subsec:np_dimer_uniform_field}}

\begin{figure}[H]
\subfloat[]{\includegraphics[scale=0.5]{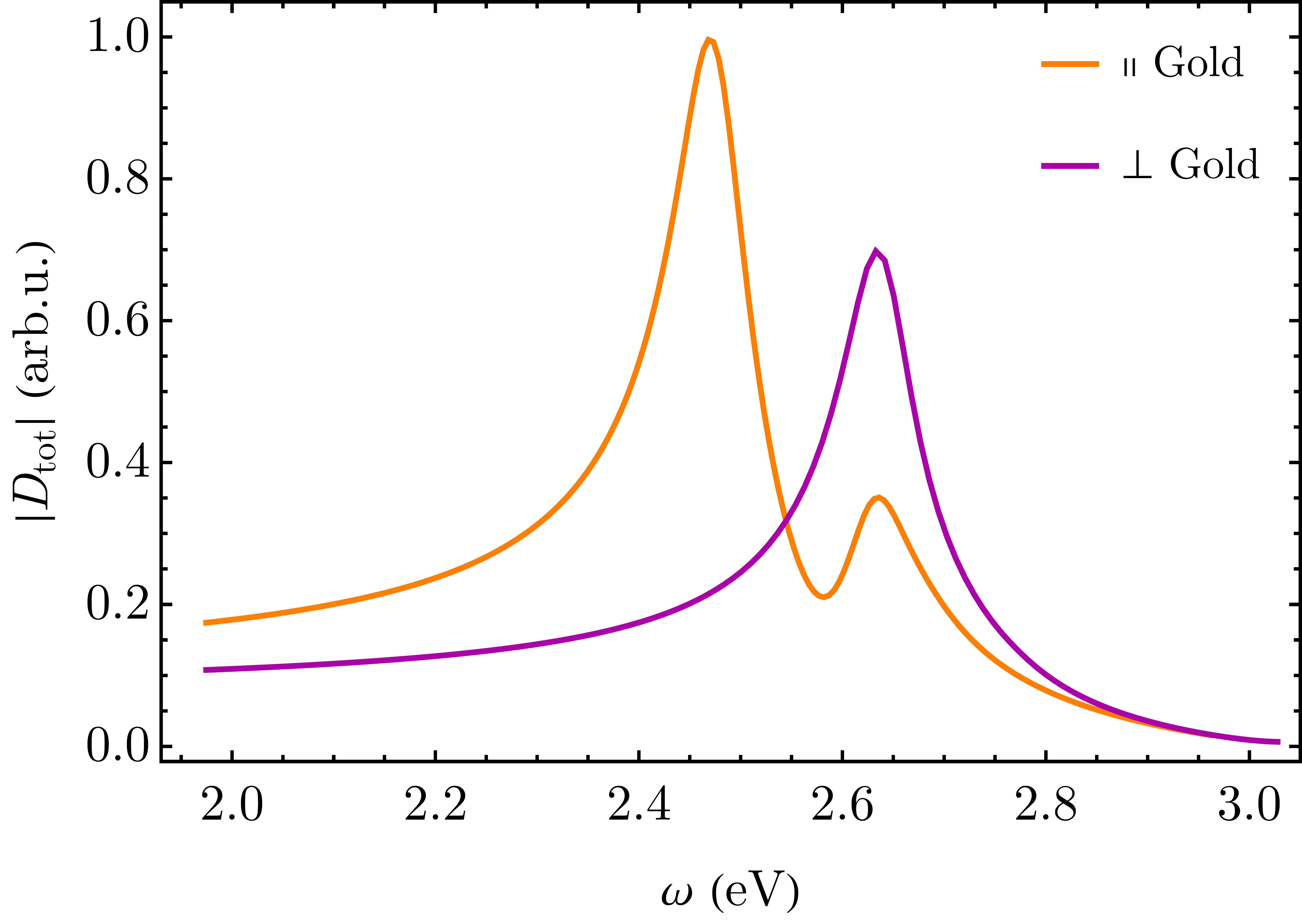}

\label{fig:SI_rad_decay_bisph_uni_field_gold_para_perp}}\subfloat[]{\includegraphics[scale=0.5]{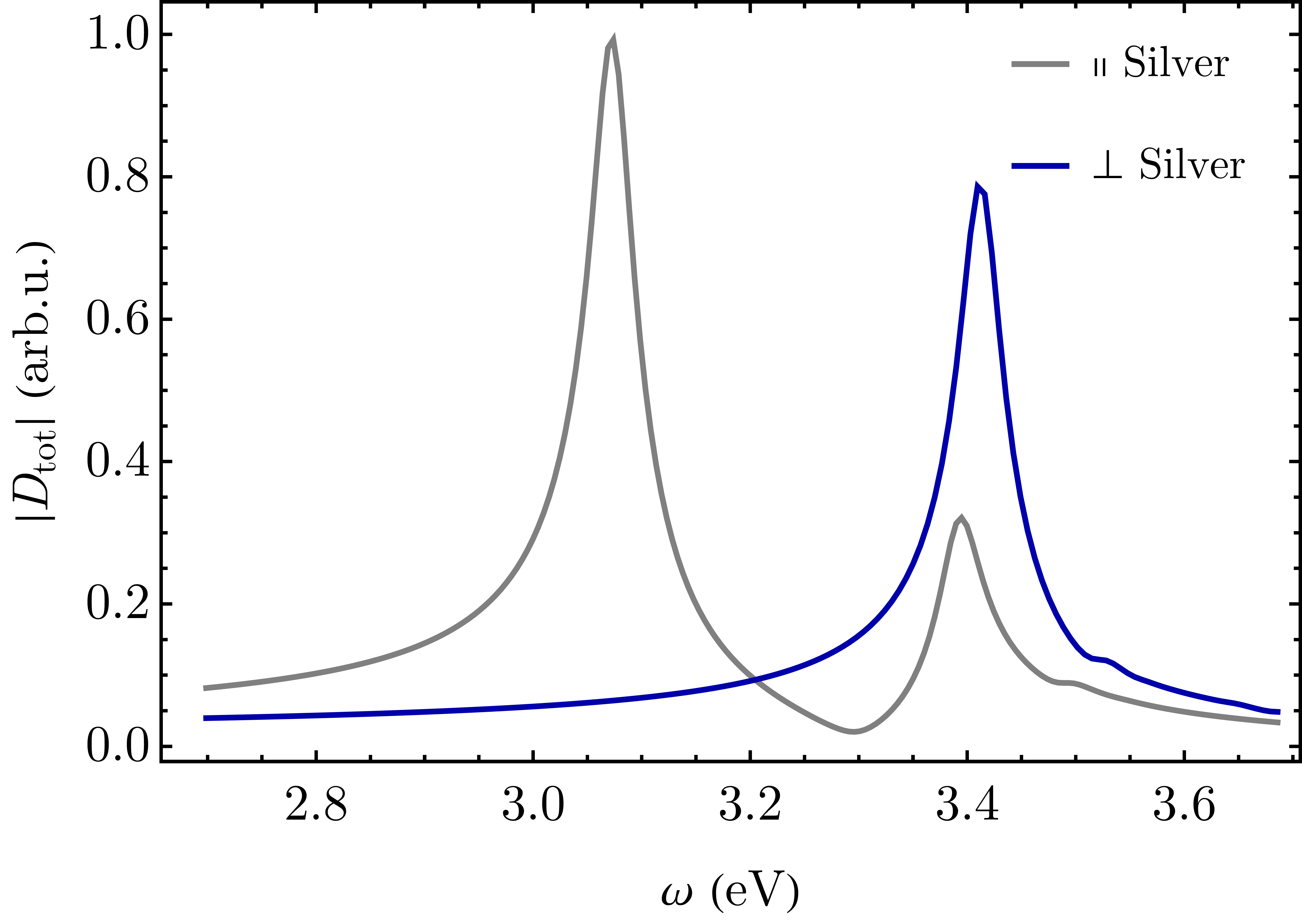}\label{fig:SI_rad_decay_bisph_uni_field_silver_para_perp}}

\caption{Normalized total dipole moment of the system as functions of driving
frequency for (a) gold and (b) silver nanoparticle dimer of $10$
nm radius and $2$ nm gap size in a uniform electric field field.
Both polarization of incident field- parallel and perpendicular relative
to the intersphere axis are shown. }
\label{fig:SI_rad_decay_bisph_uni_field_gold_silver_para_perp}
\end{figure}

\subsection{Absorption lineshape from self-consistent method for a molecule near
a silver sphere\label{subsec:SI_self_cons_silver}}

Here we show the absorption lineshape calculated from Eq. (\ref{eq:A_omega})
for a dipole emitter near a spherical silver nanophere (radius $10$
nm), oriented perpendicular to the sphere surface and parallel to
the incident field. The transition frequency $\omega_{0}$ is taken
$3.37$ eV, in resonance with the dipolar plasmon frequency of the
silver nanosphere. As before, the relaxation rate $\Gamma^{0}$ of
the free molecule is taken to be twice the radiative relaxation rate,
corresponding to an emission yield of $0.5$ for the free molecule. 

\begin{figure}[H]
\subfloat[]{\includegraphics[scale=0.5]{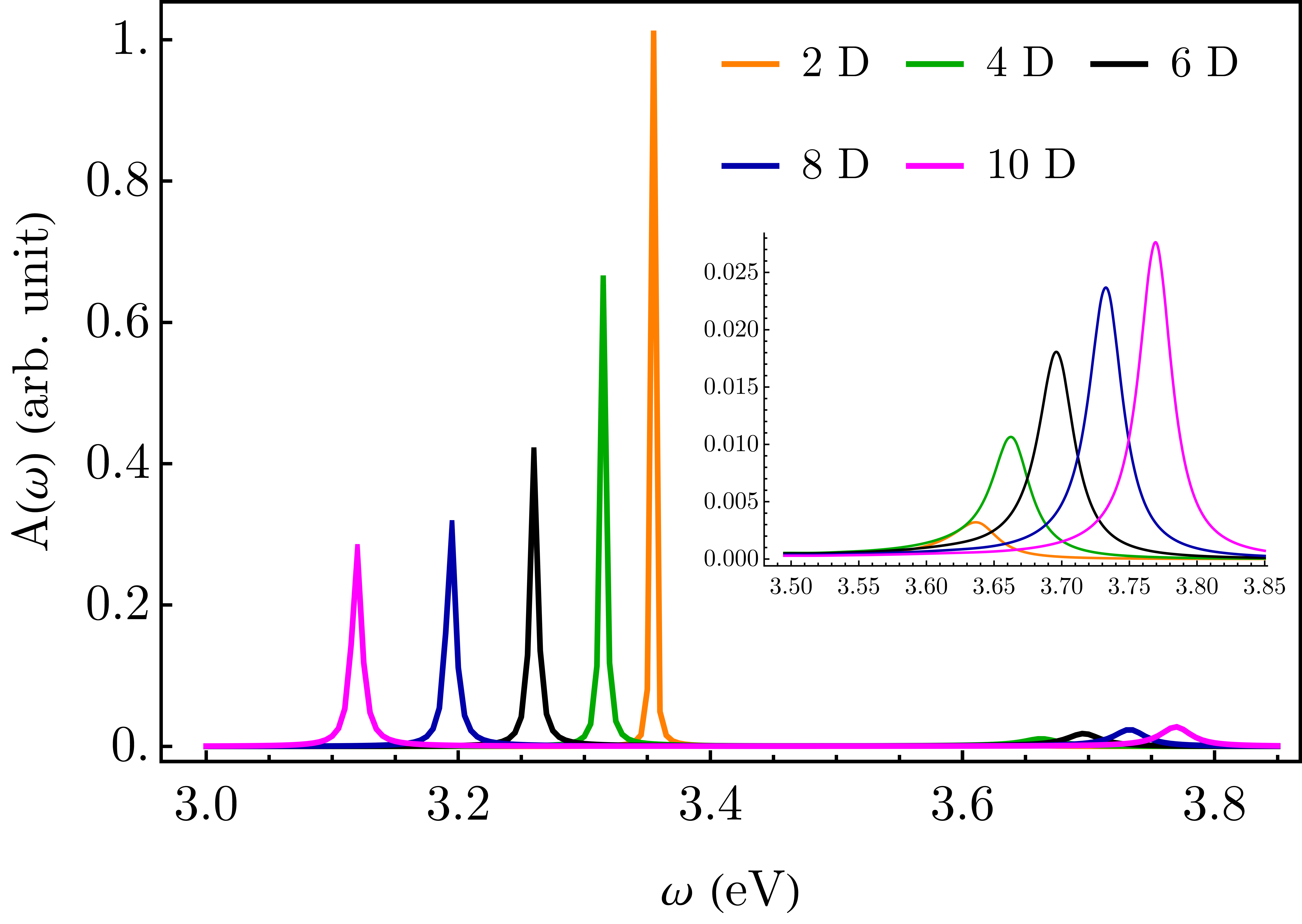}

\label{fig:SI_SC_Abs_Lnshp_sphere_perp_silevr_2D_10D_a0_10_r1_0.5}}\subfloat[]{\includegraphics[scale=0.5]{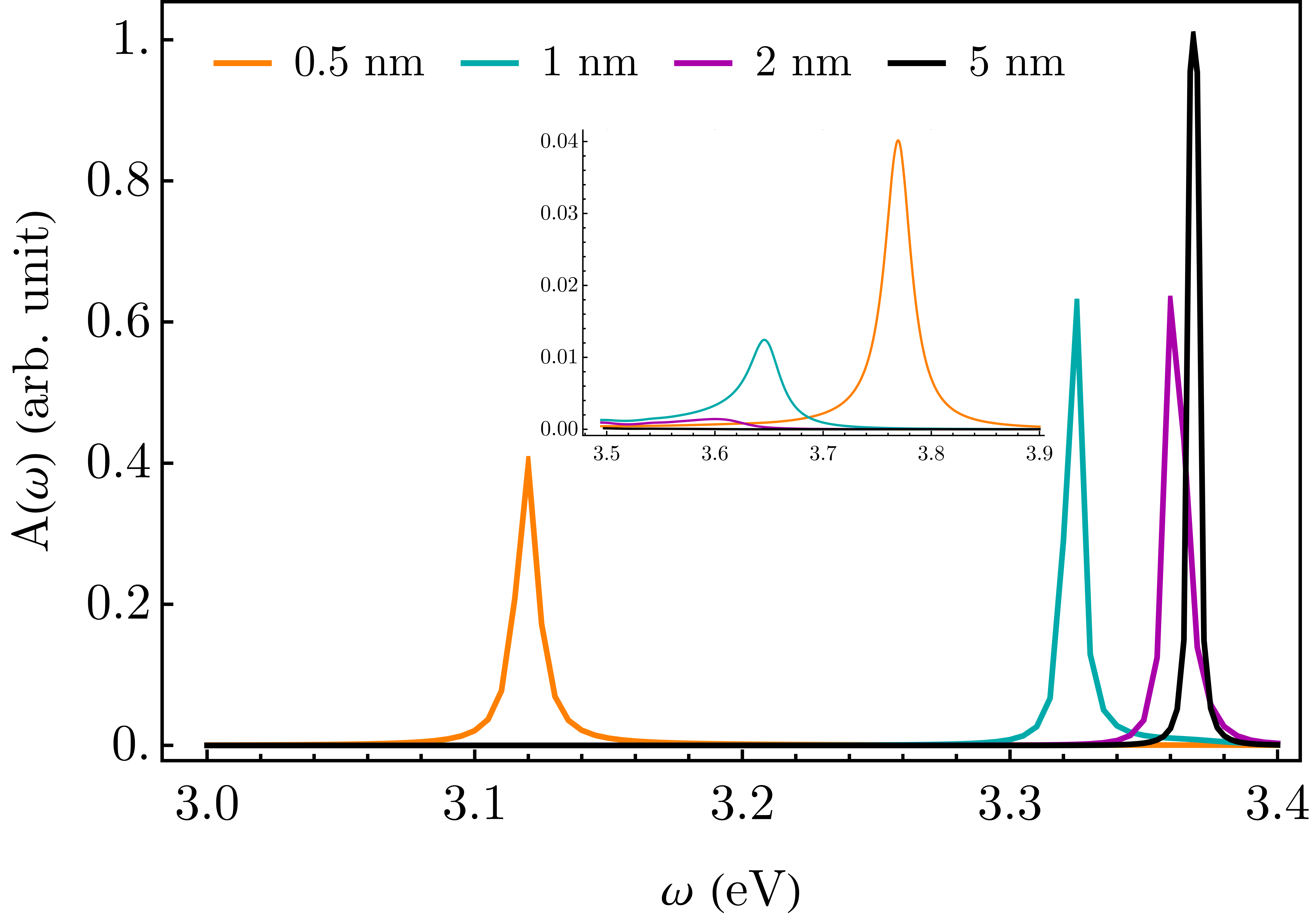}\label{fig:SI_SC_Abs_Lnshp_sphere_perp_silver_10D_a0_10_r1_0.5_5}}

\caption{The absorption lineshape, $A(\omega)$, as a function of the driving
frequency $\omega$ for a molecule near a silver nanosphere of radius
$10$ nm, perpendicular to the sphere surface. The relaxation rate
of the free molecule is taken to be twice the corresponding radiative
relaxation rate (the latter are $6.29\times10^{6}$ $\text{s}^{-1}$,
$2.52\times10^{7}$ $\text{s}^{-1}$, $5.66\times10^{7}$ $\text{s}^{-1}$,
$1.01\times10^{8}$ $\text{s}^{-1}$ and $1.57\times10^{8}$ $\text{s}^{-1}$
for $D_{12}=2,4,6,8,10$ D respectively). (a) An emitter with varying
dipole moment placed at a distance $0.5$ nm from the sphere surface.
The inset shows a closeup of the small higher energy peak structure.
(b) An emitter with a transition dipole moment $10$ D is placed at
different distances from the sphere surface. The inset displays the
small higher energy peak structures (not shown in the main figure).}
\label{fig:alter_treatment_SC_sphere_silver}
\end{figure}

\subsection{Plasmon frequency of a spheroidal metal nanoparticle as altered by
the shape\label{subsec_SI:plasmon_freq_spheroids}}

The optical response of a small spheroid particles is characterized
by plasmon oscillations along the long and short axes. Following Ref.
(\citen{link_shape_2000}) we refer to them as longitudinal and transversal,
respectively. The dipolar plasmon resonance frequencies are shown
as functions of the spheroidal aspect ratio in Figure \ref{fig:plasmon_freq_aspect_ratio_spheroid_gold}.
\begin{center}
\begin{figure}[H]
\includegraphics[scale=0.65]{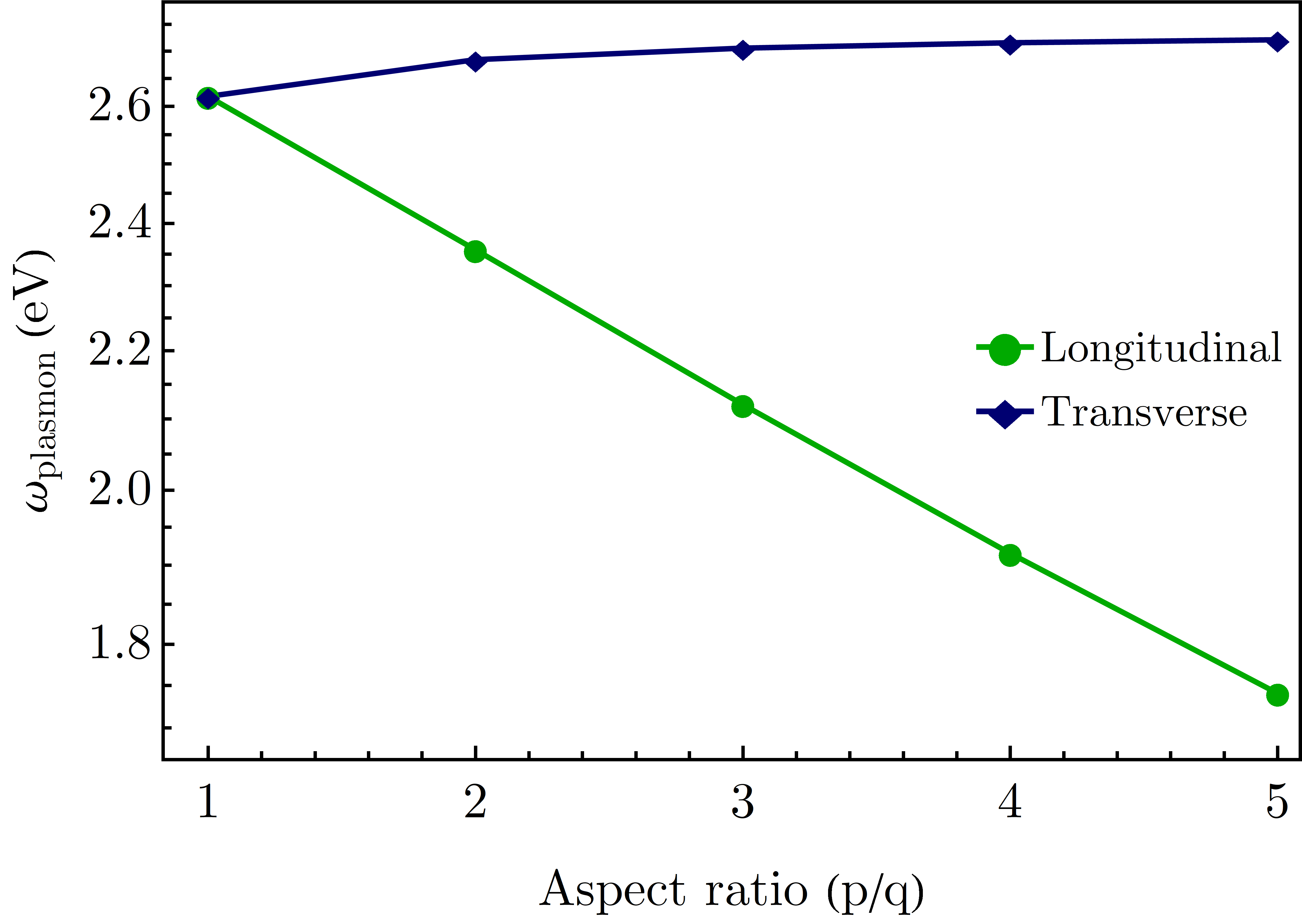}

\caption{Dipolar plasmon resonance frequencies $\omega_{plasmon}$ of a gold
nanoparticle with prolate spheroidal geometry as a function of aspect
ratio $p/q$ for longitudinal (green circles) and transverse (blue
diamonds) plasmon resonance.}
\label{fig:plasmon_freq_aspect_ratio_spheroid_gold}
\end{figure}
\par\end{center}

\bibliography{bibliography/library}